\documentclass[10pt]{article}
\usepackage[english]{babel}
\usepackage{amsmath}
\usepackage{amssymb}
\usepackage{amsfonts}
\usepackage{amsthm}
\usepackage{graphicx}
\usepackage{cool}
\usepackage{coolstr}
\usepackage{coollist}
\usepackage{forloop}
\usepackage{tikz}
\usepackage{cite}
\usepackage{hyperref}

\theoremstyle{plain}

\newtheorem*{proposition*}{Proposition}

\newtheorem*{corollary*}{Corollary}
\newtheorem*{Corollary*}{Important corollary}

\theoremstyle{definition}

\theoremstyle{remark}

\newtheorem*{remark*}{Remark}

\newtheorem*{example*}{Example}

\newtheorem*{examples*}{Examples}

\numberwithin{equation}{section}
\numberwithin{lemma}{section}
\numberwithin{theorem}{section}
\numberwithin{hypothesis}{section}
\numberwithin{definition}{section}
\numberwithin{example}{section}
\numberwithin{corollary}{section}
\numberwithin{remark}{section}
\usepackage[notcite,notref,color]{showkeys}
\graphicspath{{./Figures/}}

\begin{document}

\title{Impact of the tangential traction for radial hydraulic fracture}
\author{D.~Peck$^{(1,*)}$ \& G. Da~Fies$^{(2)}$\\
{\it $^{(1)}$Department of Mathematics, Aberystwyth University, }\\
{\it Aberystwyth, Wales, United Kingdom}
\\
{\it $^{(2)}$Rockfield Ltd, Swansea, UK}
\\
{\it $^{(*)}$ Corresponding author: dtp@aber.ac.uk}
}
\date{}
\maketitle

\begin{abstract}
The radial (penny-shaped) model of hydraulic fracture is considered. The tangential traction on the fracture walls is incorporated, including an updated evaluation of the energy release rate (fracture criterion), system asymptotics and the need to account for stagnant zone formation near the injection point. The impact of incorporating the shear stress on the construction of solvers, and the effectiveness of approximating system parameters using the first term of the crack tip asymptotics, is discussed. A full quantitative investigation of the impact of tangential traction on solution is undertaken, utilizing an extremely effective (in-house build) adaptive time-space solver.  
\end{abstract}


\section{Introduction}

Hydraulic fracture (HF) involves a fluid driven crack propagating in a solid material. This process is widely studied, due to it's appearance in nature, for example in subglacial drainage and the flow of magma in the Earth's crust, as well as it's use in energy technologies, most notably geothermal energy, unconventional hydrocarbon extraction and in the relatively new process of carbon sequestration. While many advanced models exist of this phenomena, the 1D models of hydraulic fracture developed in the 1950's and 1960's: PKN, KGD and radial (penny-shaped), still maintain their relevance. 
This is particularly true when it comes to examining the roles certain physical effects play in determining the fracture behaviour. 

One approach to updating the 1D models is the recent drive to better describe the behaviour of the fluid which drives the fracture. This has previously been considered as either purely Newtonian or as following a power-law description (see eg. \cite{Peck2018b,Perkowska2016}), however recent works attempt to incorporate a truncated power-law \cite{Lavrov2015}, Herschel-Bulkley law \cite{Kanin2021ARH}, or a Carreau fluid description \cite{Wrobel2020} into HF models. Other major developments in this area have involved approaches which provide a better description the influence of proppant (particles within the fluid) on the apparent viscosity of the fluid \cite{Wrobel2019b} and near front behaviour \cite{BESSMERTNYKH2020107110}, as well as incorporation of turbulence within the fracture fluid \cite{Dontsov2017b,Zolfaghari2019}, plasticity or porosity of the fracture walls \cite{Wrobel2022a,Wrobel2022b,SELVADURAI2021103472}, investigations of the impact of toughness heterogeneity \cite{VariTough,DONTSOV2021108144}, 
amongst others. Of crucial importance for this paper however, is the recent incorporation of shear stress induced by the fluid into the 1D models of HF \cite{Wrobel2017,Shen2018}.

The incorporation of hydraulically induced tangential traction on the fracture walls into the PKN and KGD models was provided in \cite{Wrobel2017}. One crucial result was that, when the shear stress was accounted for, there was no longer a difference in aperture asymptotics between the viscosity and toughness dominated regimes. Given the high dependence of most modern algorithms for modeling hydraulic fracture on these asymptotic terms (see eg. \cite{Peck2018a,Perkowska2016,Peirce2008}), this suggested that significant simplifications could be made to the numerical modeling of hydraulic fracture. In addition, incorporating the hydraulically induced tangential traction can also have a noticeable effect on fracture redirection, as outlined in  \cite{Perkowska2017,Wrobel2019a}, and unstable crack propagation \cite{Shen2020}.

It should also be noted however that the original paper on the incorporation of tangential traction into hydraulic fracture models \cite{Wrobel2017} was not without controversy, sparking significant discussion about whether the tangential traction on the fracture walls needs to be accounted for when modeling hydraulic fracture \cite{Linkov2018a,Linkov2018b,Wrobel2018a}. To ensure the presented paper addresses the key aspects of this discussion, here a full quantitative analysis of the time-dependent case is provided in Sect.~\ref{Sect:4}.

The paper is arranged as follows. The problem formulation of the radial model incorporating the tangential traction is outlined in Sect.~\ref{Sect:ProbForm}, including the updated elasticity equation, fracture criterion and system asymptotics for the viscosity dominated regime, as well as modifying the shear stress formulation at the injection point. Next, in Sect.~\ref{Sect:3} the self-similar formulation is used to examine the effect of the updated formulation on the construction of the algorithm, most notably the effect of the changed system asymptotics. Finally, in Sect.~\ref{Sect:4} a full quantitative investigation of the impact of the shear stress for the time dependent formulation is conducted, and the applications for which it may play a role are discussed. A summary of the most important results is given in the concluding Sect.~\ref{Sect:Conclusions}.

\section{Problem formulation}\label{Sect:ProbForm}

\subsection{Governing equations}

We consider the case of a radial hydraulic fracture, driven by a Newtonian fluid. The system is considered in cylindrical coordinates $\{r,\theta , z\}$. The crack dimensions are given by $l(t), w(r,t)$, describing the fracture radius and aperture respectively. The fracture is driven by a point source located at the origin, with known pumping rate: $Q_0 (t)$. Due to the axisymmetric nature of the problem, the solution will be independent of $\theta$, and only $0\leq r \leq l(t)$ needs to be considered. 

The fluid mass balance equation is as follows:
\begin{equation}
\frac{\partial w}{\partial t} + \frac{1}{r} \frac{\partial}{\partial r}\left( r q \right) + q_l = 0 , \quad 0<r<l(t).
\label{Nobelfluidmass}
\end{equation}
where $q_l (r,t)$ is the fluid leak-off function, representing the volumetric fluid loss to the rock formation in the direction perpendicular to the crack surface per unit length of the fracture. Throughout this paper we will assume it to be predefined and bounded at the fracture tip.

Meanwhile $q(r,t)$ is the fluid flow rate inside the crack, for a Newtonian fluid, is given by the Poiseuille law:
\begin{equation}
q =-\frac{w^{3}}{M}\frac{\partial p}{\partial r} ,
\label{NobelPoiseville}
\end{equation}
where the constant $M=12\mu$ is the fluid consistency index. 

The elasticity relation defining the deformation of the rock needs to be updated to incorporate the effect of tangential traction on the crack faces, with the derivation provided in the supplementary material (first provided by the authors in \cite{PeckThesis}, with a similar form also derived independently in \cite{Shen2018}). The elasticity equation takes the form: 
\begin{equation}
 p(r,t)= - \frac{1}{l(t)} \int_0^1 \left[ k_2 \frac{\partial w(\rho l(t))}{\partial \rho} - k_1 l(t) \tau (\rho l(t)) \right] {\cal M}\left(\frac{r}{l(t)},\rho\right) \, d\rho , \quad 0\leq r < l(t) ,
 \label{newPre2}
\end{equation}
with its inverse:
\begin{equation} \label{Nobel_InvElast}
 \begin{aligned}
k_2 w(r,t) + &k_1 \int_{r}^{l(t)} \tau \left(s ,t\right) \, ds = \quad \quad \quad \quad \quad \\
& \frac{4}{\pi^2} l(t) \left[ \underbrace{\int_0^1 \frac{\partial p(y l(t),t)}{\partial y} {\cal K}\left(y, \frac{r}{l(t)}\right) \, dy}_{w_1(r,t)} + \underbrace{\sqrt{1-\left(\frac{r}{l(t)}\right)^2} \int_0^1 \frac{\eta p(\eta l(t) , t)}{\sqrt{1-\eta^2}} \, d\eta}_{w_2(r,t)} \right]  ,
 \end{aligned}
 \end{equation}
where the kernel functions are given by:
\begin{equation}
{\cal M}\left[\tilde{r} , \rho \right] = \begin{cases} \frac{1}{\tilde{r}} \EllipticK{\frac{\rho^2}{\tilde{r}^2}} + \frac{\tilde{r}}{\rho^2 - \tilde{r}^2} \EllipticE{ \frac{\rho^2}{\tilde{r}^2}} , & \tilde{r}>\rho \\  \frac{\rho}{\rho^2 - \tilde{r}^2} \EllipticE{ \frac{\tilde{r}^2}{\rho^2}} , & \rho>\tilde{r} , \end{cases}
\end{equation}
\begin{equation} \label{Almighty_Kernel_K}
{\cal K}(y,\tilde{r}) = y\left[ \IncEllipticE{\arcsin(y)}{\frac{\tilde{r}^2}{y^2}} - \IncEllipticE{\arcsin(\psi )}{\frac{\tilde{r}^2}{y^2}}\right] , \quad \psi =\min\left(\frac{y}{\tilde{r}},1\right),
\end{equation}
with $E\left( \phi \, | \, m \right)$ denoting the incomplete elliptic integral of the second kind, while:
\begin{equation}
k_1 = \frac{1-2\nu}{\pi (1-\nu)}, \quad k_2 = \frac{E}{2\pi (1-\nu^2)} .
\end{equation}
Note that if we take $k_1 = 0$ (ie. $\nu=0.5$), this is identical to the `classical' elasticity equation. 

We can also utilize the elasticity equation to parameterise the fracture regime, as outlined in \cite{VariTough}. Note that in \eqref{Nobel_InvElast}, the fracture aperture $w$ can be represented as the sum of the term denoted $w_2$, which represents the impact of the material toughness $K_{Ic}$, and $w_1$, representing the contribution of the (viscous) fluid pressure, alongside some final shear term. Consequently, we can define the associate volumes
\begin{equation}
V_v (t) = 2\pi \int_0^{l(t)} r w_1 (r,t) \, dr , \quad V_T (t) = 2\pi \int_0^{l(t)} r w_2 (r,t) \, dr .
\end{equation}
The ratio of these two terms
\begin{equation} \label{defn_delta}
\delta (t) = \frac{V_T(t)}{V_v (t)},
\end{equation}
will provide a (rough) measure of the extent to which fracture evolution is governed by the fluid viscosity or the material toughness. This can therefore be used to parameterise whether the fracture is within the viscosity ($0\leq \delta \ll 1$), transient ($\delta \sim 1$), or toughness ($1\gg \delta$) dominated regime, which will prove useful when conducting the time-dependent investigation. Note that for the radial model this will change over time, as the fracture transitions from the (initially) viscosity dominated to the toughness dominated regime as it grows (see e.g.\ \cite{Savitski2002,Lecampion2017a,Dontsov2017} for details of the fracture regimes). For more details of the parameterisation by $\delta(t)$, see \cite{VariTough}.

These equations are supplemented by the boundary condition at $r=0$, which defines the intensity of the fluid source, $Q_0$:
\begin{equation}
 \lim_{r\to 0} r q(r,t) = \frac{Q_0 (t)}{2\pi} ,
 \label{Nobel_SourceIntense}
\end{equation}
alongside the tip boundary conditions:
\begin{equation}
w(l(t),t)=0 , \quad q(l(t) , t) = 0.
 \label{Nobel_TipBC}
\end{equation}
We assume that there is a preexisting fracture, starting with appropriate non-zero initial conditions for the crack opening and length:
\begin{equation}
w(r,0) = w_* (r) , \quad l(0) = l_0 ,
 \label{Nobel_InitCond}
\end{equation}

Finally the global balance equation takes the form:
\begin{equation}
\int_0^{l(t)} r\left[ w(r,t) - w_* (r) \right] \, dr \, + \int_0^t \int_0^{l(t)} r q_l (r,\tau) \, dr \, d\tau = \frac{1}{2\pi} \int_0^t Q_0 (\tau) \, d\tau .
\label{Nobel_fluidbalance1}
\end{equation}

In addition to the above, we employ a new dependent variable named the fluid velocity, $v$, defined by:
\begin{equation}
v(r,t) = \frac{q(r,t)}{w(r,t)} = -\frac{w^2 (r,t)}{M}\frac{\partial p}{\partial r} ,
\label{NobelPVInit}
\end{equation} 
It has the property that, provided the fluid leak-off $q_l$ is finite at the crack tip:
\begin{equation}
\lim_{r\to l(t)} v(r,t) = v_0 (t) < \infty ,
\end{equation}
which, given that the fracture apex coincides with the fluid front (no lag), allows for fracture front tracing through the so-called speed equation \cite{Linkov2011}:
\begin{equation}
\frac{dl}{dt} = v_0 (t).
 \label{Nobel_SpeedEq}
\end{equation}
Note that this replaces boundary condition \eqref{Nobel_TipBC}$_2$, which now immediately follows from \eqref{Nobel_TipBC}$_1$, \eqref{NobelPVInit}-\eqref{Nobel_SpeedEq}. This Stefan-type condition has previously been employed in 1D hydraulic fracture models, the advantages of which (alongside technical details) are shown in \cite{Kusmierczyk,Perkowska2016,Wrobel2013,Wrobel2015,Wrobel2017}. Of crucial importance is the fact that the fracture tip can now be considered in terms of the finite variable $v$, with clearly defined leading asymptotic coefficient $v_0$, eliminating the singular term $q$ from computations entirely. These singular terms are however closely related to the fluid velocity \eqref{NobelPVInit}, and as such can easily be obtained in post-processing.

\subsection{The shear stress at the fracture inlet}\label{The_wall_jet}

The normal and tangential stress on the fracture walls, created by the fluid pressure, follows directly from lubrication theory (see for example \cite{Tsai}), in this case being given by:
\begin{equation}
\sigma_0 =- p , \quad \tau (r,t) = - \frac{1}{2} w(r,t) \frac{\partial p(r,t)}{\partial r} .
\label{taudef}
\end{equation}
It should be noted that this representation of the shear stress is singular at both the crack tip ($r=l(t)$) and the fracture opening ($r=0$). While the former singularity is physically meaningful for defining the total flux within the fracture, following the same principals as that for the stress at the crack tip in linear elastic fracture mechanics, the singularity at $r=0$ should be properly addressed. 
 \begin{figure}[t]
  \centering
  \begin{tikzpicture}[scale=1.3]
    \draw[black] (-4,2.5) .. controls (1,1.35) and (2,0.9) .. (2.85,0); 
  \draw [black,thick,dotted,->] (-4,0) -- (-4,3); 
      \node at (-4,3.25) {$z$}; 
  \draw [black,->] (-4,0) -- (3.5,0); 
      \node at (3.75,0) {$r$}; 
      \draw [fill, blue] (-4,0) circle [radius=0.05]; 
  \draw[black,->] (-0.4,1.75) -- (0.5,1.47); 
  \node at (0.2,1.8) {$\tau$}; 
  \draw[black,->] (-1,1.1) -- (-1,1.6); 
    \draw[black,->] (-0.8,1.05) -- (-0.8,1.55); 
      \draw[black,->] (-0.6,1) -- (-0.6,1.5); 
       \node at (-0.8,0.8) {$p$};  
 \draw[black,<->] (-4.2,0.05) -- (-4.2,2.45); 
  \node at (-4.45,1.25) {$w$}; 
  \draw[black,<->] (-3.9,-0.1) -- (2.85,-0.1); 
  \node at (-0.5,-0.4) {$l$}; 
  \draw[blue,->] (-4,0.05)  .. controls (-4,0.8) and (-3.6,1) .. (-3,1.4); 
  \draw [blue,->] (-3,1.4) .. controls (-2.5,1.75) and (-2,2) .. (-1.5,1.8); 
   \draw [blue,->] (-1.5,1.8) -- (1,1.06); 
    \draw [blue,->] (1,1.06) .. controls (2,0.7) and (2.35,0.4) .. (2.65,0.05); 
    \draw [red,->] (-3.85,2) -- (-3.85,1.5); 
    \draw [red,->] (-3.85,1.5) .. controls (-3.85,1) and (-3.25,1.25) .. (-3,1.5); 
      \draw [red,->] (-3,1.5) .. controls (-2.6,1.75) and (-2.5,1.95) .. (-3,2.1); 
        \draw [red] (-3,2.1) .. controls (-3.75,2.3) and (-3.8,2.45) .. (-3.85,2); 
  \end{tikzpicture}
  \caption{Exaggerated depiction of the primary streamlines within a quarter-segment of a penny-shaped hydraulic fracture, which determine the tangential traction on the fracture walls. The red line indicates the longest streamline within the stagnant zone (wall-jet effect), while the blue line indicates the longest streamline connecting the fluid source (blue dot at $r=0$) to the fracture tip.}
   \label{Jet_effect_fig}
\end{figure}
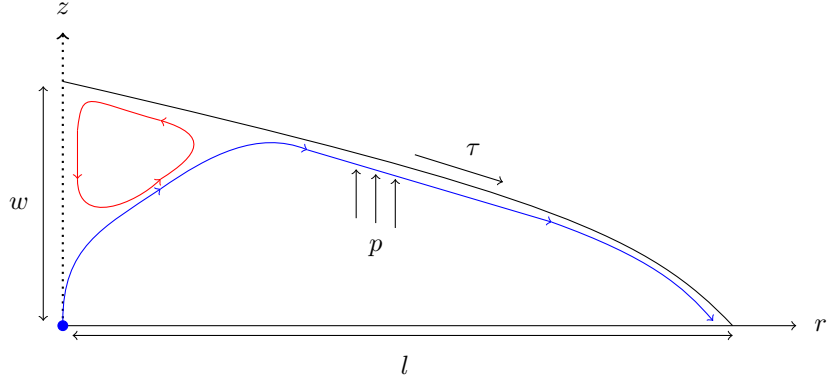
  
There is a clear explanation for the singularity at the fracture opening. HF models typically treat the fluid source as a singularity at the fracture inlet $(r,\theta,z)=(0,\theta,0)$. Tangential traction is induced by fluid traveling in a single (turbulence-free) streamline from this source directly to the fracture wall, and along this wall to the fracture front. However, this behaviour is a clear violation of established rules for fluids in such situations, where it has been demonstrated that instead stagnant regions will form in the region where the fluid source makes contact with the fracture wall $(r,\theta,z)=(0,\theta,\pm w(0,t)$, preventing fluid from the source from reaching these points (see Fig.~\ref{Jet_effect_fig}). These secondary streamlines will typically be stable, even though it arises from turbulent effects acting on the fluid, however its precise form will depend upon both the problem geometry and fluid properties (Reynold's number). This can be thought of as a form of the `wall jet' effect, analogous to the behaviour of a rocket exhaust hitting the ground (reviews can be found in \cite{Gauntner1970SurveyOL,Launder1983}). 

Consequently, while the singularity at the fracture front needs to be maintained to properly model the radial geometry, the formulation needs to updated to eliminate this non-physical singularity at $r=0$. There are three primary options for doing so:
\begin{itemize}
 \item {\bf Incorporating the wellbore} will (artificially) cut-off the current left-hand boundary ($r=0$), with the fluid flow instead ending some distance away from the origin (the half-width of the wellbore), and thus remove the singularity. This has previously been incorporated for the classical radial model, for example in \cite{Lecampion2017a} where it effectively predicted experimental results.
 \item {\bf Fixing the opening height} by adding an additional boundary condition such that $w(0,t)=w_*(0)$, a constant, where $w_* (r)$ is the initial fracture profile \eqref{Nobel_InitCond}. This could be enforced numerically, and would eliminate the effect of the tangential traction at the crack opening.
 \item {\bf Modifying the tangential traction formulation} to eliminate the singularity at $r=0$ from \eqref{taudef}. Unfortunately, there is no simple formula to describe the effect of these stagnant zones on the tangential traction induced on the fracture walls. Subsequently, this requires a more general modification, allowing multiple `possible' forms of the shear stress to be considered.
\end{itemize}
As the aim of this paper is to incorporate the tangential traction into the general radial model, rather than for some specific application, we will take the third option and modify the formulation. This has the added benefit of being the most generalised approach, allowing for a different forms of the tangential traction to be investigated. Note however that the other two approaches could be utilized for specific applications, if it were preferable.

In order to control the extent to which the shear stress is changed away from the point $r=0$, we introduce the updated formulation of the tangential stress on the fracture wall: 
\begin{equation}\label{New_tau}
\tau (r,t) = -\frac{1}{2} \frac{\chi (r,t)}{l(t)} w(r,t) \frac{\partial p(r,t)}{\partial r},
\end{equation}
where the particular form of $\chi$ is not fixed (to allow for various possible formulations to be considered), but is always a continuous function such that
\begin{equation}\label{New_rtsar}
 \chi (r,t) \sim r , \quad r \to 0, \qquad \chi (r,t) =  l(t) , \quad r \to l(t)  .
\end{equation}
In this paper we will mimic $\chi$ in the form
\begin{equation} \label{rstar_defn}
\chi (r,t) = l(t) \left[ 1 - \left( 1 - \frac{r}{l(t)} \right)^\beta \right] ,
\end{equation}
where $\beta \geq 1$ is a predefined constant. While we will assume here that $\beta$ is predefined, it will be directly linked to the size of the stagnant zones and can therefore, in principle, be chosen to match the expected behaviour of the tangential traction for a particular problem. An examination of the effect of the choice of $\beta$ on the fracture profile is provided in Sect.~\ref{Sect:InjectionBeta}.

This formulation therefore allows the potential effect of the `wall jet' behaviour to be accounted for, incorporating all expected behaviour of the phenomena, while leaving the tangential traction unchanged away from the fluid inlet. Crucially, the shear stress remains identical to the standard formulation as $r\to l(t)$, so does not effect the evaluation of the crack tip asymptotics or Energy Release Rate. 

In addition, this new formulation resolves the issues related to the fracture inlet asymptotics, creating a fully consistent formulation that can account for the varying possible effects of the stagnant zones at the crack opening. As a result, irresepective of the form of $\chi$, the asymptotics at the crack opening remain identical to those in the case without tangential traction  \cite{Peck2018a}:
\begin{equation}
 \begin{aligned}
w(r,t) &= w_0^{(0)} + w_1^{(0)} r + O\left( r^2 \log(r) \right) , \quad & \\
p(r,t) &= p_0^{(0)} \log (r) + p_1^{(0)} + O\left( r \right) ,\quad &r \to 0, \\
\tau(r,t) &= \tau_0^{(0)} + \tau_1^{(0)} r + O\left(r\log(r)\right) . \quad & \\
\end{aligned}
\end{equation}

\subsection{Crack tip asymptotics}\label{Sect:Crack_tip_Asymptotics}

In the classic radial model the basic modes of fracture propagation are related to the energy dissipation throughout the fracture, and thus can influence the tip asymptotics. Typically, fractures will begin in the viscosity dominated regime and transition to the toughness dominated regime over time, although the particular regime depends upon the system parameters (particularly $K_{Ic}$ and $\mu$). These two modes have been extensively studied, and have qualitatively different asymptotic behaviour, leading to a singular perturbation problem when transitioning between the cases. In the revised HF formulation however this problem is eliminated, as the introduction of the shear stress ensures that the tip asymptotics remain the same irrespective of the regime.

The revised crack tip asymptotics are the same irrespective of the regime, and coincide with those for the toughness dominated regime in the classical model (assuming no fluid lag)  \cite{Wrobel2017}:
\begin{equation}
w(r,t) = w_0(t) \sqrt{1-\tilde{r}}+w_1(t) \left(1-\tilde{r}\right)+w_2(t) \left(1-\tilde{r}\right)^{\frac{3}{2}} \log\left(1-\tilde{r}\right)+\hdots, \quad \tilde{r} = \frac{r}{l(t)} \to 1,
 \label{wasym1}
\end{equation}
\begin{equation}
p(r,t) = p_0(t) \log\left(1-\tilde{r}\right)+p_1(t)+p_2(t)\sqrt{1-\tilde{r}}+p_3(t) \left(1-\tilde{r}\right)\log\left(1-\tilde{r}\right)+\hdots, \quad 
\tilde{r} = \frac{r}{l(t)} \to 1,
\label{pasym1}
\end{equation}
additionally, we immediately have the following asymptotics for the fluid velocity and shear stress:
\begin{equation}
v(r,t) = v_0(t) + v_1 (t) \sqrt{1-\tilde{r}} + \hdots , \quad \tilde{r} = \frac{r}{l(t)} \to 1,
\label{vasym1}
\end{equation}
\begin{equation}
\tau (r,t) = \frac{\tau_0}{\sqrt{1-\tilde{r}}} + \tau_1 + \hdots , \quad \tilde{r}=\frac{r}{l(t)}\to 1,
\label{tasym1}
\end{equation}
where:
\begin{equation}
v_0(t) = \frac{w_0^2 (t) p_0(t)}{M l(t)} , \quad v_1(t) = \frac{w_0^2 (t) p_2 (t) + 4 w_0 (t) w_1 (t) p_0 (t)}{2M l(t)} , 
\label{Nobel_v0InitBruv}
\end{equation}
\begin{equation}
\tau_0 (t) = \frac{w_0 (t) p_0(t)}{2 l(t)} , \quad \tau_1 (t) = \frac{w_0(t) p_2 (t) + 2w_1 (t) p_0 (t)}{4 l(t)} .
\end{equation}  
This yields the relation between the coefficients:
\begin{equation}
v_0 (t) = \frac{2}{M} w_0 (t) \tau_0 (t), \quad v_1 (t) = \frac{2}{M}\left[ w_0 (t) \tau_1 (t) + w_1 (t) \tau_0 (t) \right].
\end{equation}

Note that by evaluating the elasticity equation \eqref{Nobel_InvElast} at the crack tip, noting the asymptotics above, we obtain:
\begin{equation} \label{Nobel_SI_Int}
k_2 w_0 (t) + k_1 w_0 (t) p_0 (t) = \frac{4\sqrt{2}}{\pi^2} l(t) \int_0^1 \frac{\eta p(\eta l(t) , t)}{\sqrt{1-\eta^2}} \, d\eta,
\end{equation}
which replaces the standard integral definition of the stress intensity factor.

Finally, combining the speed equation \eqref{Nobel_SpeedEq} with \eqref{Nobel_v0InitBruv} yields:
\begin{equation}
 \frac{dl}{dt} =  \frac{w_0^2 (t) p_0(t)}{M l(t)} ,
\end{equation}
which can be integrated directly to determine the crack length:
\begin{equation}
l(t) = \sqrt{ l^2 (0) + \frac{1}{M} \int_0^t w_0^2 (s) p_0(s) \, ds}.
\end{equation}

\subsection{Energy release rate}\label{Sect:ERR}

It has previously been shown that the crack tip asymptotics play a crucial role in the behaviour of a hydraulic fracture \cite{Garagash2000,Savitski2002}. As such these must be examined in more detail, which is achieved through an examination of the Energy Release Rate (ERR), accounting for the effect of tangential traction. An updated form of Linear Elastic Fracture Mechanics to provide the Energy Release Rate accounting for tangential traction is provided in \cite{Piccolroaz2021}, while a summary of results specific to the radial model from \cite{Wrobel2017,Perkowska2017} are provided below.

We have that
\begin{equation}\label{Energy_Release_Rate}
K_{Ic}^2 = K_I^2 + 4(1-\nu) K_I K_f .
\end{equation}

The form of the first term of the apertures asymptotic representation \eqref{wasym1} is as follows:
\begin{equation}
w_0 (t) = \gamma\sqrt{l(t)} \left(K_I (t) + K_f (t)\right) , \quad K_f = B^{-1} \sqrt{M v_0 (t) p_0 (t)l(t)}, \quad B=\frac{2\sqrt{2}}{\sqrt{\pi}}(1-\nu) ,
 \label{w02}
\end{equation}
where:
\begin{equation}
\gamma = \frac{8}{\sqrt{2\pi}} \frac{(1-\nu^2)}{E} , 
\end{equation}
Here the term $K_f$ is denoted the shear stress intensity factor. 

\begin{equation}
K_I = \frac{K_{Ic}}{\sqrt{1+4(1-\nu)\bar{\omega}}} , \quad K_f = \frac{K_{Ic}\bar{\omega}}{\sqrt{1+4(1-\nu)\bar{\omega}}} , \quad \bar{\omega} = \frac{p_0}{G-p_0} ,
 \label{Nobel_StessIntense}
\end{equation}
where $G$ is the shear modulus and $p_0$ is the first term of the pressures asymptotic representation at the fracture front \eqref{pasym1}. As such we can represent \eqref{w02} in the following form:
\begin{equation}
w_0 (t) = \sqrt{l(t)}  \frac{\gamma (1+\bar{\omega})}{\sqrt{1+4(1-\nu)\bar{\omega}}} K_{Ic} .
 \label{Nobel_w03}
\end{equation}
It is clear from the above and \eqref{Nobel_StessIntense}$_3$ that we must have:
\begin{equation} \label{p0_bounds}
0 < p_0 (t) < G .
\end{equation}

Combining the above with the speed equation \eqref{Nobel_SpeedEq}, we have:
\begin{equation}\label{237}
\frac{1}{\gamma^2 l(t) K_{Ic}^2 (t) } v_0 (t)  = \frac{p_0 (t)}{M} F\left(  p_0 (t) \right) ,
\end{equation}
where:
\begin{equation}
F\left( p_0 (t) \right) = \frac{G^2}{\left[ G - p_0 (t) \right] \left[ G + \left(3-4\nu \right) p_0 (t) \right]} .
\label{FdefOld}
\end{equation}
It is worth noting that in \eqref{237} the right-hand side is a monotonically increasing function from zero (when $p_0 = 0$) to infinity (when $p_0 = G$). Consequently, the solution for $p_0$ is unique, and can be found as a function of $v_0$, $K_{Ic}$ and $l(t)$ (or similarly for $v_0$).

Using the above notation, we can also rewrite \eqref{Nobel_w03} as:
\begin{equation}
w_0 (t) = \gamma K_{Ic} \sqrt{l(t) F\left(p_0 (t)\right)}.
\label{Nobel_w0344}
\end{equation}
Note that unlike with \eqref{237}, the right-hand side of \eqref{Nobel_w0344} is not monotonic with respect to $p_0$. Note that $F(0)=1$, while the right-hand side subsequently decreases until $p_0 (t) = (1-2\nu) G / (3-4 \nu)$, before beginning to increase and tending to infinity as $p_0 \to G$.

\section{Effect on algorithm construction}\label{Sect:3}

Incorporating the tangential traction, in particular the updated fracture criterion \eqref{Energy_Release_Rate} and system asymptotics (see Sect.~\ref{Sect:Crack_tip_Asymptotics}-\ref{Sect:ERR}), fundamentally alters the construction of algorthims for generating solutions to the radial model. We investigate the consequenes of this change using the self-similar formulation, as this simple case allows for the clearest results. It is not possible to obtain a power-law type solution, so instead an exponential variant must be obtained, similar to that utilized in \cite{Spence1985}. We normalise the problem as
\begin{equation} \label{Normalisation1}
\tilde{r} = \frac{r}{l(t)} , \quad \tilde{t}=\frac{t}{t_n},  \quad t_n=\frac{M}{k_2},
\end{equation}
where $\tilde{r}\in \left[0,1\right]$, before utilizing the following separation of variables
\begin{equation}
\tilde{Q}_0 (\tilde{t}) = \hat{Q}_0 e^{2 \Upsilon \tilde{t}} ,
\label{SSdef1Nobel}
\end{equation}
for some chosen constant $\Upsilon$. The full normalised and self-similar problem formulations are provided in the supplementary material. It is important to note that the self-similar equations still feature the Poisson's ratio $\nu$, self-similar fracture toughness $\hat{K}_{Ic}$, self-similar injection rate $\hat{Q}_0$ and parameter $\beta$ describing the shear near the fracture inlet \eqref{New_tau} - \eqref{rstar_defn}, while the remaining material constants are eliminated from the governing equations. The values of the self-similar constants used in simulations (unless stated otherwise) are provided in Table.~\ref{Table:SSConst}. For the remainder of this section, the `$\wedge$' symbol will be used to denote self-similar parameters (e.g. $\hat{w}(\tilde{r})$ for the self-similar aperture).

\begin{table}[b!]
 \centering
 \begin{tabular}{c|c|c|c}
 $\nu$  & $\hat{Q}_0$ & $\Upsilon$ & $\beta$ \\
 \hline 
 &&&\\
 0.3 & 1 & 1/3 & 1 \\
 &&&\\
 \hline 
 \end{tabular}
 \caption{Values of the parameters used in self-similar computations. Here $\beta$ defines the behaviour of the shear stress at the injection point \eqref{New_tau} - \eqref{rstar_defn}.}
 \label{Table:SSConst}
\end{table}

Solutions are obtained using an approach based on the ``universal algorithm'', first introduced in \cite{Wrobel2015}, which is an explicit solver combining rigorous use of the system asymptotics and implementation of the speed equation to trace the fracture front \eqref{Nobel_SpeedEq}, amongst other novelties. This method was previously used by the authors for the radial model \cite{Peck2018a,Peck2018b}, and the reader is directed there for the details of the algorithms construction (alongside \cite{Perkowska2016}). This method stands in contrast to the implicit level set method more common in the literature (see eg. \cite{Dontsov2017,Peirce2008} or the recent open-source general solver \emph{PyFrac} \cite{Zia2019b}), which is typically far more flexible but achieves a lower level of accuracy (for a more complete comparison, see e.g.  \cite{Zia2019a,Linkov2019}). The solver utilized here for the self-similar scheme achieves an exceptionally low level of error against both analytical benchmarks and convergence-based error tests (below $10^{-4}$ across the entire domain when taking $N=300$ nodal points, see \cite{Peck2018a}).

\subsection{Transition from viscosity to toughness dominated regimes}

Typically, when obtaining the solution for the radial model, one of the most important aspects to incorporate is the transition from the viscosity dominated regime to the toughness dominated mode as the fracture develops (a detailed overview of the differing fracture regimes can be found in e.g.\ \cite{Savitski2002,Lecampion2017a,Dontsov2017}). However, as the updated system asymptotics no longer vary between the two regimes when the tangential traction is incorporated, this transition will now occur automatically. 

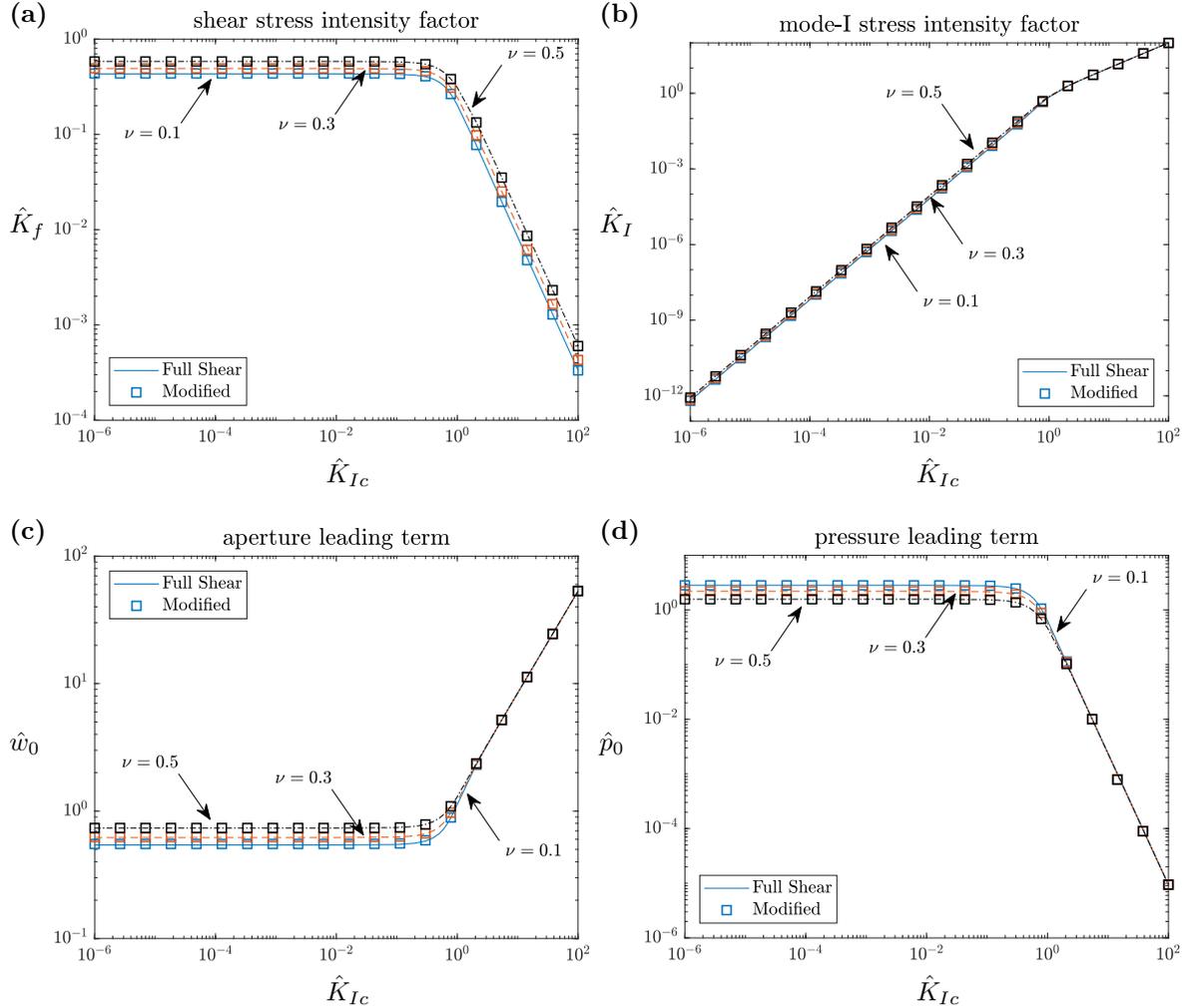
\begin{figure}[b!]
 \center
 \includegraphics[width=0.45\textwidth]{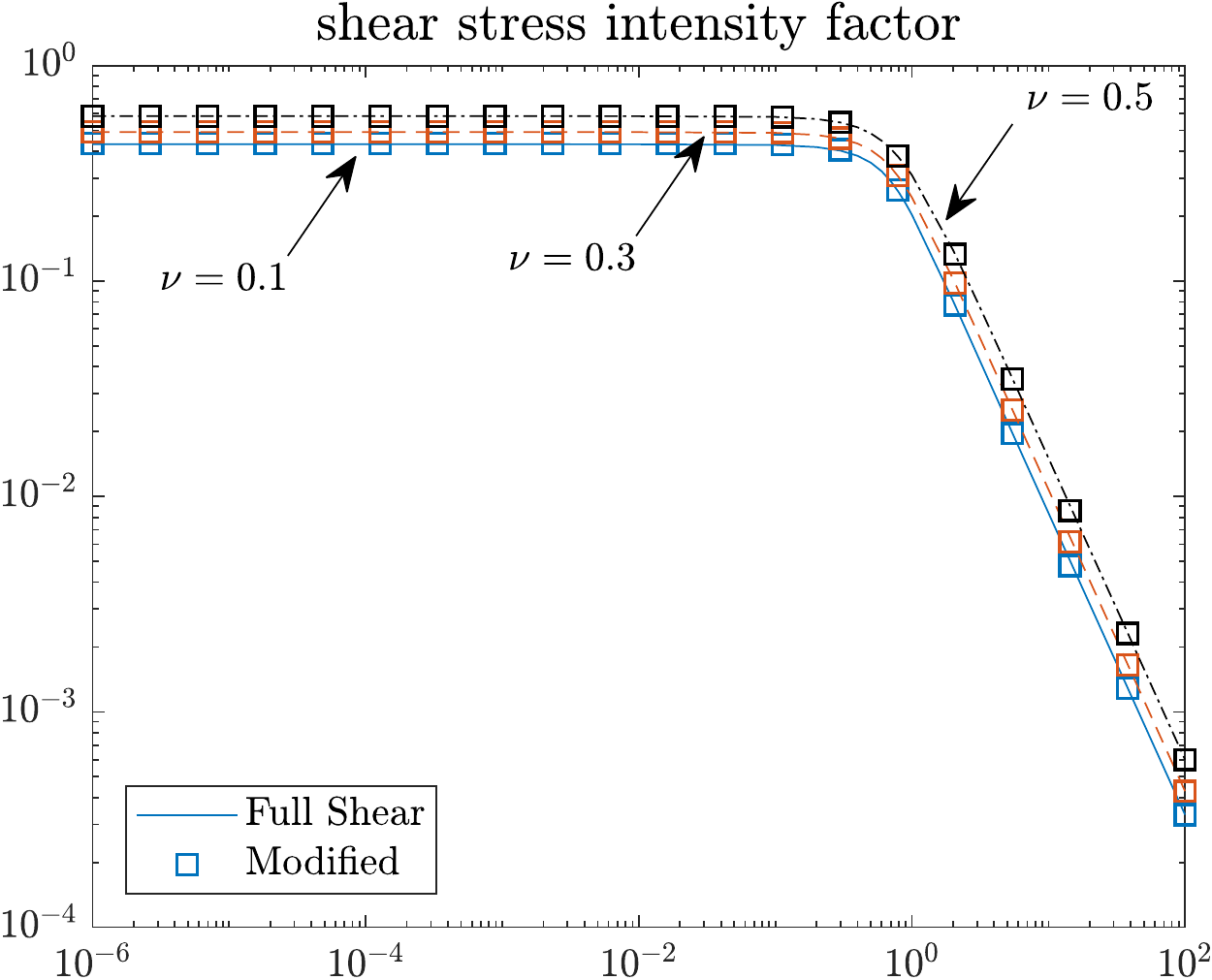}
 \put(-220,160) {{\bf{(a)}}}
 \put(-220,80) {$\hat{K}_f$}
 \put(-100,-15) {$\hat{K}_{Ic}$}
 \hspace{6mm}
 \includegraphics[width=0.45\textwidth]{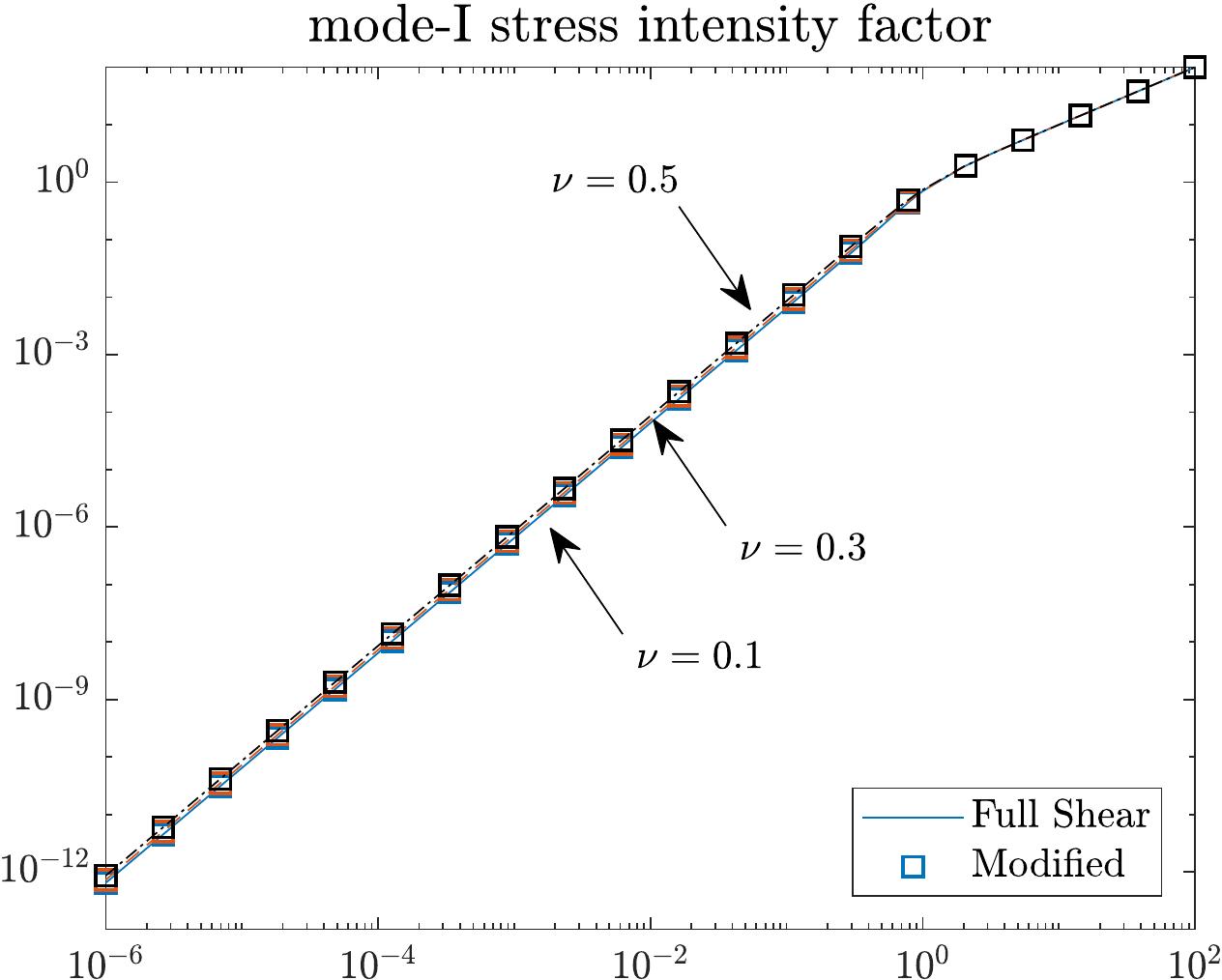}
 \put(-220,160) {{\bf{(b)}}}
 \put(-220,80) {$\hat{K}_I$}
 \put(-100,-15) {$\hat{K}_{Ic}$}

\vspace{4mm}

 \includegraphics[width=0.45\textwidth]{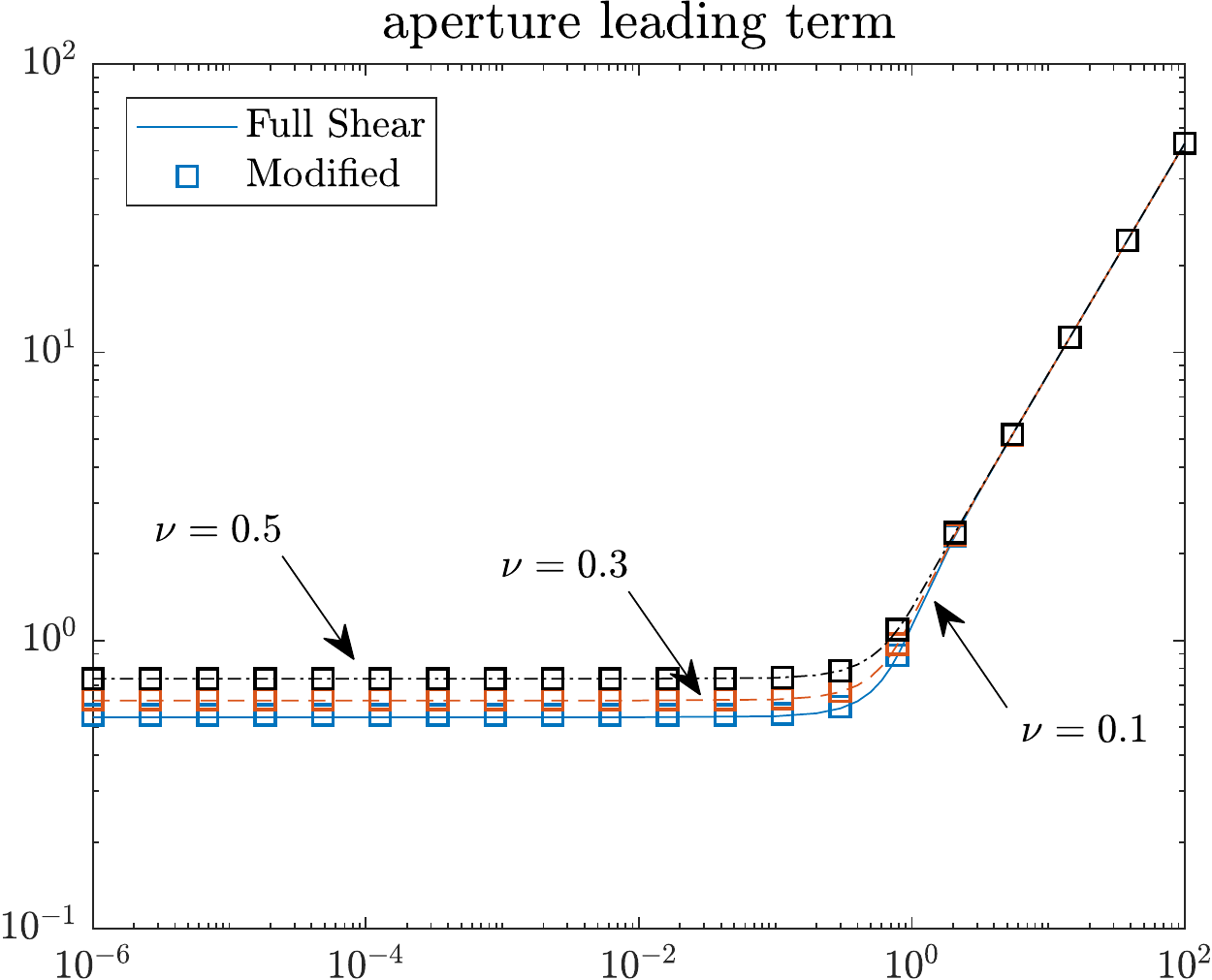}
 \put(-220,160) {{\bf{(c)}}}
 \put(-220,80) {$\hat{w}_0$}
 \put(-100,-15) {$\hat{K}_{Ic}$}
 \hspace{6mm}
 \includegraphics[width=0.45\textwidth]{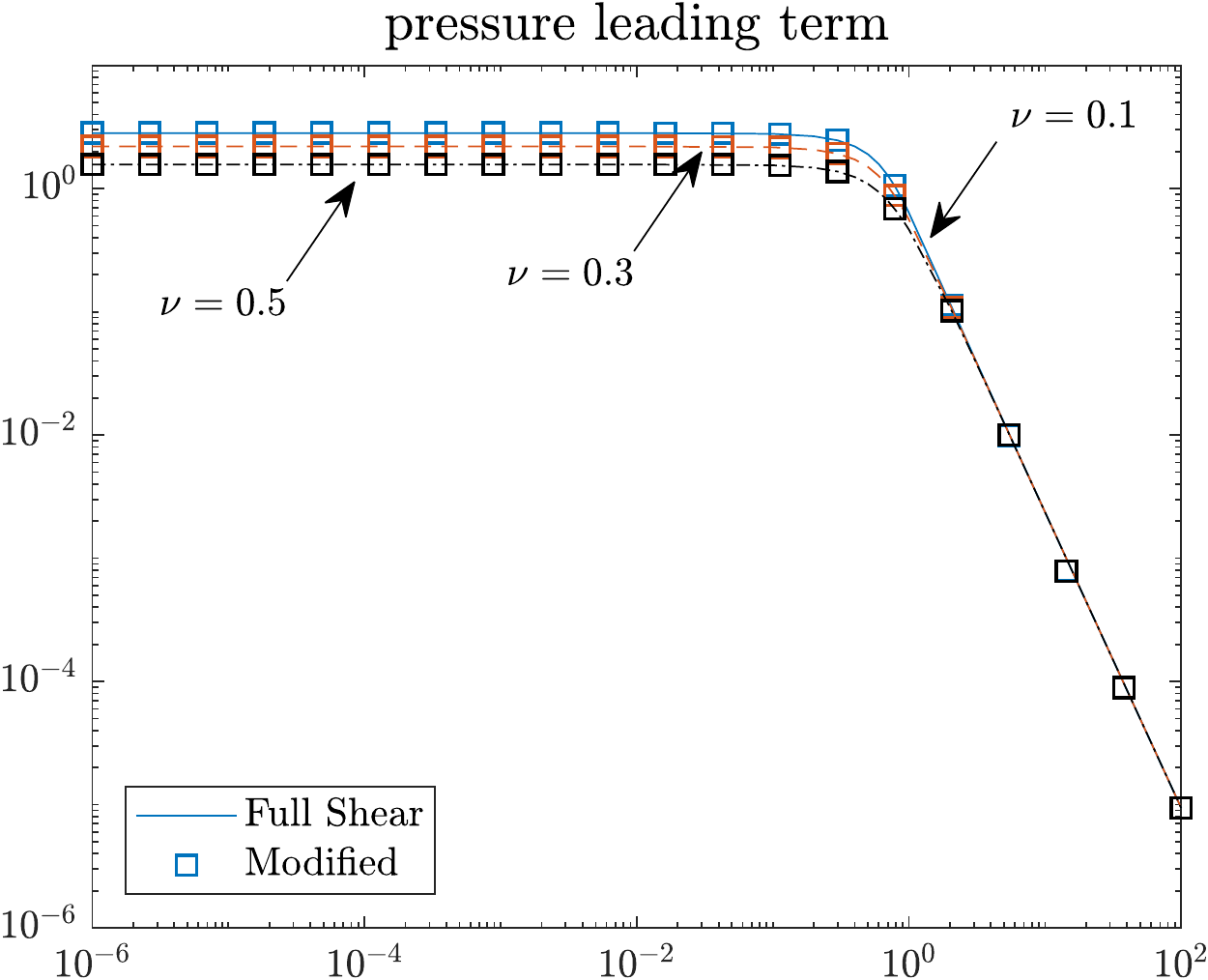}
 \put(-220,160) {{\bf{(d)}}}
 \put(-220,80) {$\hat{p}_0$}
 \put(-100,-15) {$\hat{K}_{Ic}$}
 \caption[The relationship between the self-similar material toughness and the system stress intensity factors]{The relationship between the self-similar material toughness $\hat{K}_{Ic}$ and the system stress intensity factors. Here we show the self-similar forms of: (a) the shear stress indensity factor $\hat{K}_f$, (b) the mode-I stress intensity factor $\hat{K}_I$, and the leading term of the system asymptotics for (c) the aperture $w$, (d) the pressure $p$.}
 \label{Nobel_AutoSwitch}
\end{figure}

As this ``automatic switch'' is a result of the updated asymptotics \eqref{wasym1} - \eqref{tasym1} and fracture criterion \eqref{Energy_Release_Rate}, a modified form of the problem can be considered that avoids having to fully incorporate the updated elasticity equation \eqref{Nobel_InvElast}. To demonstrate this, we consider two variants of the problem
\begin{enumerate}
 \item {\bf Full shear:} This is the full radial model incorporating the tangential traction induced on the fracture walls. Note that in this section we will take $\beta=1$ in the shear stress formulation \eqref{New_tau} - \eqref{rstar_defn}, signifying the minimum potential impact of the shear stress on the fracture behaviour. 
 \item {\bf Modified variant:} This is a reduced form of the radial model with shear stress, but reducing the need to incorporate the updated elasticity equation. There are two possible approaches to achieving this. The first is to neglect the additional term of the elasticity equation (equivalent to taking $k_1=0$), similar to that done for KGD in \cite{Wrobel2017}. For the radial model however, this approach leads to inconsistencies in the asymptotics. For this reason, we instead favour a partial incorporation, in which the updated integral definition of the stress intensity factor is utilized \eqref{Nobel_SI_Int}, but the additional term of the elasticity equation is not. This avoids asymptotic inconsistencies, whilst also avoiding incorporating the elasticity equation in full. This won't effect the `automatic switch', as we continue to utilize the updated fracture criterion 
and system asymptotics. 
\end{enumerate}

The values of the stress intensity factors (mode-I and shear), and the leading asymptotic coefficients for the aperture and pressure, for varying $\hat{K}_{Ic}$ are provided in Fig.~\ref{Nobel_AutoSwitch}. The transition between viscosity and toughness dominated regimes can clearly be seen (starting near to $\hat{K}_{Ic}=1$). It is interesting however to note that, in the viscosity dominated regime, the coefficient $\hat{p}_0$ is almost exactly $\pi (1-\nu)$ (with it being exact for $\hat{K}_{Ic}\equiv 0$), and behaves in a monotonic fashion with increasing $\hat{K}_{Ic}$. The combination of near-constant $\hat{p}_0$ in the viscosity dominated regime and increasing $\hat{K}_{Ic}$, leads to $\hat{w}_0$ monotonically increasing with $\hat{K}_{Ic}$, overcoming the non-monotonic behaviour observed in \eqref{Nobel_w0344}.

It is also apparent from Fig.~\ref{Nobel_AutoSwitch} that the modified formulation is an effective substitute when computing the local parameters describing the crack tip, with there being no noticeable difference between the full shear/modified variants\footnote{For example, asymptotic coefficient $\hat{w}_0$ has a relative difference between the `full shear' and `modified' variants of $2$\% or below in the viscosity dominated regime with $\nu=0.1$, and below $1$\% for $\nu=0.3$, both of which rapidly decrease when entering the toughness dominated regime. For $\hat{p}_0$, the difference is negligible (of order $10^{-10}$ for $\hat{K}_{Ic}=10^{-4}$) except at the point of transition between viscosity and toughness dominated regimes, where there is a maximum relative difference is just below $1$\%.}. Consequently, incorporating the tangential traction can have a benefit in reducing algorithm complexity. The more complicated form of the elasticity equation can be incorporated solely through the updated integral definition of the stress intensity factor without significantly impacting the result, and instead only the updated asymptotics and fracture criterion incorporated, to simplify the modeling of hydraulic fracture during viscosity-toughness transition.

\subsection{The fracture tip vs near-tip asymptotics}

While incorporating the updated system asymptotics has a notable benefit on simplifying algorithm construction, it may have a detrimental effect on how effectively the first term of the crack tip asymptotics approximate key problem parameters. This is because the updated system asymptotics for the viscosity dominated regime now only describe the behaviour at the fracture tip, while experimental results indicate that the near-tip behaviour remains the same as `classical' asymptotics for the viscosity dominated regime \cite{Bunger2008}. This is crucial to understand, as in the case without tangential traction the leading asymptotic terms for the aperture and pressure are highly effective at approximating the solution (see e.g. \cite{Savitski2002}), and form the basis of many semi-analytical approximations (see e.g. \cite{Dontsov2016}). Consequently, differing fracture tip and near-tip behaviour may reduce the effectiveness of these approaches, and need to be accounted for.

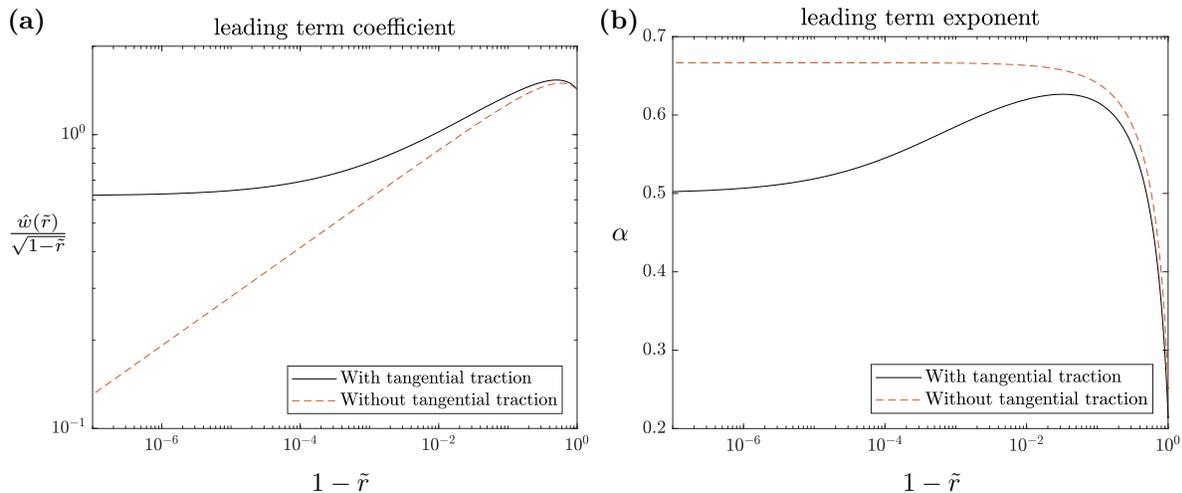
\begin{figure}[t!]
 \centering
 \includegraphics[width=0.45\textwidth]{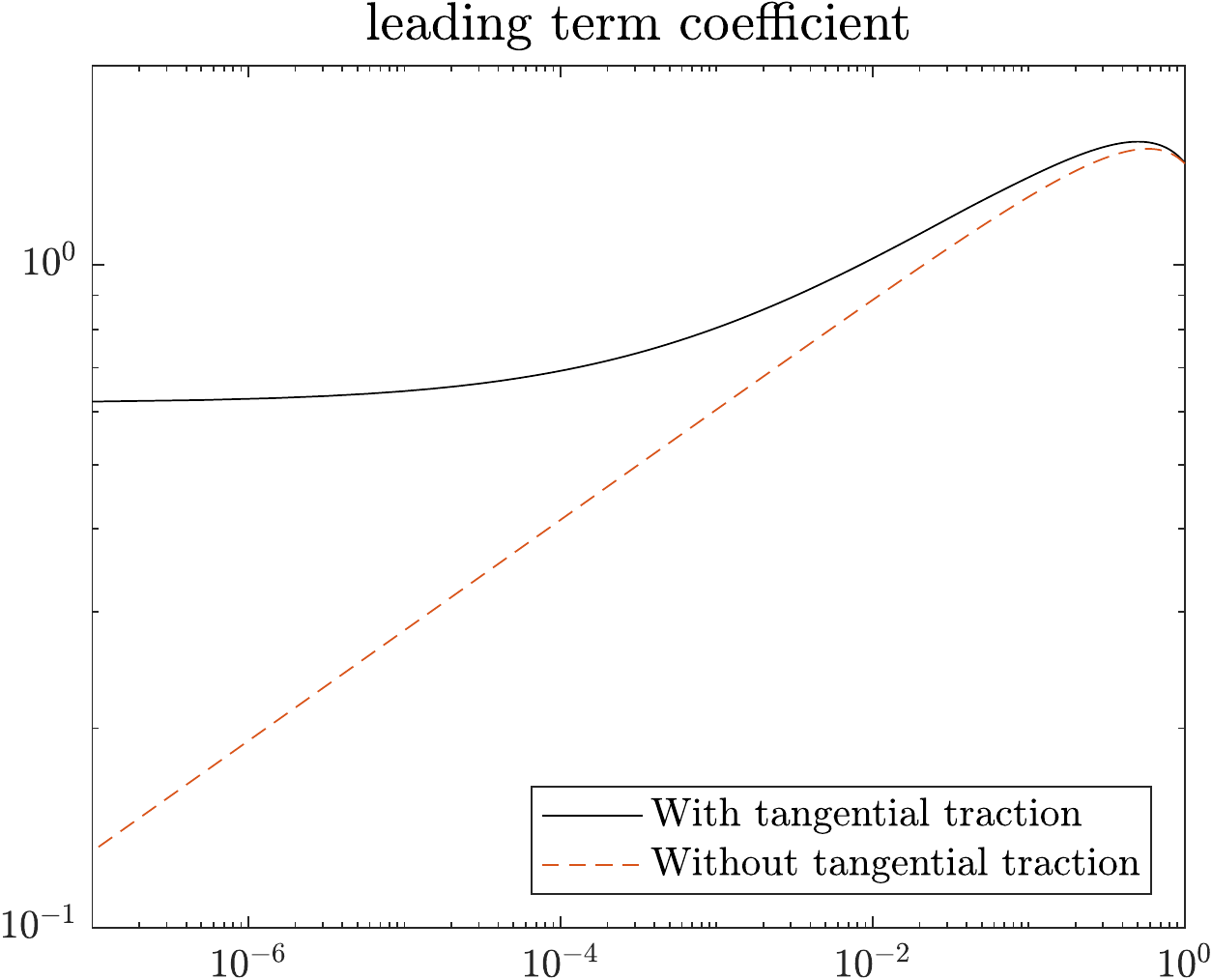}
 \put(-220,160) {{\bf (a)}}
  \put(-220,80) {$\frac{\hat{w}(\tilde{r})}{\sqrt{1-\tilde{r}}}$}
 \put(-105,-15) {$1-\tilde{r}$}
\hspace{6mm}
 \includegraphics[width=0.45\textwidth]{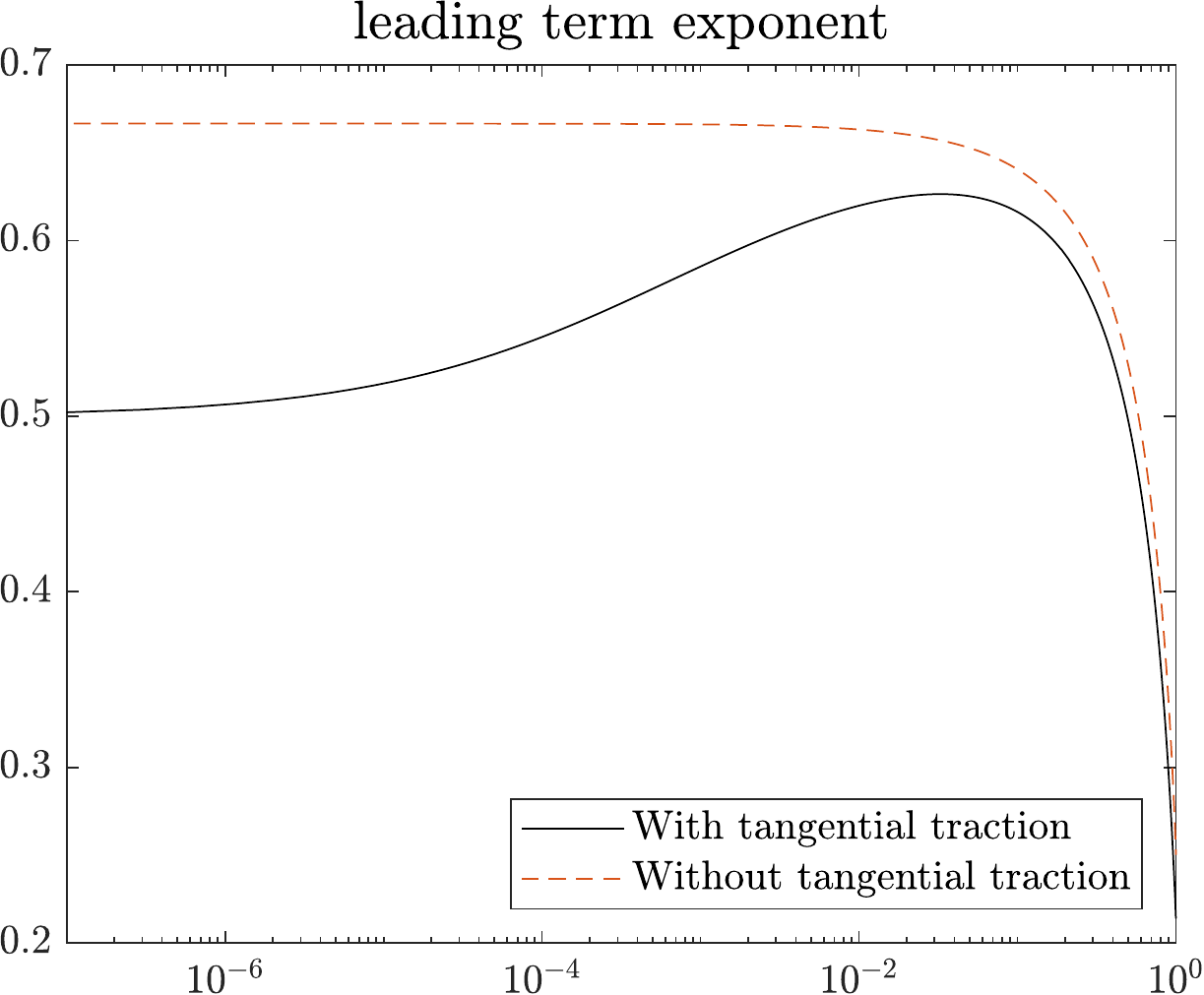}
 \put(-220,160) {{\bf (b)}}
 \put(-215,80) {$\alpha$}
  \put(-105,-15) {$1-\tilde{r}$}
 \caption{(a) Log-log plot of the aperture over the leading tip asymptote $\hat{w}(\tilde{r})/\sqrt{1-\tilde{r}}$ in the viscosity dominated regime ($\hat{K}_{Ic}=0$) with (black) and without (red) tangential traction for $\nu=0.3$. (b) The exponent of the first-term asymptotics $(1-\tilde{r})^{\alpha (\tilde{r})}$ \eqref{alpha_defininininin}-\eqref{alpha_deriviv} which best describes the behaviour of the aperture at point $\tilde{r}$. }
 \label{Fig:DefAlpha}
\end{figure}

To investigate whether there is any divergence in the crack tip and near-tip behaviour of the leading asymptotic term of the aperture, we consider the exponent, denoted $\alpha$
\begin{equation}\label{alpha_defininininin}
\hat{w}(\tilde{r}) \approx \hat{w}_0 \left(1-\tilde{r}\right)^{\alpha}.
\end{equation}
We consider this for fixed points in space $\tilde{r}$, to determine the associated constant $\alpha$ which best describes the behaviour of the aperture. It can be demonstrated that this exponent, $\alpha (\tilde{r})$, is given by
\begin{equation}\label{alpha_deriviv}
\alpha (\tilde{r}) = -\frac{1}{\log\left(1-\tilde{r}\right)}\int_{\tilde{r}}^1 \frac{1}{\hat{w}(\xi)}\frac{d\hat{w}}{d\xi} \, d\xi . 
\end{equation}
The deviation of this parameter away from the value at the crack tip ($\alpha=1/2$) gives an indication of the extent to which the aperture can be described by it's leading crack-tip asymptotic term along the fracture front. We compute $\alpha$ for each $\tilde{r}$ numerically, using spline-based approaches, for both the `classical' case the case with tangential traction (including the full elasticity equation for completeness). An example for the viscosity dominated regime ($\hat{K}_{Ic}=0$) is provided in Fig.~\ref{Fig:DefAlpha}, with all other material constants as in Table.~\ref{Table:SSConst}. It is immediately apparent that, while in the case without tangential traction the tip asymptotics will provide a highly accurate description of the solution behaviour even beyond the near-tip region, the crack tip asymptotics are not as effective at approximating the whole fracture when the shear stress is accounted for. In the case with tangential traction the exponent $\alpha$ has deviated from the tip solution by $16$\% for $\tilde{r}=0.999$, and by $23$\% for $\tilde{r}=0.99$, while the deviation is less than $0.5$\% for $\tilde{r}=0.99$ when the shear stress is neglected. This trend for the viscosity dominated regime holds true irrespective of the value of Poisson's ratio $\nu$ being considered, although will become less significant when transitioning to the toughness dominated regime (for which the asymptotics between the two cases are unchanged). 

\begin{figure}[t!]
 \centering
 \includegraphics[width=0.5\textwidth]{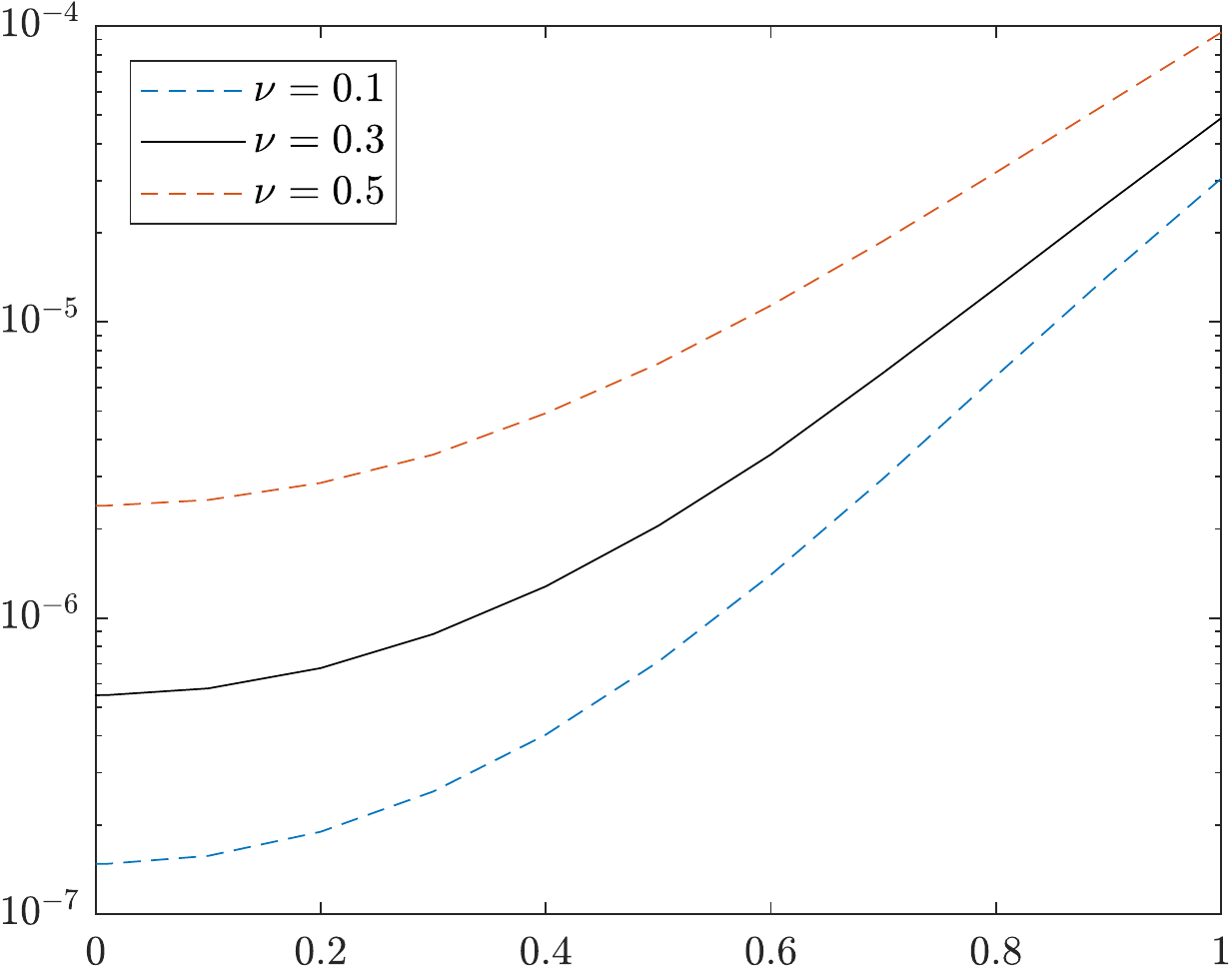}
 \put(-110,-15) {$\hat{K}_{Ic}$}
 \put(-245,85) {$\tilde{r}_\alpha$}
 \caption{The smallest distance from the fracture front $\tilde{r}_\alpha$ \eqref{rstar_01} where the exponent $\alpha (\tilde{r})$ of the near-tip aperture asymptotics is 1\% greater than the exponent at the fracture tip ($\alpha=0.5$). }
 \label{Fig:Alpha_r}
\end{figure}

To better examine this behaviour, let us consider the smallest distance away from the crack tip where the exponent of the near-tip aperture asymptotics $\alpha (\tilde{r})$ is $1$\% greater than that of the crack tip asymptotics ($\alpha=1/2$). We label this new length $\tilde{r}_\alpha$:
\begin{equation} \label{rstar_01}
\tilde{r}_\alpha =\min \left\{ 1- \tilde{r}\in [0,1 ] \, : \, \alpha (\tilde{r}) >0.505 \right\}.
\end{equation}
The plot of $\tilde{r}_\alpha$ over $\hat{K}_{Ic}$, for various values of the Poisson's ratio $\nu$, is given in Fig.~\ref{Fig:Alpha_r}. It is immediately apparent that the near-tip asymptote begins to deviate from the crack-tip exponent exceptionally close to the fracture front in the viscosity dominated regime, with it occurring when $1-\tilde{r}<10^{-5}$ for all Poisson's ratio $\nu$ when $\hat{K}_{Ic}=0$. The crack tip asymptote however provides a far better approximation of the near-tip behaviour with increasing $\hat{K}_{Ic}$, with the distance $\tilde{r}_{\alpha}$ where the exponent differs by $1$\% being of order $10^{-4}$ for all $\nu$ when $\hat{K}_{Ic}=1$. This trend is not surprising, as the tip asymptotics in the toughness dominated regime are unchanged from the classical case, and have been confirmed to correspond to the near-tip asymptotics in experiments \cite{Bunger2008}.

We conclude that the crack tip asymptotics do not correspond to the near-tip asymptotics even a short distance from the front in the viscosity dominated regime when tangential traction is incorporated. This adds additional difficulty to the modeling of problems incorporating this effect, and must be accounted when constructing such algorithms or semi-analytical solutions.


\section{Analysis of the time-dependent formulation}\label{Sect:4}

Having investigated the effect of incorporating the tangential traction on the construction of numerical solvers, we can now move towards an examination of the quantitative effect of the tangential traction in the time-dependent case.

The numerical solver used to obtain time-dependent results is outlined in \cite{DaFiesThesis}. It follows a similar ``universal algorithm'' methodology to that for the self-similar case, utilizing the fluid velocity \eqref{NobelPVInit} as a process parameter, tracing the fracture front using the associated Stefan-type condition \eqref{Nobel_SpeedEq}, and employing rigorous use of the system asymptotics \eqref{wasym1} - \eqref{tasym1} to properly treat any singular points at all stages of the algorithm. The algorithm is also adaptive in both the spatial and temporal dimensions, ensuring a high level of accuracy over the whole domain\footnote{All simulations were run to the level of accuracy necessary to confirm the stated results.}. The reader is referred to \cite{DaFiesThesis} for further details.

Throughout the investigation, the parameter $\delta$ introduced in \eqref{defn_delta} will be utilized to parameterise the fracture regime (viscosity, transient or toughness dominated). An initial examination against the reference case of HF in shale will be conducted, before examining the impact of different parameters on the significance of the shear stress for a variety of applications.

\subsection{Quantitative impact of the shear stress}

%
%

\subsubsection{The reference case - hydraulic fracturing of shale rock}\label{Sect:Reference}

We first consider the quantitative effect of the tangential traction for the case of a hydraulic fracture in shale, as encountered in numerous (typically energy-related) applications. The reference values for the material constants and process parameters are provided in Table.~\ref{Table:MatConst}, with the values of the Young's modulus $E$ and Poisson's ratio $\nu$ taken in line with values typically encountered during hydraulic fracturing in rock, and the material toughness $K_{Ic}$ from the range given in \cite{2015JB012756}. The pumping rate and viscosity may vary widely between sites, and even stages of the HF process, so convenient but reasonable values were taken for simplicity. Finally, the shear-related constant $\beta$ \eqref{New_tau} - \eqref{rstar_defn} was chosen to minimise the effect of the tangential traction, to avoid unfairly biasing the result.

\begin{table}[b!]
 \centering
 \begin{tabular}{c|c|c|c|c|c}
 $E$  & $\nu$ & $\mu$  & $Q_0$ & $K_{Ic}$  & $\beta$ \\
 \hline 
 &&&&&\\
 $2.81 \times 10^{10}$ [Pa] & $0.25$ & $1\times 10^{-3}$ [Pa s] & $6.62\times 10^{-2}$ [m$^3$ / s]  & $1 \times 10^6$ [Pa m$^{\frac{1}{2}}$] & $1$ \\
 &&&&&\\
 \hline 
 \end{tabular}
 \caption{Reference values of the material constants and process parameters used in simulations. Note that the pumping rate $Q_0 (t)$ is taken as constant, while $\beta$ defines the behaviour of the shear stress at the injection point \eqref{New_tau} - \eqref{rstar_defn}.}
 \label{Table:MatConst}
\end{table}

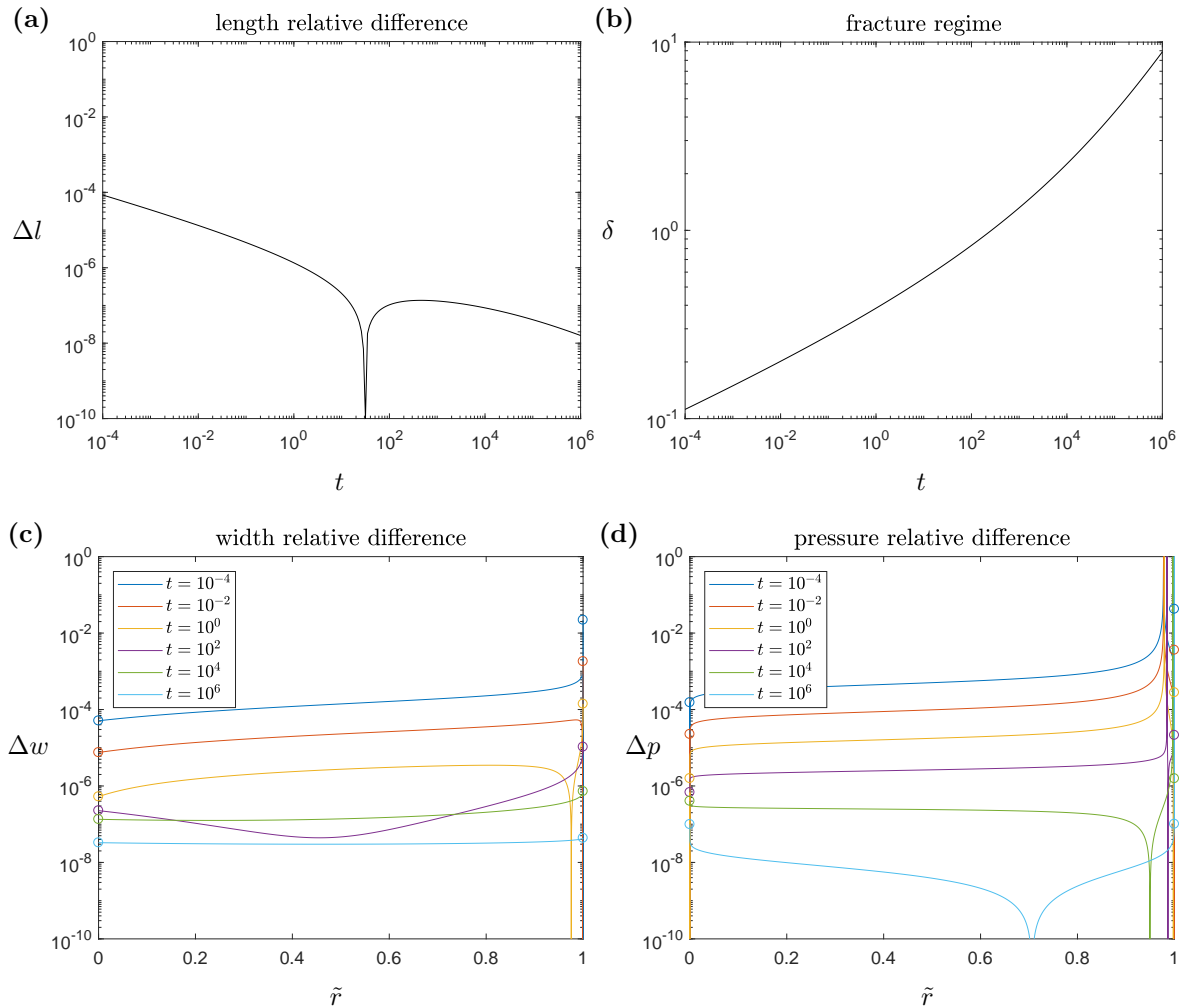
\begin{figure}[t!]
 \center
 \includegraphics[width=0.45\textwidth]{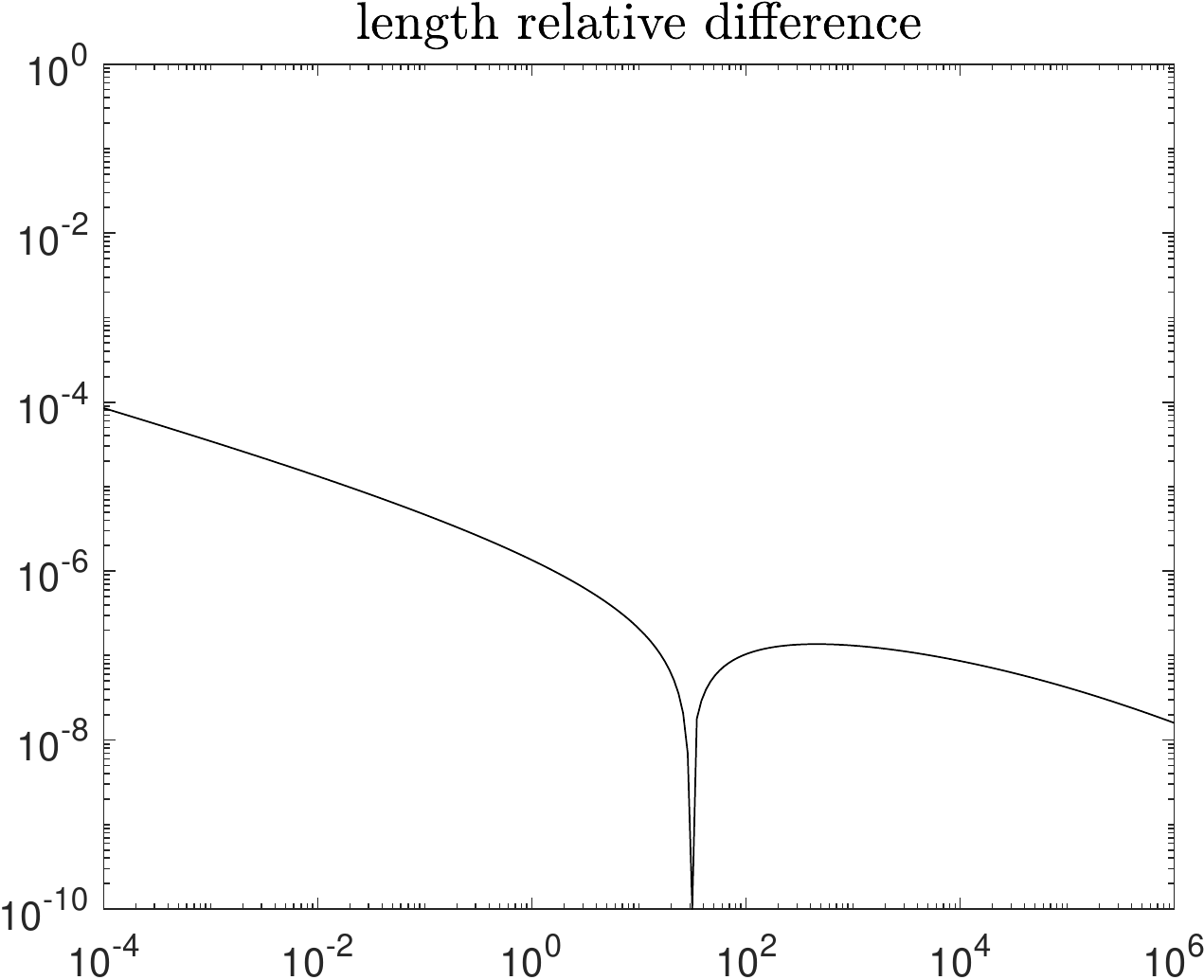}
 \put(-220,160) {{\bf{(a)}}}
 \put(-220,80) {$\Delta l$}
 \put(-98,-15) {$t$}
 \hspace{6mm}
   \includegraphics[width=0.442\textwidth]{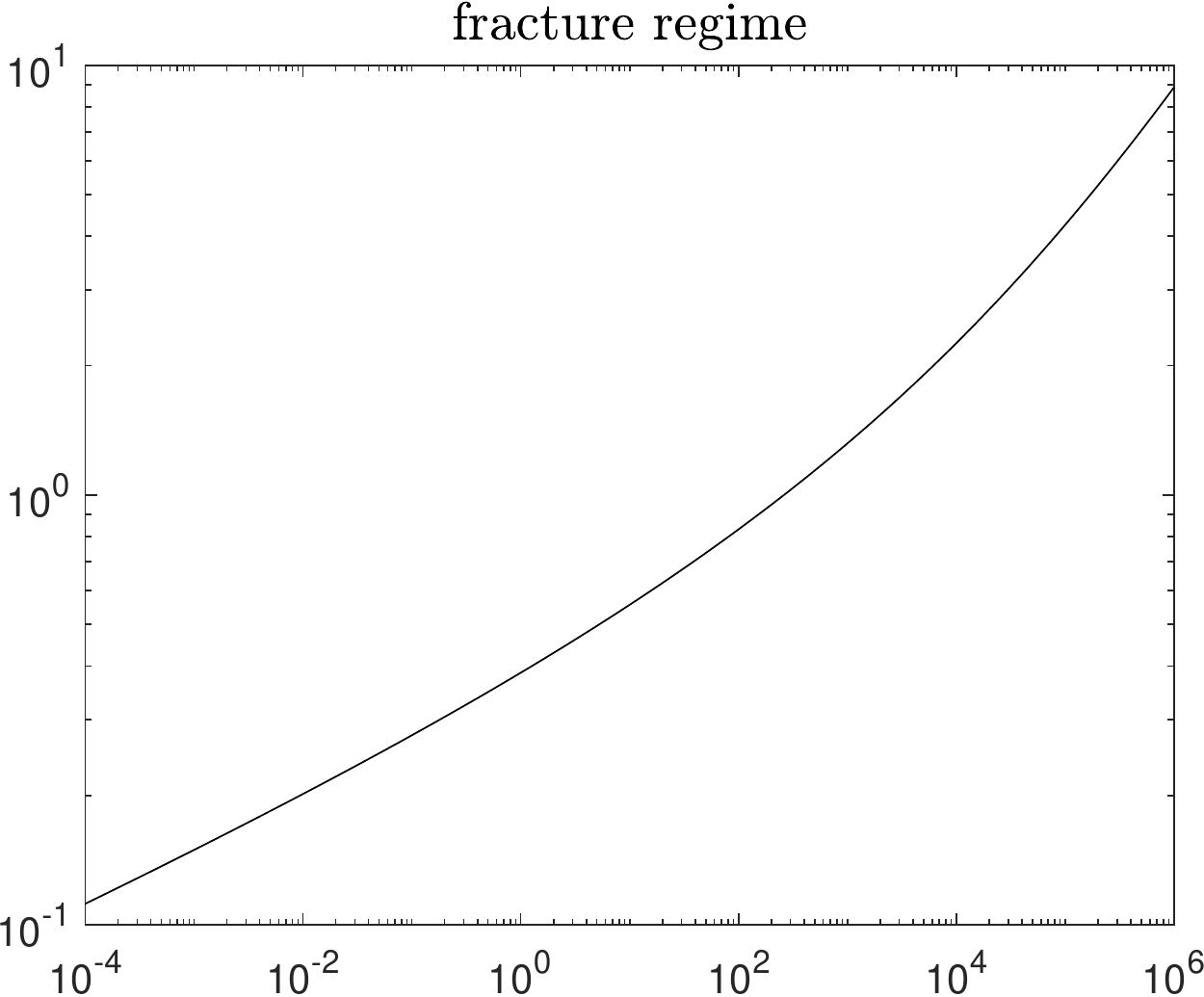} 
 \put(-220,160) {{\bf{(b)}}}
 \put(-217,80) {$\delta$}
 \put(-98,-15) {$t$}

\vspace{4mm}

\includegraphics[width=0.45\textwidth]{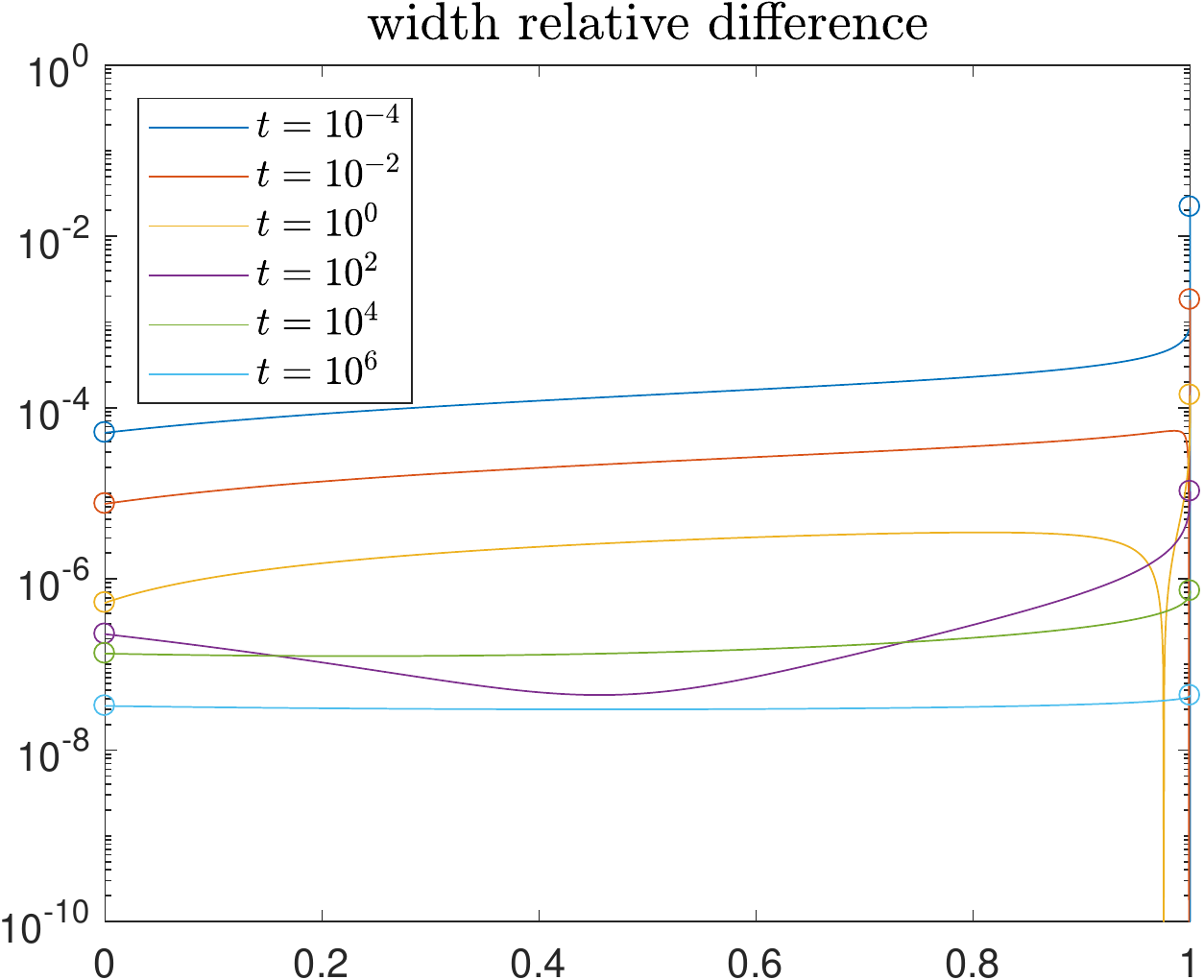}
 \put(-220,160) {{\bf{(c)}}}
 \put(-220,80) {$\Delta w$}
 \put(-98,-15) {$\tilde{r}$}
 \hspace{6mm}
\includegraphics[width=0.45\textwidth]{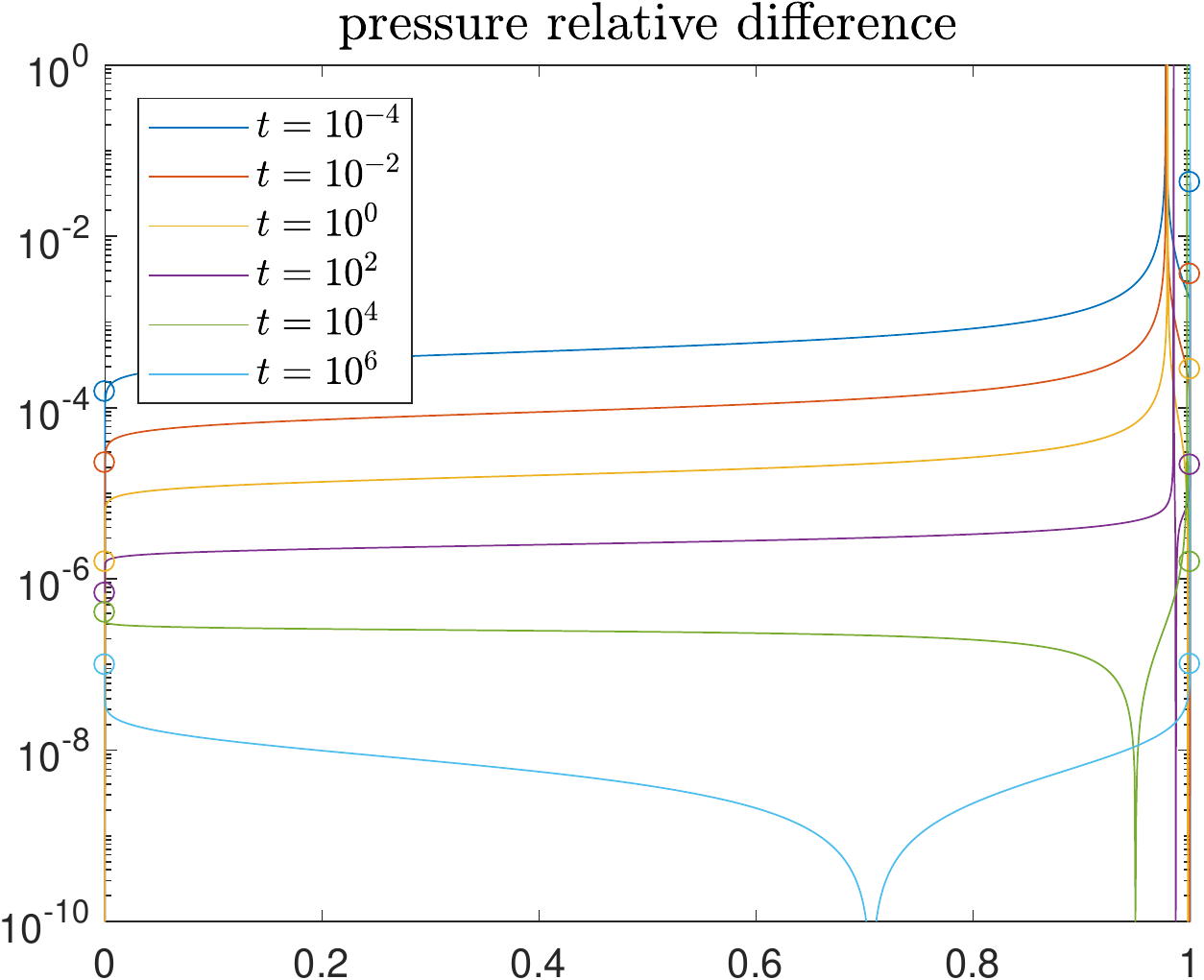}
 \put(-220,160) {{\bf{(d)}}}
 \put(-210,80) {$\Delta p$}
  \put(-98,-15) {$\tilde{r}$}
 \caption{The relative difference, $\Delta$, of the {\bf (a)} the crack (half-)length $l(t)$, {\bf (c)} the aperture $w(r,t)$, {\bf (d)} the pressure $p(r,t)$, between the case with and without tangential traction on the fracture walls for the reference case of HF in shale rock (material constants in Table.~\ref{Table:MatConst}). Here time $t$ [s] is not normalised, while the crack length $\tilde{r}$ is normalised over the length \eqref{Normalisation1}$_1$. In {\bf (b)} $\delta (t)$ which parameterises whether the system is in the viscosity ($\delta \ll 1$) or toughness ($\delta \gg 1$) dominated regime \eqref{defn_delta}.}
 \label{Time_Reference}
\end{figure}

The relative difference, $\Delta$, for the fracture (half-)length $l(t)$, the aperture $w(r,t)$ and fluid pressure $p(r,t)$ between the case with and without tangential traction are provided in Fig.~\ref{Time_Reference}, alongside the values of $\delta(t)$ parameterising the regime. It can be seen that the aperture achieves a difference larger than $1$\% at the crack tip for time $t=10^{-4}$, however this is only at the tip and dissipates rapidly over time. Over the remainder of the domain, and for the crack length, the relative difference is of order $10^{-4}$ or below even at $t=0.0001$ seconds\footnote{\label{Footnote1}Note that throughout Sect.~\ref{Sect:4} we are evaluating over such small times or high values of the viscosity in order to demonstrate what would be required to obtain a non-negligible impact of the shear stress within the current model. To accurately model these scenarios modifications should be made to the model, most notably incorporating the fluid lag (see e.g. \cite{LECAMPION20074863}).}, and decreases to order $10^{-7}$ away from the crack tip within $100$ seconds. From Fig.~\ref{Time_Reference}b, it can be seen that $100$ seconds is approximately the time when the crack begins transitioning to the toughness dominated regime, meaning that the effect of the shear becomes negligible even before this transition occurs.

\subsubsection{Effect of the material/process parameters}

\begin{figure}[t!]
 \center
 \includegraphics[width=0.45\textwidth]{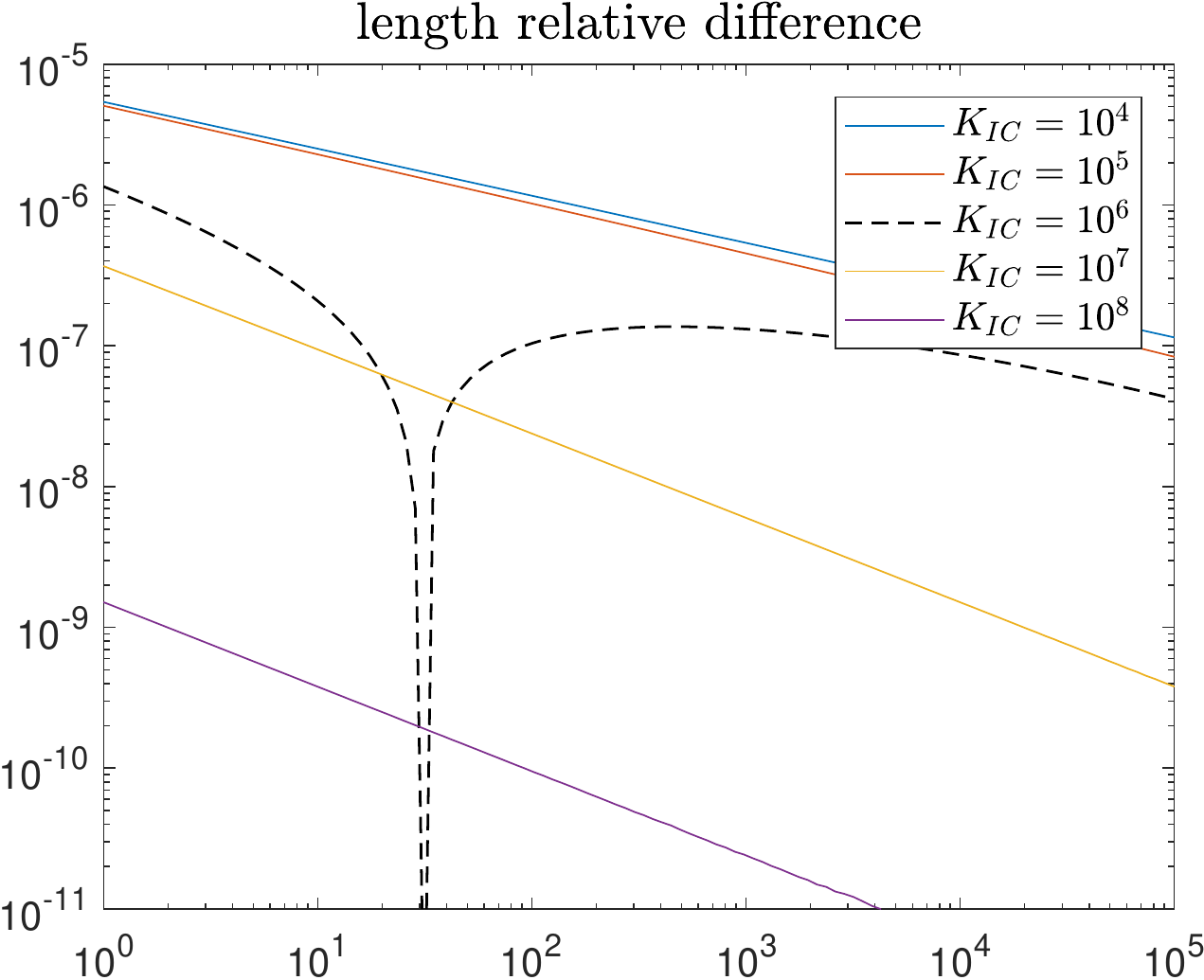}
 \put(-220,160) {{\bf{(a)}}}
 \put(-220,80) {$\Delta l$}
 \put(-98,-15) {$t$}
 \hspace{6mm}
 \includegraphics[width=0.45\textwidth]{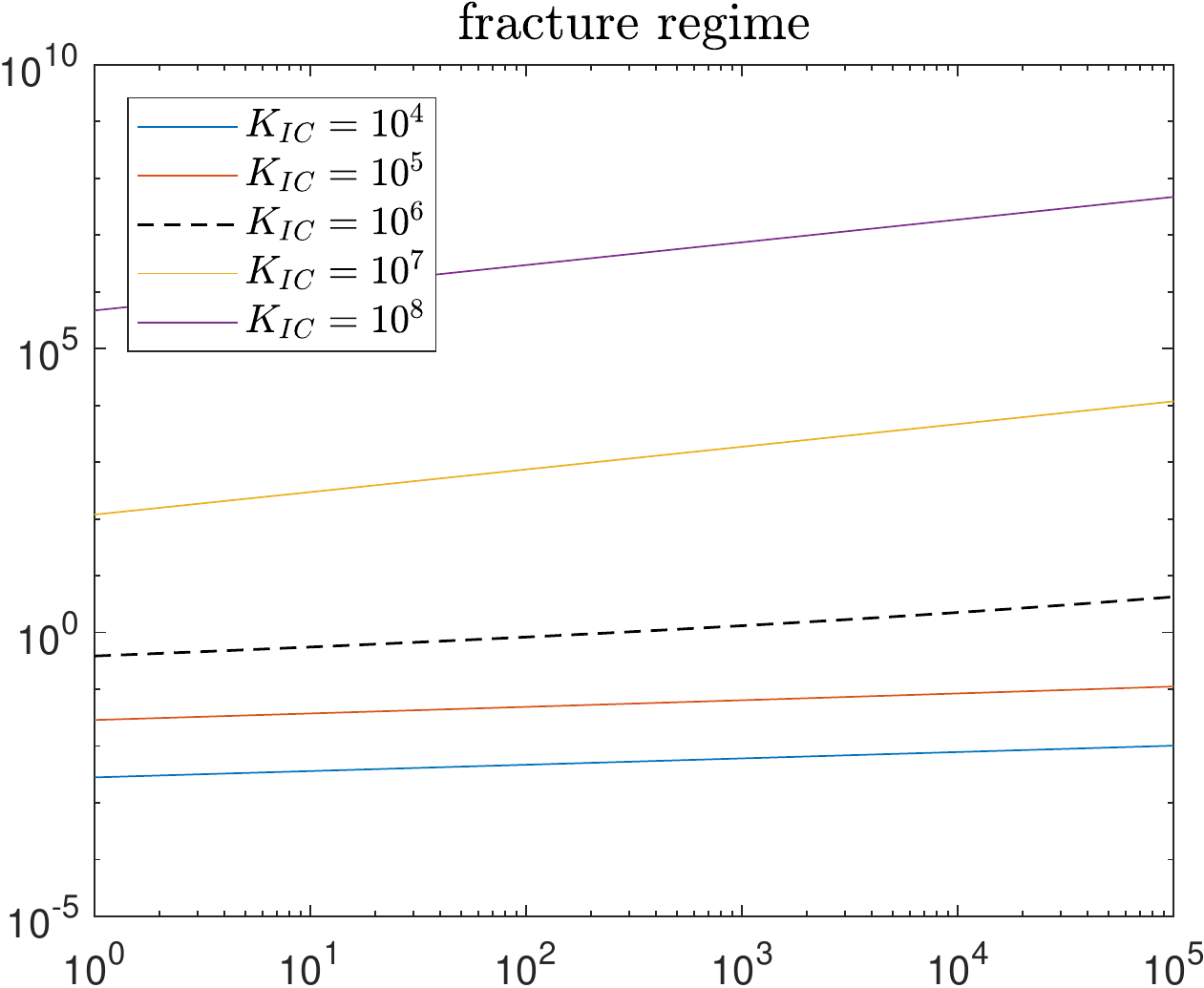}
 \put(-220,160) {{\bf{(b)}}}
 \put(-210,80) {$\delta $}
 \put(-98,-15) {$t$}
 \caption{The effect of the material toughness $K_{Ic}$  [Pa m$^{\frac{1}{2}}$] on the impact of the shear stress. All other material parameters are taken as in Table.~\ref{Table:MatConst}: {\bf (a)} the relative difference in the crack length $l(t)$, against the case without tangential traction, {\bf (b)} $\delta (t)$ which parameterises whether the system is in the viscosity ($\delta \ll 1$) or toughness ($\delta \gg 1$) dominated regime \eqref{defn_delta}.}
 \label{Time_varKIC}
\end{figure}

With the reference case now established, we can consider a wider range of process parameters to determine whether the traction may be impactful in any other contexts. Noting that the relative difference over the crack length in Fig.~\ref{Time_Reference}a is consistently of the same order as that of the aperture and pressure (Fig.~\ref{Time_Reference}b,c) over almost the entire domain (except the crack tip) at each point in time, only the relative difference of the fracture length will be provided in the remaining subsections for the sake of brevity\footnote{The authors computed the average of the relative differences over the crack length for the aperture and fluid pressure for each simulation in the remainder of the paper, and confirmed that they are of identical order.}.  Additionally, in all subsequent figures the relative difference for the reference case in Sect.~\ref{Sect:Reference}, is shown on each figure as a dashed black line. Note that we are focusing on a narrower temporal range in Fig.~\ref{Time_varKIC} and subsequent figures ($t\in [0,10^5 ]$ seconds) compared to Fig.~\ref{Time_Reference} ($t\in [10^{-4},10^6]$ seconds), to focus on the most important area of effect.

\begin{figure}[t!]
 \center
 \includegraphics[width=0.45\textwidth]{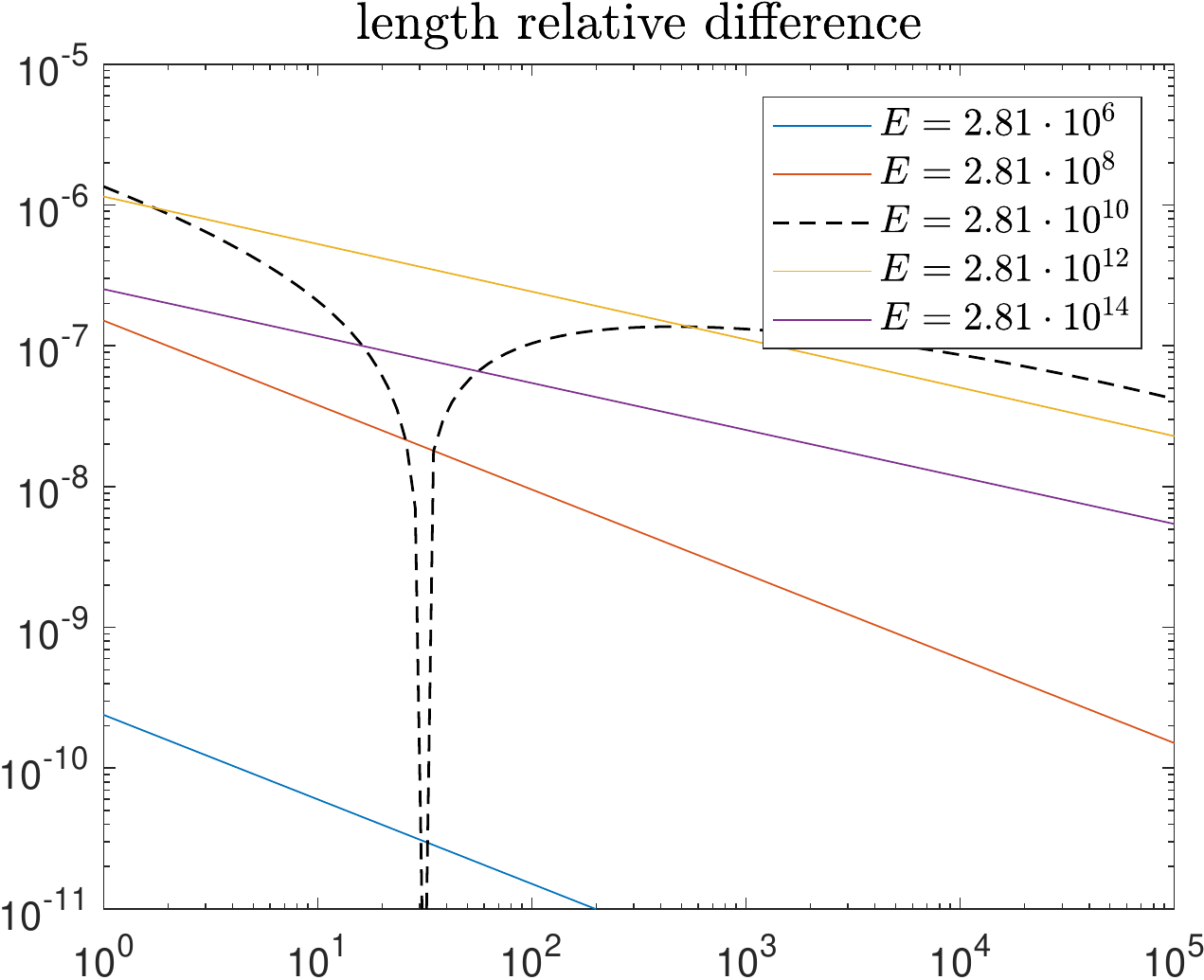}
 \put(-220,160) {{\bf{(a)}}}
 \put(-220,80) {$\Delta l$}
 \put(-98,-15) {$t$}
 \hspace{6mm}
 \includegraphics[width=0.45\textwidth]{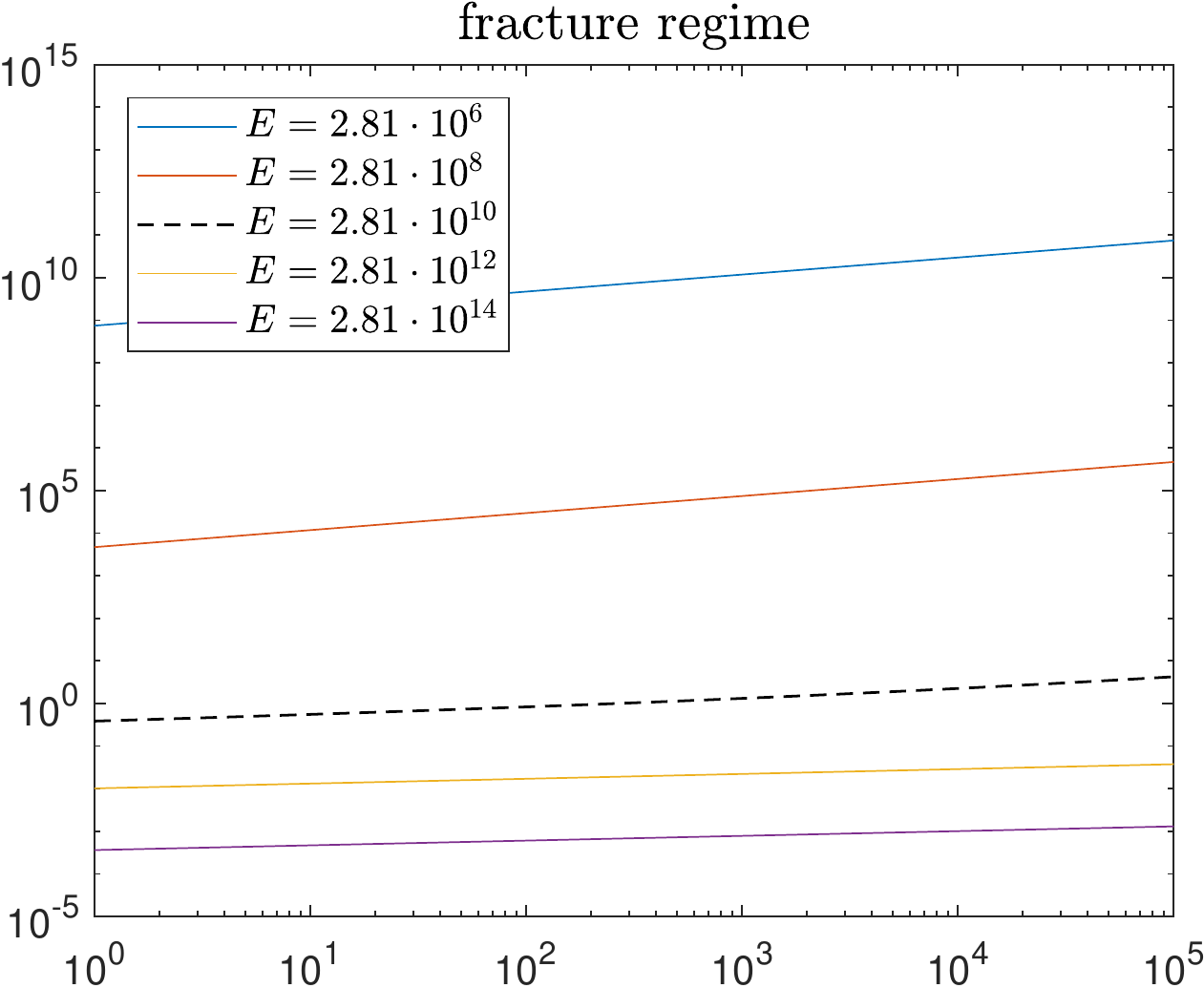} 
 \put(-220,160) {{\bf{(b)}}}
 \put(-210,80) {$\delta $}
 \put(-98,-15) {$t$}
 \caption{The effect of the Young's modulus $E$ [Pa] on the impact of the shear stress. All other material parameters are taken as in Table.~\ref{Table:MatConst}: {\bf (a)} the relative difference in the crack length $l(t)$, against the case without tangential traction, {\bf (b)} $\delta (t)$ which parameterises whether the system is in the viscosity ($\delta \ll 1$) or toughness ($\delta \gg 1$) dominated regime \eqref{defn_delta}.}
 \label{Time_varE}
\end{figure}

We begin by examining the effect of varying the fracture toughness $K_{Ic}$. The relative difference $\Delta l$ obtained for a variety of toughness' are provided in Fig.~\ref{Time_varKIC}. It can be seen that having a lower fracture toughness does increase the effect of the shear, but only up to a certain point. For both $K_{Ic}=10^4$  Pa $\cdot$ m$^{\frac{1}{2}}$ and $K_{Ic}=10^5$ Pa $\cdot$ m$^{\frac{1}{2}}$ the relative difference is almost identical. This is because taking a significantly lower toughness places it further into the viscosity dominated regime, where the material toughness has a significantly smaller effect on the crack evolution. Meanwhile, increasing the toughness significantly decreases the impact of the shear, with the difference clearly tending to zero in the limiting case of an immobile crack. We can conclude that changing the toughness alone will not cause the effect of tangential traction to be significant.

This trend continues when considering the Young's modulus $E$, which is shown in Fig.~\ref{Time_varE}. Here, taking a very low value of the Young's modulus ($<2.81\cdot 10^{8}$ Pa) results in the fracture starting in the toughness regime, where the effect of the tangential traction is negligible. Conversely, while having a higher Young's modulus does lead to the fracture remaining the viscosity dominated regime for a longer time period, this does not always increase the effect of the tangential traction. Instead, for the material constants (aside from $E$) taken as in Table.~\ref{Table:MatConst}, the effect of the tangential traction appears to be maximised when the Young's modulus is between $10^{10}$ and $10^{12}$, with the relative difference decreasing with increasing Young's modulus after that point. We can conclude that the impact of the shear stress increases as $E$ decreases, but only if the system remains in the viscosity dominated regime.

\begin{figure}[t!]
 \center
 \includegraphics[width=0.45\textwidth]{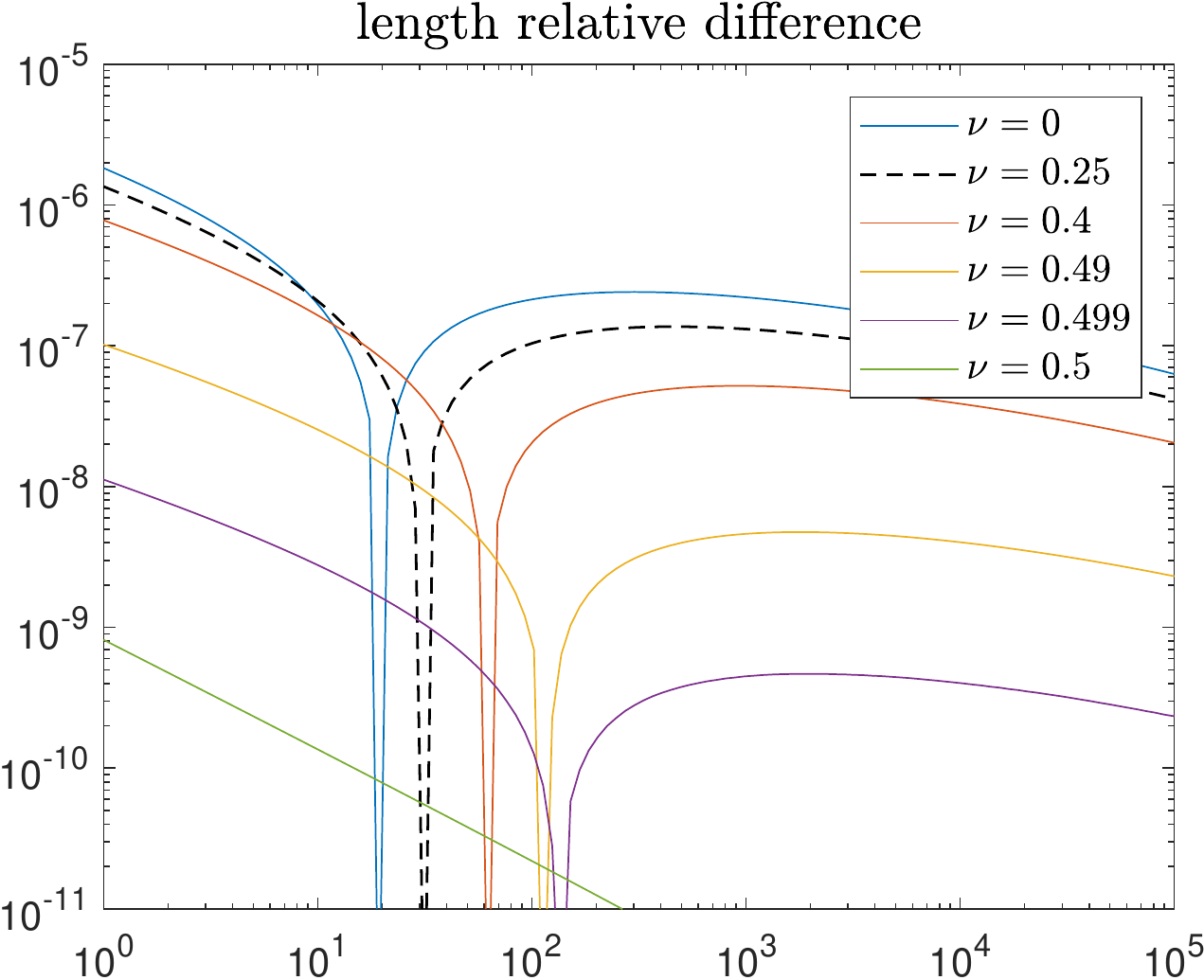}
 \put(-220,160) {{\bf{(a)}}}
 \put(-220,80) {$\Delta l$}
 \put(-98,-15) {$t$}
 \hspace{6mm}
 \includegraphics[width=0.442\textwidth]{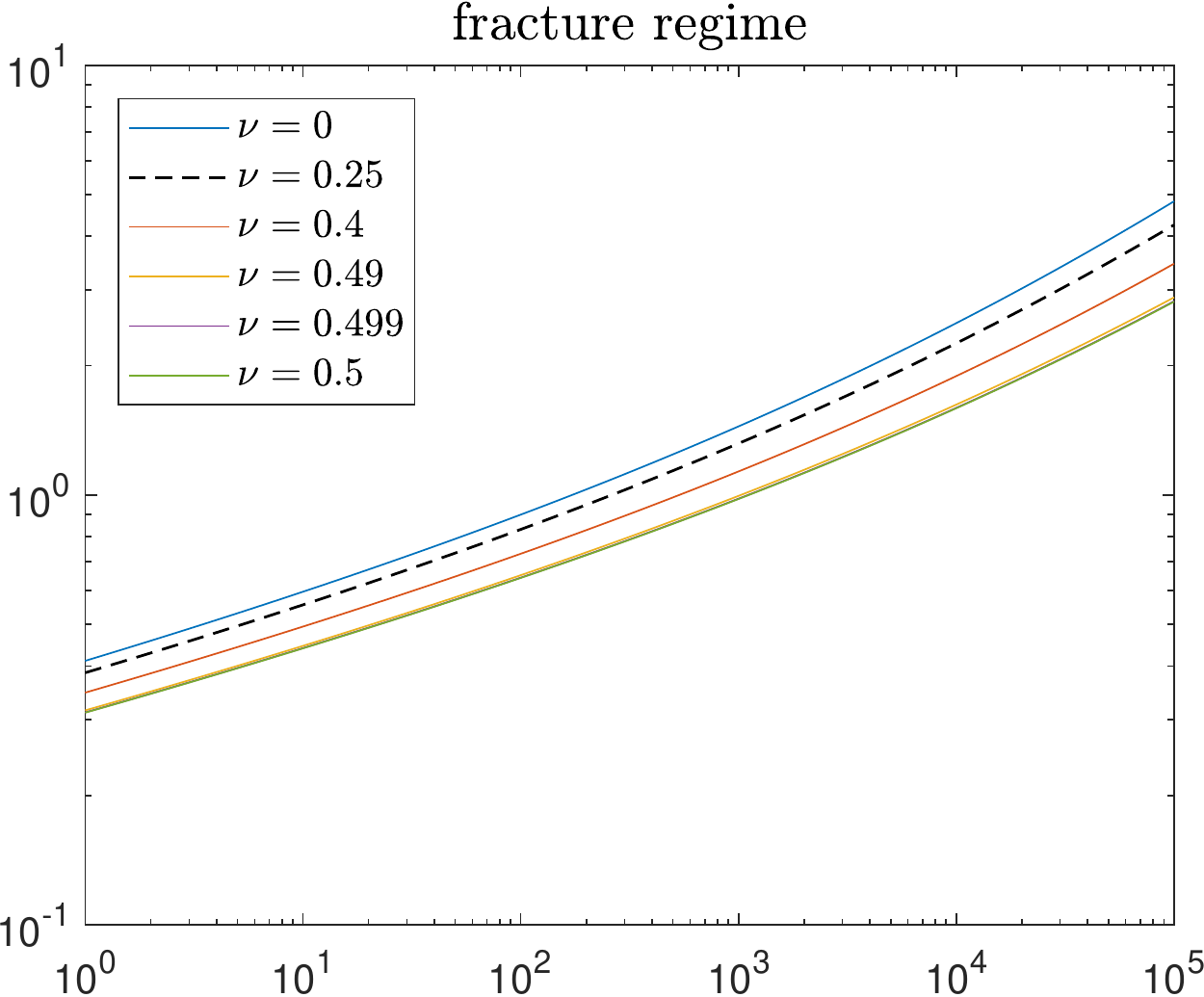}
 \put(-220,160) {{\bf{(b)}}}
 \put(-210,80) {$\delta $}
 \put(-98,-15) {$t$}
 \caption{The effect of the Poisson's ratio $\nu$ on the impact of the shear stress. All other material parameters are taken as in Table.~\ref{Table:MatConst}: {\bf (a)} the relative difference in the crack length $l(t)$, against the case without tangential traction, {\bf (b)} $\delta (t)$ which parameterises whether the system is in the viscosity ($\delta \ll 1$) or toughness ($\delta \gg 1$) dominated regime \eqref{defn_delta}.}
 \label{Time_varNu}
\end{figure}

\begin{figure}[t!]
 \center
 \includegraphics[width=0.45\textwidth]{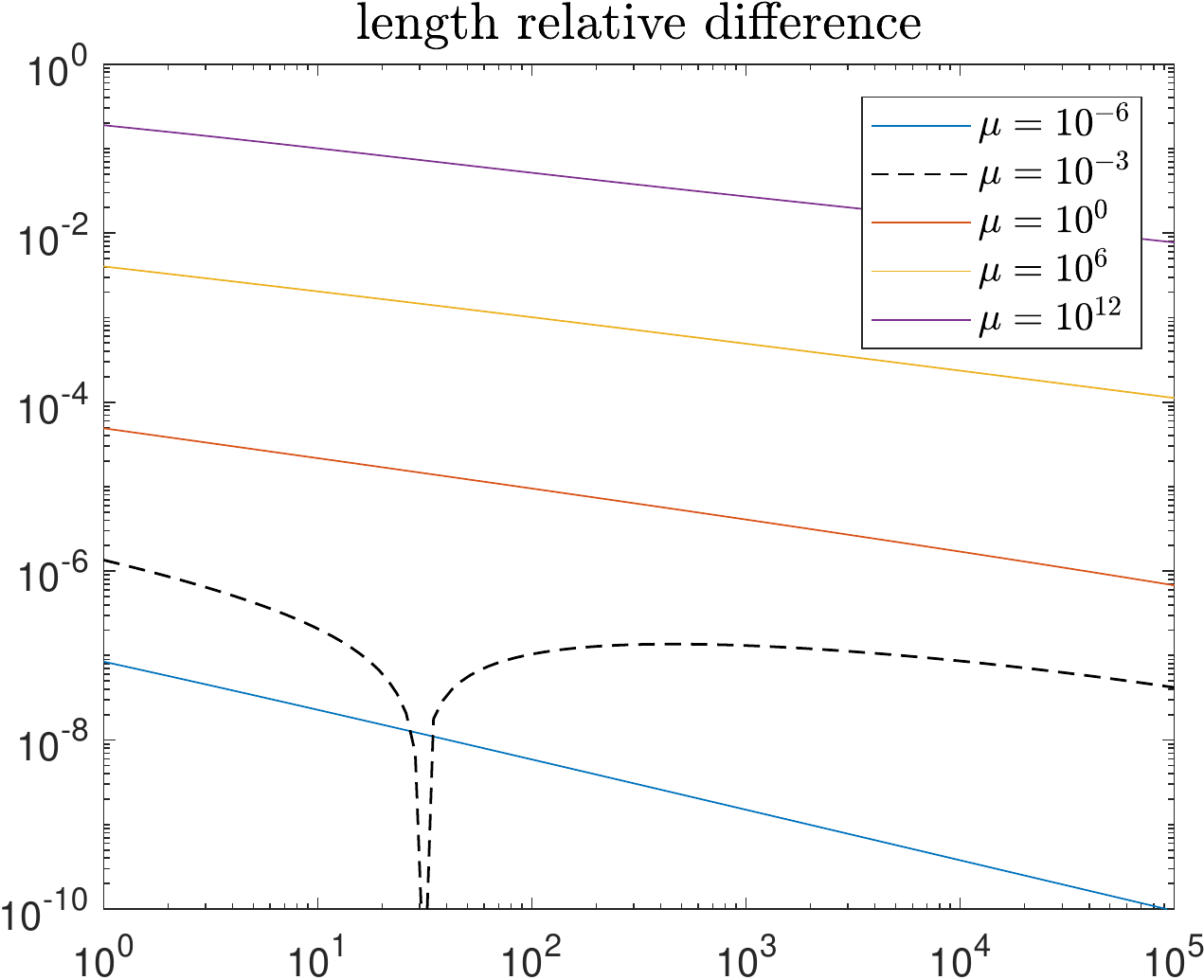}
 \put(-220,160) {{\bf{(a)}}}
 \put(-220,80) {$\Delta l$}
 \put(-98,-15) {$t$}
 \hspace{6mm}
 \includegraphics[width=0.45\textwidth]{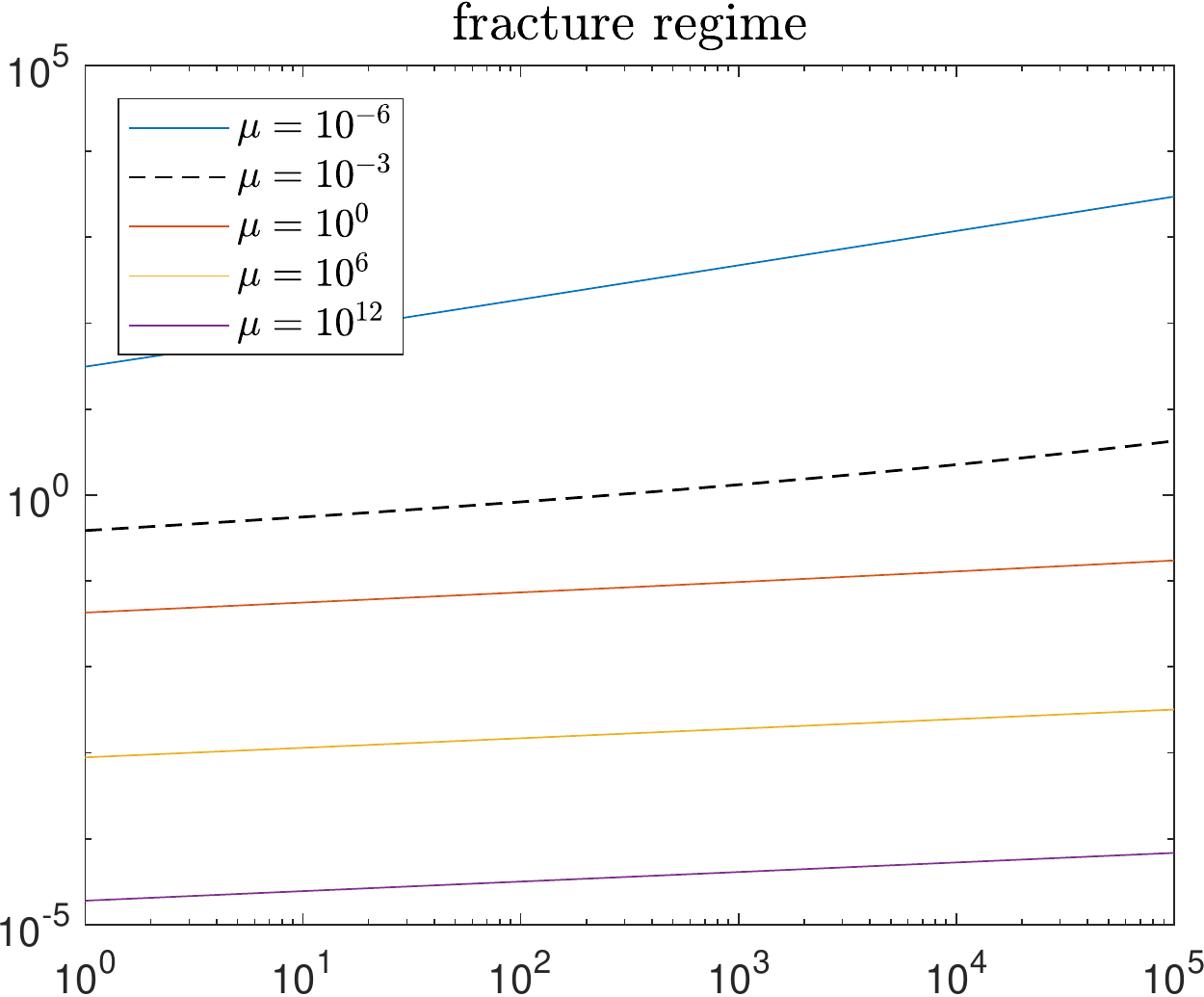}
 \put(-220,160) {{\bf{(b)}}}
 \put(-210,80) {$\delta $}
 \put(-98,-15) {$t$}
 \caption{The effect of the fluid viscosity $\mu$ [Pa s] on the impact of the shear stress. All other material parameters are taken as in Table.~\ref{Table:MatConst}: {\bf (a)} the relative difference in the crack length $l(t)$, against the case without tangential traction, {\bf (b)} $\delta (t)$ which parameterises whether the system is in the viscosity ($\delta \ll 1$) or toughness ($\delta \gg 1$) dominated regime \eqref{defn_delta}.}
 \label{Time_varMu}
\end{figure}

Next, we examine the effect of varying the Poisson's ratio $\nu$, with the relative differences provided in Fig.~\ref{Time_varNu}. Here, it is clear that when the Poisson's ratio is low ($\nu<0.4$), the impact of the tangential traction is not significantly affected by changing $\nu$. However, this changes in the limit as $\nu\to 0.5$, with the shear stress playing a rapidly diminishing role as the Poisson's ratio increases.

In the final set of figures, Fig.~\ref{Time_varMu}, we examine the effect of changing the fluid viscosity$^{\ref{Footnote1}}$. It can be seen that this parameter plays the largest role in determining the effect of the tangential traction, with very high viscosity leading to a shear stress that can significantly effect the resulting fracture length. Taking a value of $\mu=10^{12}$ Pa {$\cdot$} s, which can be found for some forms of magma, leads to a difference that is above $1$\% even after $10^{4}$ seconds. However, outside of this particularly extreme case the effect of the tangential traction remains small, and even fluids with an exceptionally high viscosity $\mu = 10^{6}$ Pa {$\cdot$} s experiencing a relative difference below $1$\% even at $t=1$ second. 

Finally, it should be stated that the pumping rate $Q_0$ will not significantly effect the impact of the tangential traction. Increasing $Q_0$ is equivalent to decreasing the toughness $K_{Ic}$, which does not produce a sizable effect (see Fig.~\ref{Time_varKIC}). Decreasing the pumping rate $Q_0$ meanwhile, like increasing $K_{Ic}$, reduces the effect of shear stress. Consequently, altering the pumping rate can not lead to a significant impact of the tangential traction compared to the classical case.

\subsubsection{Estimate of the quantitative impact in the viscosity dominated regime}

With the impact of the tangential traction for each parameter individually now considered, it is useful to provide a method of approximating the relative effect that the tangential traction may have in a given scenario. To do this, we note from the results of the previous subsection that the shear stress remained negligible in the toughness dominated regime for all of the cases considered. Consequently, only the viscosity dominated regime needs to be considered, and the typical scalings for the viscosity dominated regime can be used to provide an estimate of the relative error for the crack length.

It can be demonstrated that in the viscosity dominated regime ($\delta \ll 1$), if the relative deviation $\Delta l$ is small ($\Delta l \ll 1$), then it behaves as
\begin{equation} \label{Est_delL}
\Delta l \approx 0.17 \left[ \frac{(1-\nu^2) \mu}{E t}\right]^{\frac{3.16}{6}} + 0.25 \left(\frac{1-2\nu}{1-\nu}\right)\left[ \frac{(1-\nu^2) \mu}{E t}\right]^{\frac{1}{3}} .
\end{equation}
Here the first term comes from the viscosity dominated scaling \cite{GaragashSummary} (see also e.g.\ \cite{Dontsov2019a,Garagash2011,Peirce2008}) accounting for the modified stress intensity factor, while the second was obtained using numerical analysis when varying the values of the parameters. In the toughness dominated regime, or where the effect of shear stress is not negligible, it can be demonstrated that this estimate will act as an upper bound on the relative difference. The regime can be approximated by noting that, in the viscosity dominated regime, the parameter $\delta(t)$ behaves as
$$
\delta \sim 0.9642 \left[ \frac{K_{Ic}^{18} (1-\nu^2)^{13}}{\mu^5 E^{13} Q_0^{3} }\right]^{\frac{1}{18}} t^\frac{1}{9} , \quad \delta \ll 1.
$$

Recall that \eqref{Est_delL} will also provide an estimate of the order of the difference in the fracture aperture and fluid pressure away from the crack tip (see Sect.~\ref{Sect:Reference}), and as such can be used to estimate the direct impact of the tangential traction for all key process parameters. This was confirmed in numerous simulations by the authors, using several different combinations of parameters that span all cases.

Consequently, this can be used to determine if the shear will likely play any direct, quantitatively significant, role in a given HF process, with the relative difference obtained for the reference example in Sect.~\ref{Sect:Reference} acting as a point of comparison. Note however that it is not possible to achieve an arbitrarily large relative deviation by decreasing the Young's modulus $E$, as seen in Fig.~\ref{Time_varE}, as this will cause a transition to the toughness dominated regime for which $\Delta l$ remains negligible.

\subsection{Effect at the injection point}\label{Sect:InjectionBeta}

The final quantitative investigation to conduct is an investigation of the parameter $\beta$, introduced into the model in Sect.~\ref{The_wall_jet} to account for the stagnant zones of fluid reducing the tangential traction near the wellbore ($r=0$). As this parameter is assumed to be predefined, rather than part of the solution, knowing the sensitivity of the solution to the value of $\beta$ is crucial in understanding the ability of the model to make accurate predictions near to $r=0$.

\begin{figure}[t!]
 \center
 \textit{Crack width near to the injection point}\par\medskip
 \includegraphics[width=0.45\textwidth]{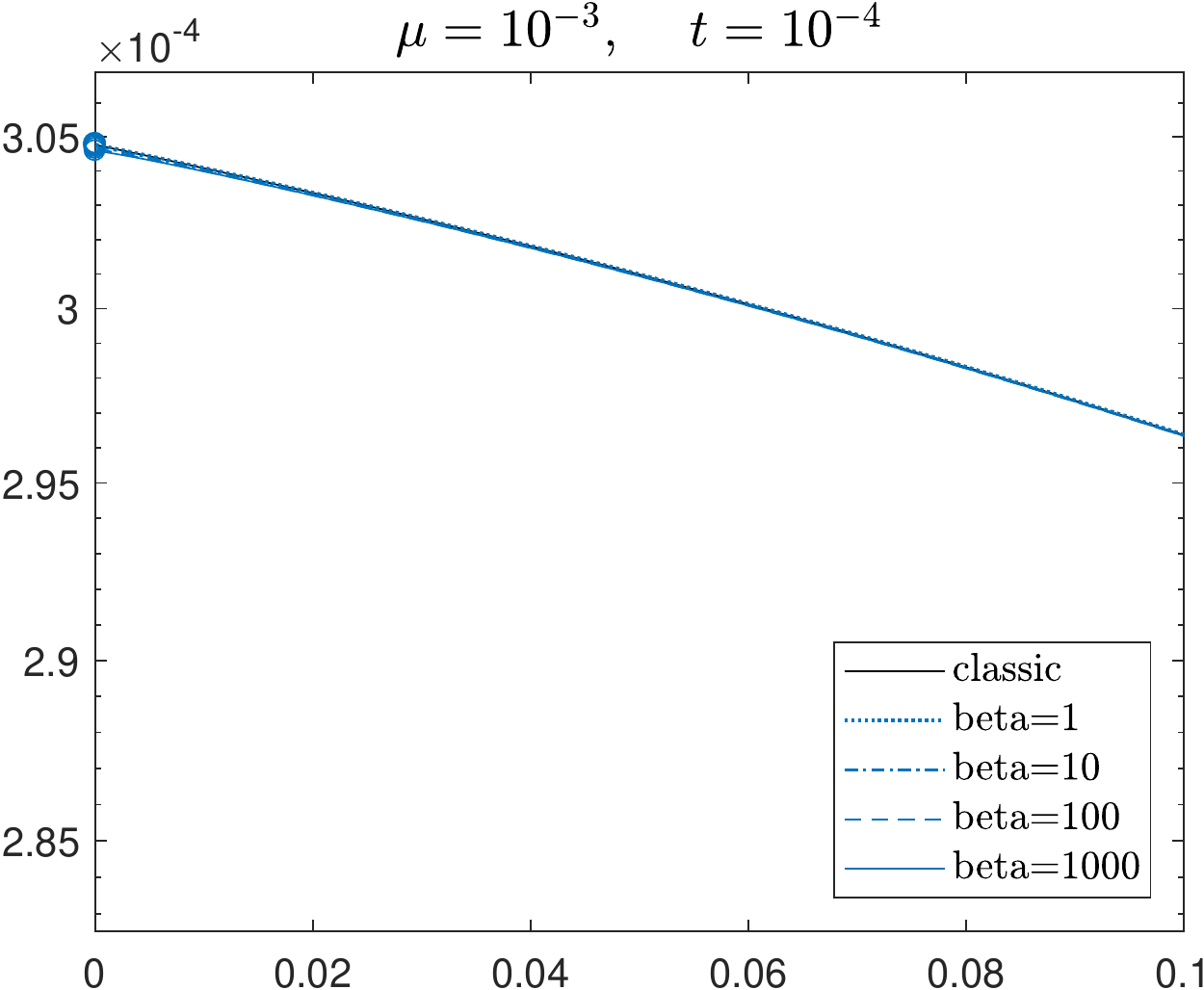}
 \put(-220,160) {{\bf{(a)}}}
 \put(-220,90) {$w$}
 \put(-98,-15) {$\tilde{r}$}
 \hspace{6mm}
 \includegraphics[width=0.45\textwidth]{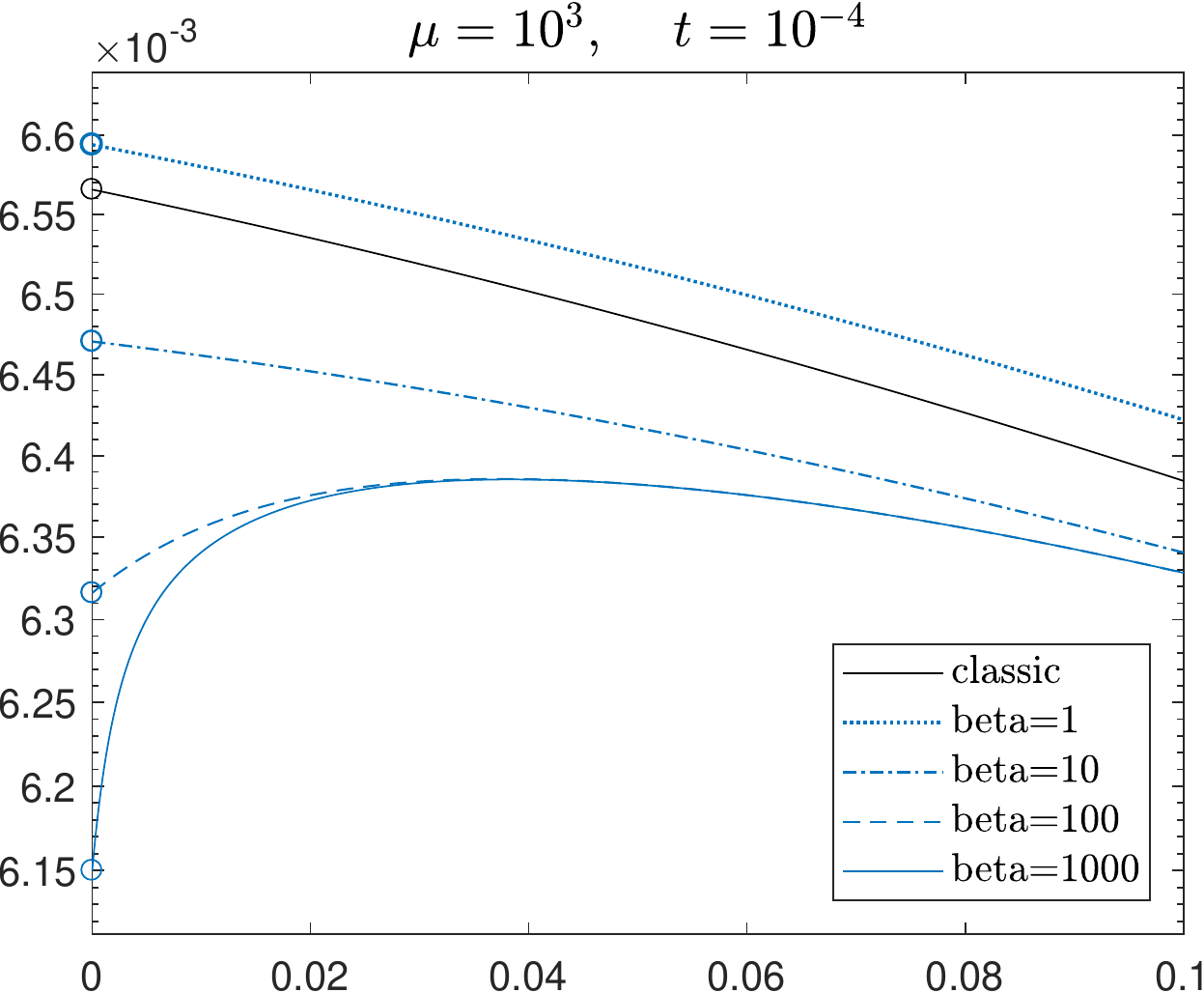}
 \put(-220,160) {{\bf{(b)}}}
 \put(-215,90) {$w$}
 \put(-98,-15) {$\tilde{r}$}

\vspace{2mm}

 \includegraphics[width=0.45\textwidth]{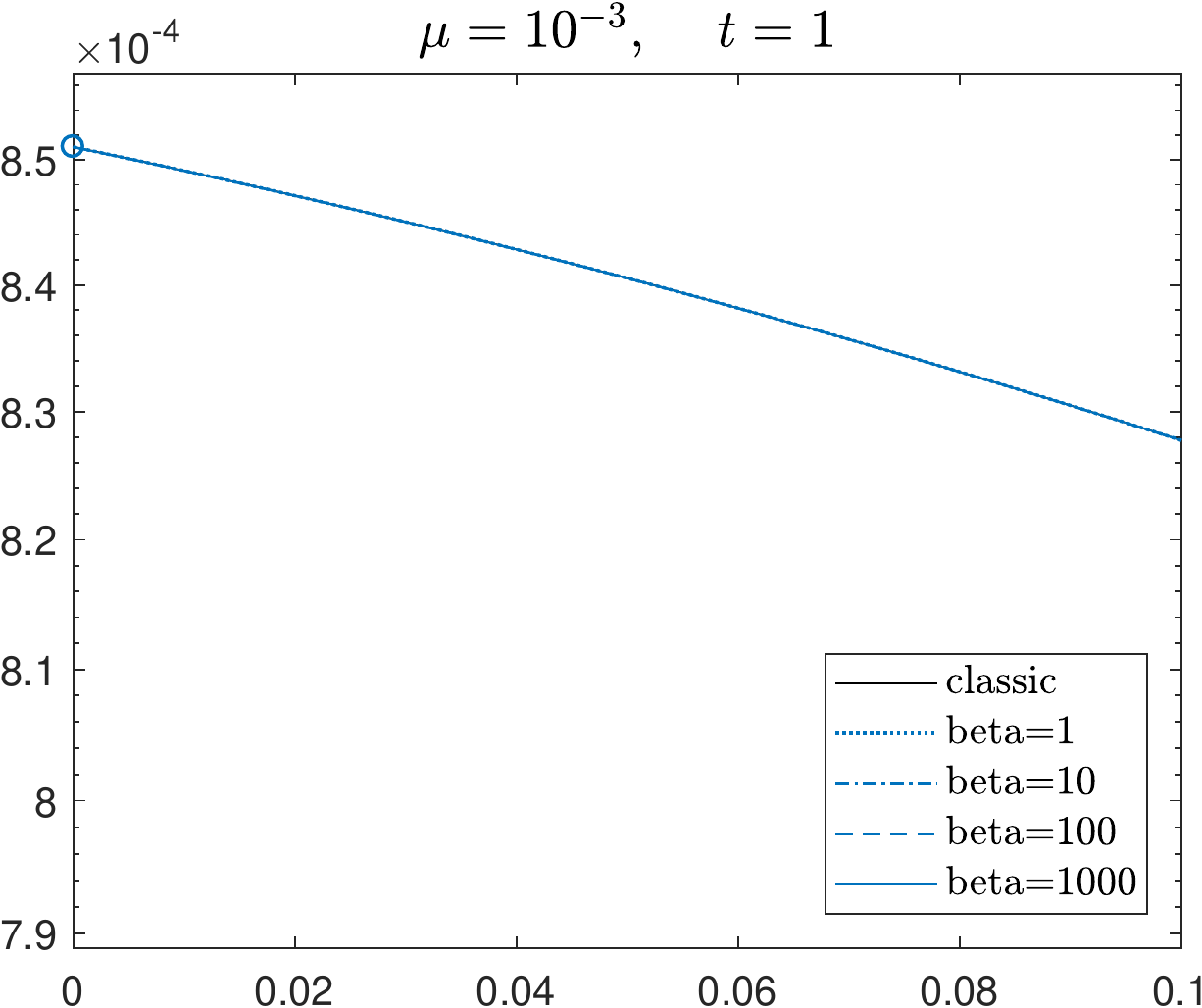}
 \put(-220,160) {{\bf{(c)}}}
 \put(-220,85) {$w$}
 \put(-98,-15) {$\tilde{r}$}
 \hspace{4mm}
 \includegraphics[width=0.45\textwidth]{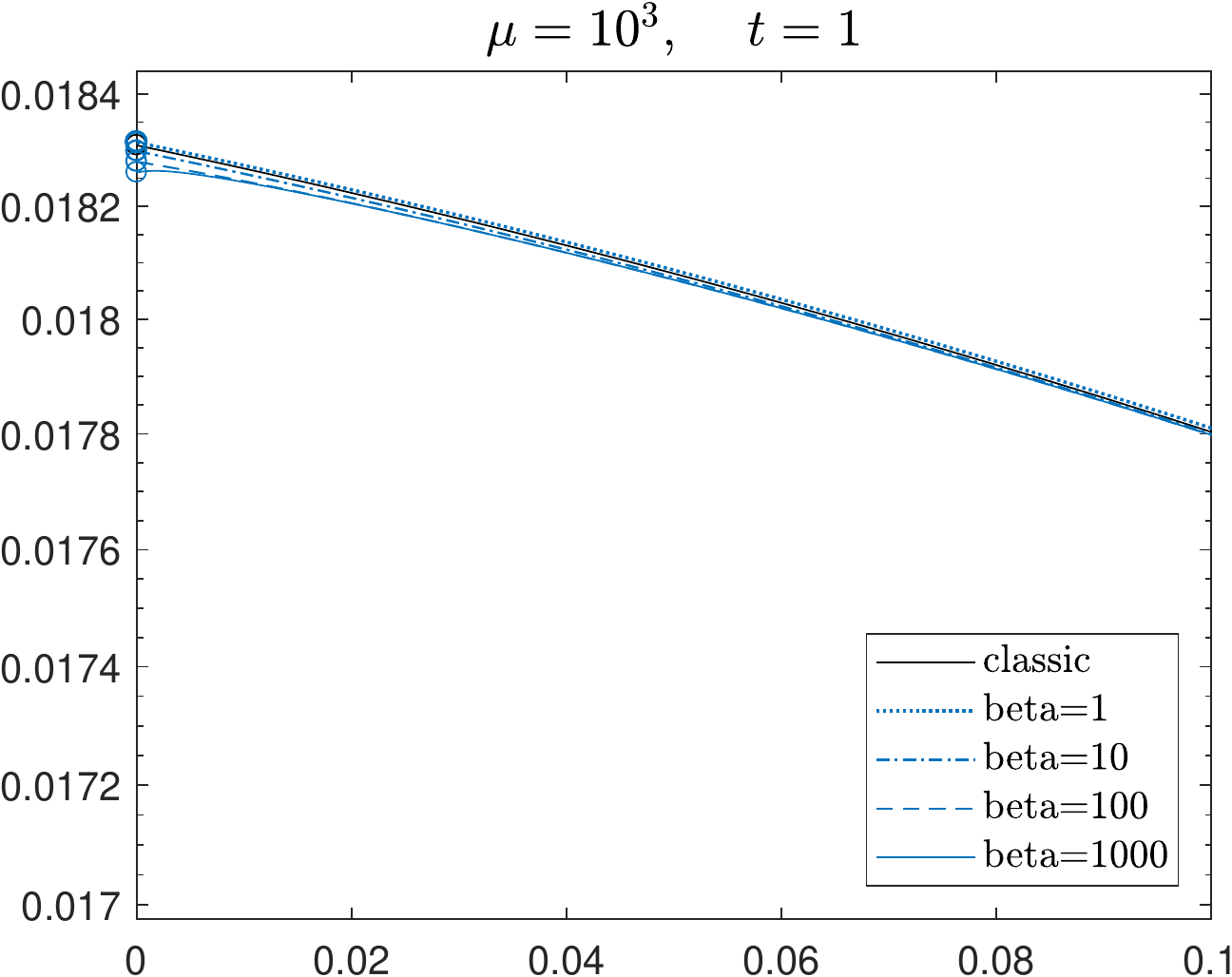}
 \put(-220,160) {{\bf{(d)}}}
 \put(-215,85) {$w$}
 \put(-98,-15) {$\tilde{r}$}
 \caption{The fracture aperture $w(r,t)$ near the wellbore for varying $\beta$ at fixed moments in time. Here we evaluate over normalised spacial variable $\tilde{r}$, taking all material parameters other than viscosity as in Table.~\ref{Table:MatConst}. We show at times {\bf (a)}, {\bf (b)} $t=10^{-4}$ s, {\bf (c)}, {\bf (d)} $t=1$ s, for viscosity {\bf (a)}, {\bf (c)} $\mu=10^{-3}$ Pa {$\cdot$} s, {\bf (b)}, {\bf (d)} $\mu = 10^3$ Pa {$\cdot$} s.}
 \label{VarBeta_1}
\end{figure}

\begin{figure}[t!]
 \center
 \textit{Tangential traction near to the injection point}\par\medskip
 \includegraphics[width=0.45\textwidth]{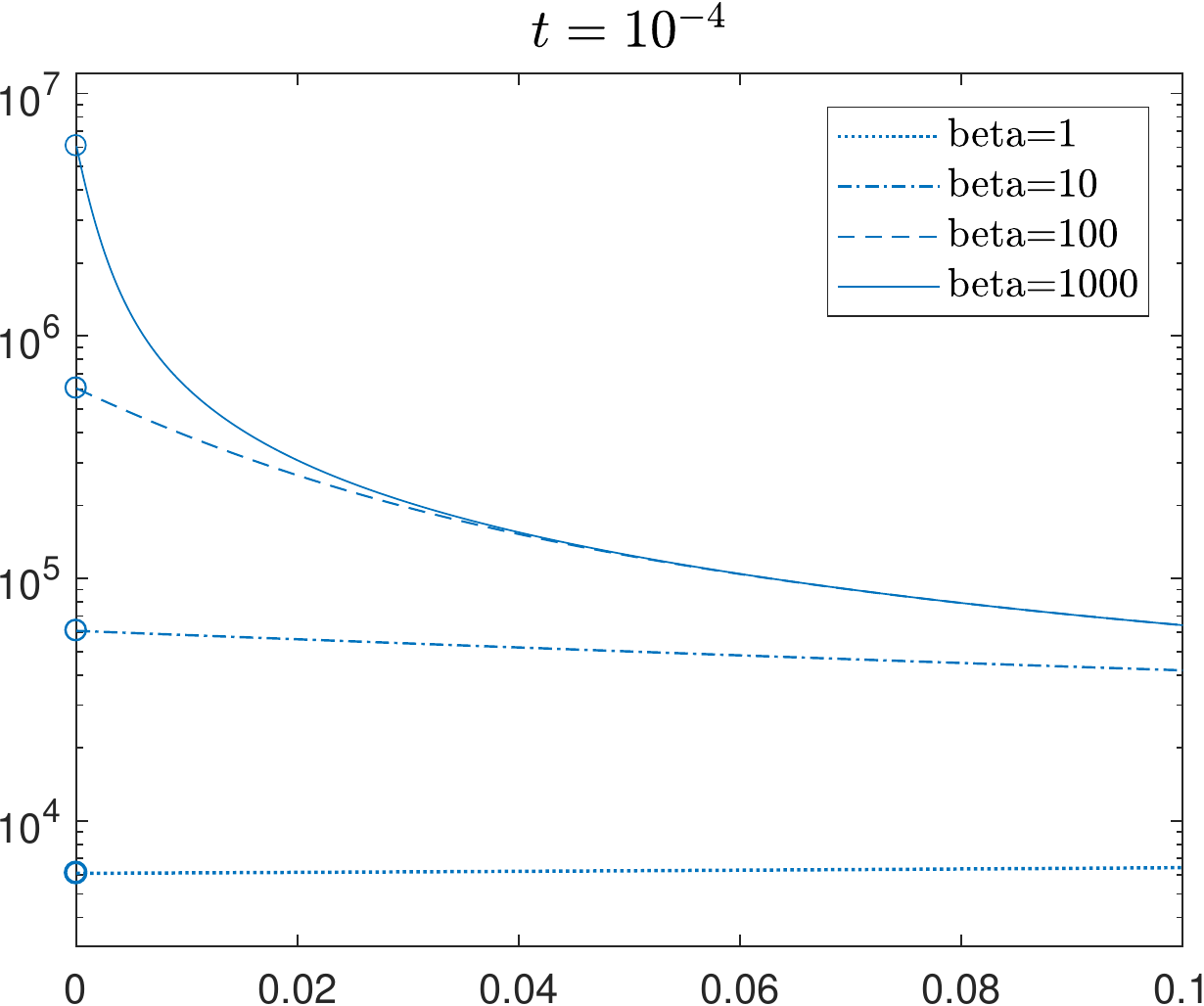}
 \put(-220,160) {{\bf{(a)}}}
 \put(-220,80) {$\tau$}
 \put(-98,-15) {$\tilde{r}$}
 \hspace{6mm}
 \includegraphics[width=0.45\textwidth]{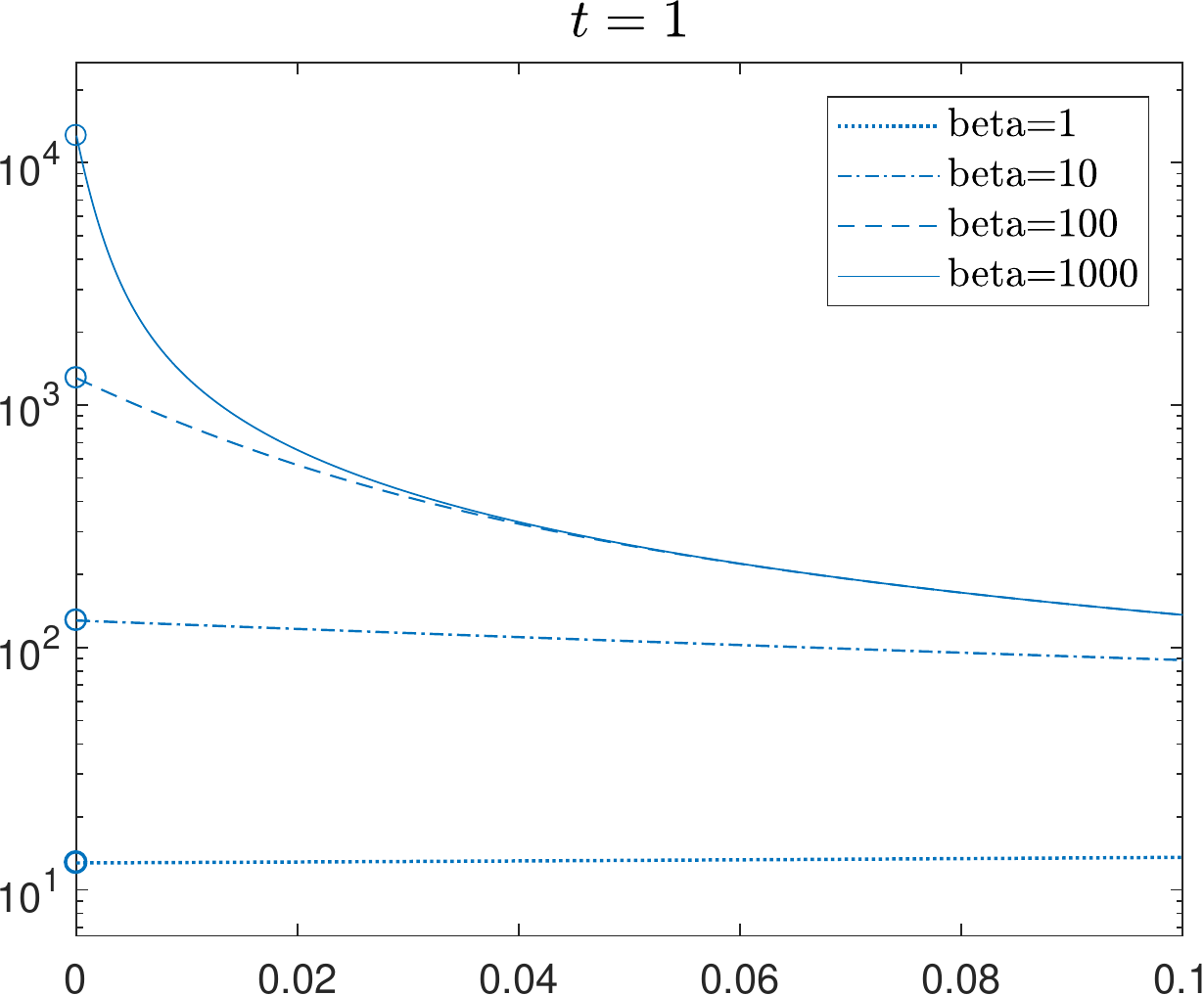}
 \put(-220,160) {{\bf{(b)}}}
 \put(-210,80) {$\tau$}
 \put(-98,-15) {$\tilde{r}$}
 \caption{The tangential traction $\tau$ for varying $\beta$ at fixed moments in time. Here we evaluate over normalised spacial variable $\tilde{r}$, taking all material parameters as in Table.~\ref{Table:MatConst}, including viscosity $\mu=10^{-3}$ {Pa {$\cdot$} s} corresponding to those in Fig.~\ref{VarBeta_1}a,c. We show at times {\bf (a)} $t=10^{-4}$ s, {\bf (b)} $t=1$ s.}
 \label{VarBeta_2}
\end{figure}

We begin by analysing the effect of this parameter on the aperture near the wellbore. The fracture opening near $r=0$ is shown for a variety of $\beta$ in Fig.~\ref{VarBeta_1}, at two different time-steps and for two different values of fluid viscosity $\mu$. The corresponding tangential traction $\tau$ is provided for the case $\mu=10^{-3}$ Pa {$\cdot$} s in Fig.~\ref{VarBeta_2}. Two trends are immediately apparent. Firstly, the effect of the parameter $\beta$ on the fracture opening is dependent upon the viscosity, with a higher fluid viscosity making the system more sensitive to the parameter $\beta$. The second clear trend is that the impact of the shear stress reduces significantly with time, in part as the tangential traction $\tau$ itself reduces rapidly with time as shown in Fig.~\ref{VarBeta_2}. There is very little difference in fracture opening behaviour when $\mu = 10^{-3}$ Pa {$\cdot$} s at $t=10^{-4}$ s, and even this difference has disappeared by $t=1$ s. Similarly, while there is a far greater difference in fracture profile for different $\beta$ when $\mu = 10^3$ Pa {$\cdot$} s, the impact of the tangential traction decreases significantly between $t=10^{-4}$ s and $t=1$ s. One interesting observation is that for the crack aperture, when $\beta = 1$ the case with shear remains above the classical case as $r\to 0$, but acts to decrease it for larger values of $\beta$.


\begin{figure}[t!]
 \center
 \includegraphics[width=0.45\textwidth]{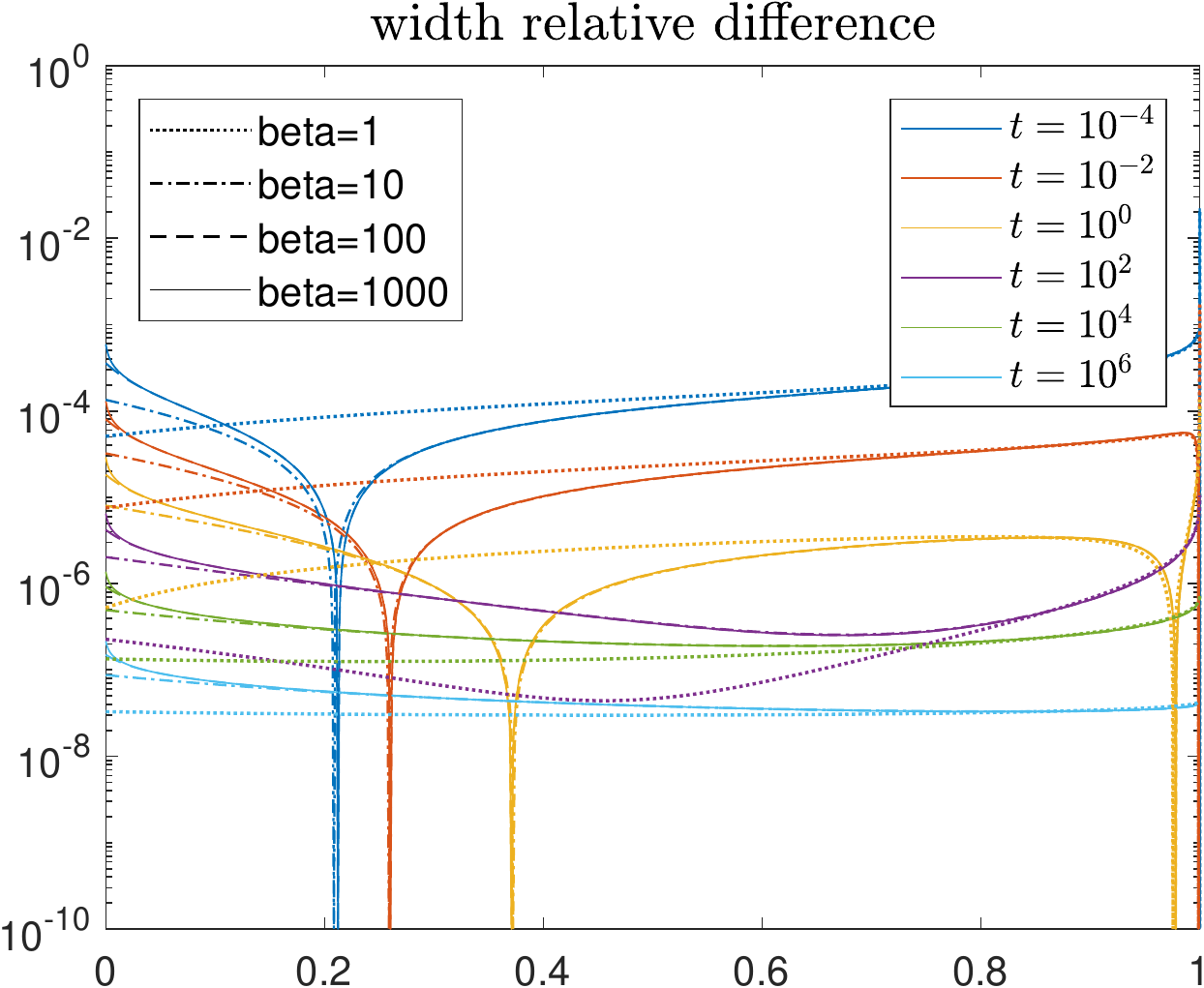}
 \put(-220,160) {{\bf{(a)}}}
 \put(-220,80) {$\Delta w$}
 \put(-98,-15) {$\tilde{r}$}
 \hspace{6mm}
 \includegraphics[width=0.45\textwidth]{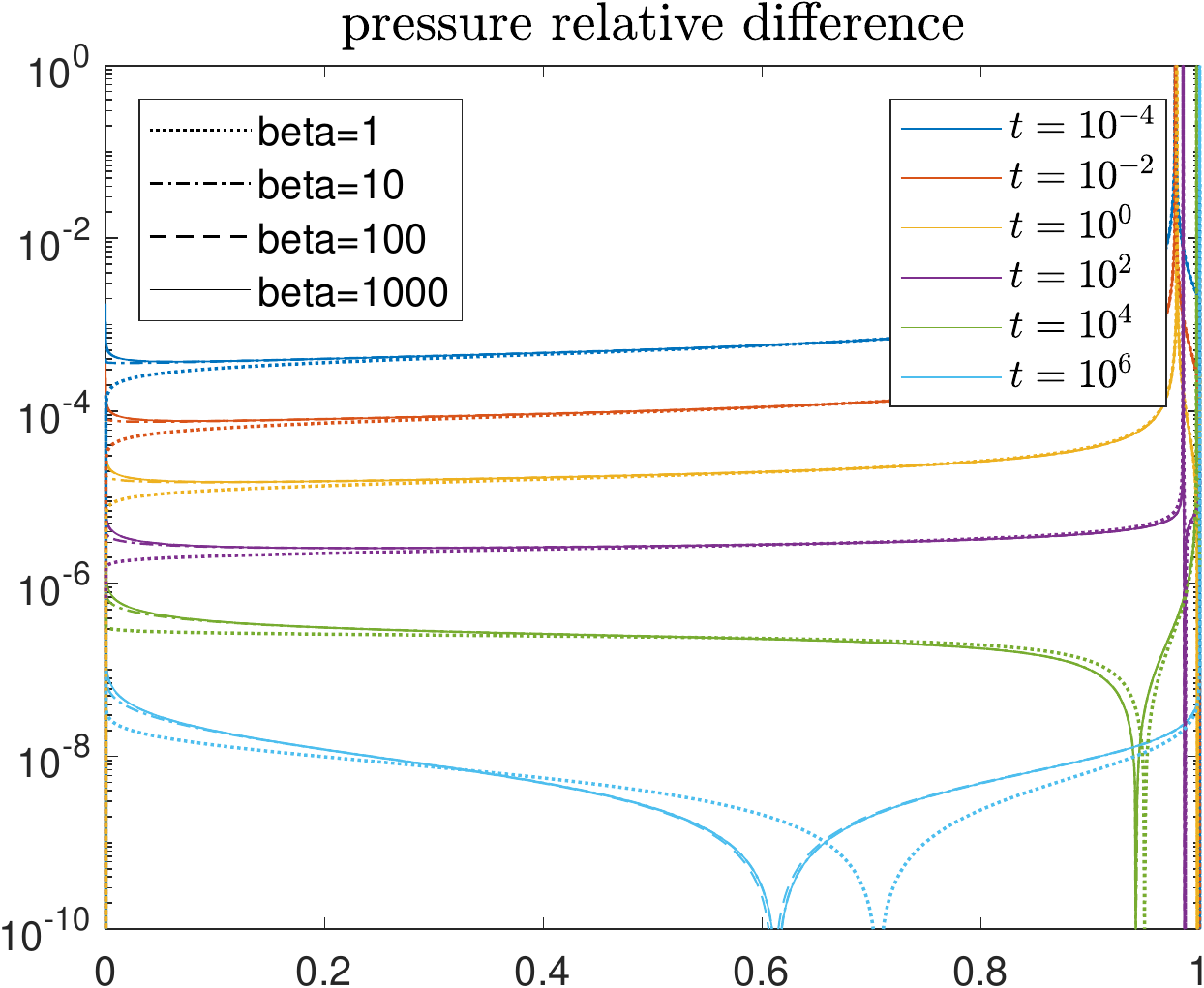}
 \put(-220,160) {{\bf{(b)}}}
 \put(-210,80) {$\Delta p$}
 \put(-98,-15) {$\tilde{r}$}

\vspace{6mm}

 \includegraphics[width=0.45\textwidth]{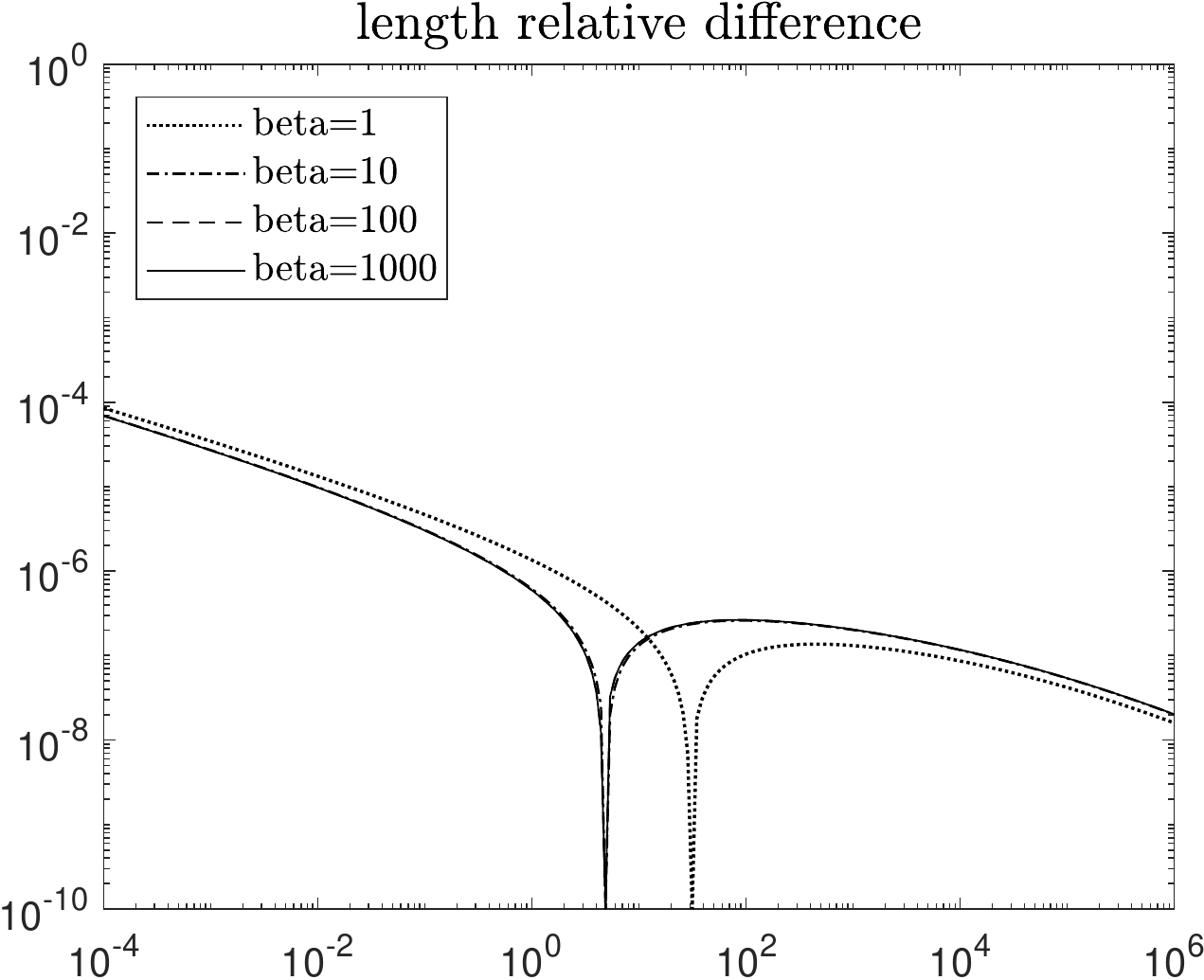}
 \put(-220,160) {{\bf{(c)}}}
 \put(-210,80) {$\Delta l$}
 \put(-98,-15) {$t$}
 \caption{The relative difference $\Delta$ of the {\bf (a)} fracture aperture $w(r,t)$ and {\bf (b)} normal fluid pressure $p (r,t)$, over normalised spacial variable $\tilde{r}$ at fixed moments in time and {\bf (c)} the crack length over time, for various $\beta$ \eqref{New_tau} - \eqref{rstar_defn}. Here all material parameters are taken as in Table.~\ref{Table:MatConst}, including viscosity $\mu=10^{-3}$ {Pa {$\cdot$} s}.}
 \label{VarBeta_4}
\end{figure}

Finally, the values of the relative difference of the fracture aperture $w$, normal fluid pressure $p$ and the crack length $l(t)$, against the case without tangential traction, are provided in Fig.~\ref{VarBeta_4}, for a variety of values $\beta$ at different points in time $t$. It can clearly be seen that the differing behaviour near the wellbore does not significantly effect the impact of the tangential traction on the key system parameters, with the relative difference at the crack tip always exceeding that at the wellbore while in the viscosity dominated regime, and negligible for the toughness dominated regime. Finally, from Fig.~\ref{VarBeta_4}c it can be seen that the impact of the tangential traction on the fracture length is largely independent of $\beta$, indicating that the effect of the stagnant zones remains local to the fracture opening, and does not impact the global parameters in a significant way.


\section{Discussion and conclusions}\label{Sect:Conclusions}

An updated formulation for the radial (penny-shaped) model of hydraulic fracture was created to account for the tangential traction on the fracture walls. This model incorporated the updated fracture criterion, system asymptotics, and accounted for the stagnant zone formation of fluid near the injection point. An examination of the impact of the shear on both the construction of numerical solvers, and the direct quantitative effect on the solution for the time-dependent case, was undertaken.

It was demonstrated that:
\begin{itemize}
 \item As the crack tip asymptotics for the key system parameters no longer vary between the viscosity and toughness dominated regimes, incorporating the tangential traction into numerical solvers eliminates the need to implement methods of transition between the different regimes (similar to that shown for the KGD model \cite{Wrobel2017}). It was also demonstrated that a modified model, utilizing the classical elasticity equation and incorporating the shear effects via the updated integral definition of the stress intensity factor, could accurately compute the updated tip parameters (asymptotic coefficients and stress intensity factors), simplifying the application of this approach. This `automatic switch' can simplify the construction of solvers handling the viscosity-toughness transition, however may make the leading term of the crack tip asymptotics less effective at approximating the system parameters (aperture, fluid pressure). 
 \item The direct impact of the shear stress on the process parameters (aperture, fluid pressure, crack length) is negligible for the vast majority of applications. There was no examined scenario for which the shear stress played any significant role in the toughness dominated regime. In the viscosity dominated regime, it was only possible that the tangential traction may influence the crack development in the case of exceptionally viscous materials, such as magmatic fracture. The model would however require some modification to accurately model such extreme cases.
 \item An estimate for the effect of the tangential traction in the viscosity dominated regime was provided \eqref{Est_delL}. This allows the order of the change in crack length $l(t)$ resulting from the traction to be approximated, which was of the same order to the average of that for the crack aperture $w(r,t)$ and fluid pressure $p(r,t)$ away from the fracture front in all simulations conducted by the authors.
 \item The stagnant zones near the injection point $r=0$ were accounted for by updating the formulation of the tangential traction $\tau$, including the introduction of a new (pre-defined) parameter $\beta$ \eqref{New_tau} - \eqref{rstar_defn}. The aperture profile was shown to have some sensitivity to this parameter for high viscosities, however it diminished rapidly with time. The impact of the tangential traction on the aperture and fluid pressure profiles always appeared to be more significant at the crack tip than that observed at the injection point for the Newtonian fluid considered here, while the impact on global parameters (such as the crack length) does not appear to be significant. 
\end{itemize}

The presented results indicate that the direct impact of shear stress is largely negligible for radial hydraulic fracture. The shear stress may play some role in HF models for use in volcanology, where exceptionally high viscosity magma plays a role, however the current model would need to be modified to provide accurate predictions in this instance. Incorporating the tangential traction does however offer some benefits for the construction of HF algorithms, due to the `automatic switch' between viscosity and toughness dominated regimes, but this has to be balanced against the reduced effectiveness of the crack tip asymptotics to approximate the system parameters.

It should be noted however that the tangential traction may still play an important role for penny-shaped fractures in special cases. For example, the impact of the stagnant zone formation will depend upon the fluid properties, and some classes of non-Newtonian fluids will need to account for this feature (for example, in plastic fluids it may influence the activation of plastic behaviour). The impact of fluid-induced shear could also be significant in cases where the solid behaves as a hyperelastic material.

It is also important to consider the secondary role that tangential traction may play in hydraulic fracture processes, in areas that this model did not account for. For instance, the tangential traction has been shown to play some role in crack redirection \cite{Perkowska2017,Wrobel2019a}, and may induce `wrinkling' in the near-tip region when plasticity is accounted for. These effects however require further investigation.

\section*{Author Contributions}
D.P. derived the initial problem formulation and constructed the self-similar solver. G.D.F. developed the time-dependent solver and performed the quantitative analysis. The final paper was prepared collaboratively between the authors. 

\section*{Acknowledgments}

The authors would also like to thank Prof. Gennady Mishuris, Dr. Michal Wrobel and Dr. Martin Dutko for their fruitful discussions when working on the paper.

\section*{Declaration of Competing Interest}
The authors have no competing interests to declare.

\section*{Funding}
The authors have been funded by Welsh Government via S\^{e}r Cymru Future Generations Industrial Fellowship grant AU224 and the European Union’s Horizon 2020 Research and Innovation Programme under the Marie Sklodowska-Curie grant agreement EffectFact No. 101008140.

\bibliographystyle{plain}
\bibliography{Bibliography}

\begin{appendix}

\section{Derivation of the updated elasticity equation}\label{Updateroo2}

The derivation of the elasticity equation accounting for tangential traction was previously provided in \cite{PeckThesis} (a similar form was also derived independently in \cite{Shen2018}), but is included here for completeness. We consider a 3D penny-shaped crack, defined in polar coordinates by the system $\{r,\theta,z\}$, with associated crack dimensions $\{l(t),w(t)\}$ as the fracture radius and aperture respectively. As the flow is axisymmetric, all variables will be independent of the angle $\theta$.\\

 The equation for the net fluid pressure on the fracture walls (i.e. $p=p_f - \sigma_0$, $\sigma_0$ is the confining stress), including the tangential stress term, is given in Cartesian coordinates $(x_1 , x_2, x_3)$ by \cite{Piccolroaz2013}:
\begin{equation}
 \begin{aligned}
p(r,t) &= \frac{E}{8\pi (1-\nu^2)} \int_{\Omega} \frac{1}{\sqrt{(x_1-\xi_1)^2+(x_3-\xi_3)^2}}\left[\frac{\partial^2 w}{\partial \xi_1^2} + \frac{\partial^2 w}{\partial \xi_3^2} \right] \, d\xi_1 d\xi_3 \\
&\quad - \frac{1-2\nu}{8\pi (1-\nu)} \int_{\Omega}  \frac{1}{\sqrt{(x_1-\xi_1)^2+(x_3-\xi_3)^2}} \left[ \frac{\partial [[p_{\xi_1}]]}{\partial \xi_1} + \frac{\partial [[p_{\xi_3}]]}{\partial \xi_3} \right] \, d\xi_1 d\xi_3 .
 \end{aligned}
\end{equation}
Here $[[x]]$ indicates the jump in $x$ (i.e. $[[p]]=p_+-p_-$), $\Omega=\left\{(x_1,x_3) : \sqrt{x_1^2 + x_3^2}\leq l(t) \right\}$ is the fracture domain, while $E$ and $\nu$ are the Young's modulus and Poisson ratio respectively.

As the problem is invariant of the angle $\theta$, the pressure term can be obtained by transforming this into radial coordinates $(r,\theta)$, integrated with respect to the corresponding variables $(\eta_1 , \eta_2)$. We obtain the relationship:
\begin{equation}
 \begin{aligned}
p(r,t) &=\frac{E}{8\pi(1-\nu^2)}   \int_0^{l(t)} \frac{\partial}{\partial \eta_1}\left(\eta_1 \frac{\partial w(\eta_1 ,t)}{\partial \eta_1}\right)  \int_0^{2\pi} \frac{1}{\sqrt{r^2 +\eta_1^2 - 2r\eta_1 \cos(\theta-\eta_2)}}  \, d\eta_2  d\eta_1 \\
&\quad  - \frac{1-2\nu}{4\pi\left(1-\nu\right)} \int_0^{l(t)} \frac{\partial}{\partial \eta_1}\left( \eta_1 \tau \left(\eta_1 ,t \right) \right)  \int_0^{2\pi}  \frac{1}{\sqrt{r^2+\eta_1^2-2r\eta_1 \cos(\theta-\eta_2)}}   \, d\eta_2 \, d\eta_1 .
 \end{aligned}
\end{equation}
It can be shown that:
\begin{equation}
 \int_0^{2\pi} \frac{1}{\sqrt{r^2 +\eta_1^2 - 2r\eta_1 \cos(\theta-\eta_2)}}  \, d\eta_2 =  \frac{4\EllipticK{\frac{4r \eta_1}{\left(\eta_1+r\right)^2}}}{|\eta_1+r|} ,
\label{int3}
\end{equation}
where $\EllipticK{x}$ is the complete elliptic integral of the first kind \cite{Abramowitz1972}.

Inserting this, before using integration by parts, gives:
\begin{equation}
 p(r,t)= - \int_0^{l(t)} \left[ k_2 \frac{\partial w}{\partial \eta_1} - k_1 \tau (\eta_1) \right] {\cal M}\left(r,\eta_1\right) \, d\eta_1 ,
 \label{newPre}
\end{equation}
substituting the dimensionless variable $\rho=\eta_1/l(t)$, we have:
\begin{equation}
 p(r,t)= - \frac{1}{l(t)} \int_0^1 \left[ k_2 \frac{\partial w(\rho l(t))}{\partial \rho} - k_1 l(t) \tau (\rho l(t)) \right] {\cal M}\left(\frac{r}{l(t)},\rho\right) \, d\rho ,
 \label{newPre}
\end{equation}
where:
\begin{equation}
 \begin{aligned}
{\cal M}(\tilde{r},\rho) &= \frac{1}{2(\tilde{r}+\rho)}\EllipticK{\frac{4\tilde{r}\rho}{\left(\rho+\tilde{r}\right)^2}} - \frac{1}{2(\tilde{r}-\rho)}\EllipticE{\frac{4\tilde{r}\rho}{\left(\rho+\tilde{r}\right)^2}} \\
&=    \frac{1}{2(\tilde{r}+\rho)}\EllipticK{1-\left(\frac{\rho-\tilde{r}}{\rho+\tilde{r}}\right)^2} - \frac{1}{2(\tilde{r}-\rho)}\EllipticE{1-\left(\frac{\rho-\tilde{r}}{\rho+\tilde{r}}\right)^2} ,
 \label{KerNew}
 \end{aligned}
\end{equation}
\begin{equation}
k_1 = \frac{1-2\nu}{\pi (1-\nu)}, \quad k_2 = \frac{E}{2\pi (1-\nu^2)} .
\end{equation}
Here $\EllipticE{x}$ is the complete elliptic integral of the second kind \cite{Abramowitz1972}. It can be shown numerically that, within the corresponding domains ($0\leq \tilde{r} \leq1$, $0\leq \rho \leq 1$),  this kernel function ${\cal M}$ is merely an alternative representation of the standard kernel for this problem \cite{Keer1977}:
\begin{equation}
{\cal M}\left[\tilde{r} , \rho \right] = \begin{cases} \frac{1}{\tilde{r}} \EllipticK{\frac{\rho^2}{\tilde{r}^2}} + \frac{\tilde{r}}{\rho^2 - \tilde{r}^2} \EllipticE{ \frac{\rho^2}{\tilde{r}^2}} , & \tilde{r}>\rho \\  \frac{\rho}{\rho^2 - \tilde{r}^2} \EllipticE{ \frac{\tilde{r}^2}{\rho^2}} , & \rho>\tilde{r} \end{cases}
\label{M1b}
\end{equation}
Additionally, it is worth noting that the constants $k_1$, $k_2$, are identical to those obtained in the KGD case \cite{Wrobel2017}.

With the elasticity equation with tangential stresses obtained, it is clear that the next step is to invert the operator and obtain the inverse relation. This is achieved by noting that we can place \eqref{newPre} in the form:
\begin{equation}
p(r,t) = - \int_0^{l(t)}  g^\prime (x) {\cal M}\left( r , x \right) \, dx , \quad 0<r < 1 ,
\end{equation}
where:
\begin{equation}
g ( r ) = \int_r^{l(t)} \left( k_1 \tau (s,t) - k_2 \frac{\partial w}{\partial s} \right) \, ds = k_2 w(r,t) + k_1 \int_{r}^{l(t)} \tau (s,t) \, ds .
\end{equation}

From this, it immediately follows that the inverse relation must be (compare with the classical result, see e.g. \cite{Savitski2002}):
\begin{equation}
k_2 w(r,t) + k_1 \int_{r}^{l(t)} \tau (s ,t) \, ds =  \frac{4}{\pi^2} l(t) \int_{r/l(t)}^1 \frac{\xi}{\sqrt{\xi^2-\left(r/l(t)\right)^2}} \int_0^1 \frac{\eta p(\eta \xi l(t),t)}{\sqrt{1-\eta^2}} \, d\eta d\xi  ,
 \label{Nobel_InvElast_wTang}
\end{equation}

Following the steps previously outlined in \cite{Peck2018a}, this can alternatively be written in the form:
{\begin{equation}
k_2 w(r,t) + k_1 \int_{r}^{l(t)} \tau \left(s ,t\right) \, ds = \quad \quad \quad \quad \quad
 \label{Nobel_InvElast_wTang2}
\end{equation}
$$
 \frac{4}{\pi^2} l(t) \left[ \int_0^1 \frac{\partial p(y l(t),t)}{\partial y} {\cal K}\left(y, \frac{r}{l(t)}\right) \, dy + \sqrt{1-\left(\frac{r}{l(t)}\right)^2} \int_0^1 \frac{\eta p(\eta l(t) , t)}{\sqrt{1-\eta^2}} \, d\eta \right]  ,
$$}
where:
\begin{equation}
{\cal K}(y,\tilde{r}) = y\left[ \IncEllipticE{\arcsin(y)}{\frac{\tilde{r}^2}{y^2}} - \IncEllipticE{\arcsin(\psi  )}{\frac{\tilde{r}^2}{y^2}}\right] , \quad \psi =\min\left(\frac{y}{\tilde{r}},1\right),
\end{equation}
with $E\left( \phi \, | \, m \right)$ denoting the incomplete elliptic integral of the second kind.

\section{Normalized form of the governing equations}\label{Append_Norm}

\subsection{Normalization}

We introduce the following normalization scheme:
\begin{equation}
 \begin{aligned}
&\tilde{r} = \frac{r}{l(t)} , \quad \tilde{t}=\frac{t}{t_n}, \quad \tilde{w}(\tilde{r},\tilde{t})=\frac{w(r,t)}{l_*}  , \quad L(\tilde{t})=\frac{l(t)}{l_*} , \quad \tilde{q}_l (\tilde{r}, \tilde{t}) = \frac{t_n}{l_*} q_l (r,t) ,& \\
&\tilde{q}(\tilde{r},\tilde{t}) = \frac{t_n}{l_*^2} q(r,t) , \quad \tilde{Q}_0(\tilde{t})=\frac{t_n}{2 \pi l_*^2 l(t)}Q_0(t), \quad \tilde{v}(\tilde{r},\tilde{t})=\frac{t_n}{l_*}v(r,t), \quad \tilde{\tau}(\tilde{r},\tilde{t}) = \frac{t_n}{M}\tau (r,t), & \\
& \tilde{p}(\tilde{r},\tilde{t}) = \frac{t_n}{M} p(r,t) , \quad \tilde{K}_{\{Ic,I,f\}} = \frac{\gamma}{\sqrt{l_*}} K_{\{Ic,I,f\}} , \quad \tilde{\chi}(\tilde{r}) = \frac{\chi(r,t)}{l(t)}, \quad t_n=\frac{M}{k_2} , &
\end{aligned}
\label{NobelNormalizations1}
\end{equation}
where $\tilde{r}\in \left[0,1\right]$ and $l_*$ is chosen for convenience.

Under the normalization scheme provided in \eqref{NobelNormalizations1}, the Poiseuille equation provides the following relation for the fluid velocity \eqref{NobelPVInit}:
\begin{equation}
\tilde{v} = -\frac{\tilde{w}^2}{L(\tilde{t})}\frac{\partial \tilde{p}}{\partial \tilde{r}} ,
\label{Npv}
\end{equation}
while the tangential (sheer) stress is now given by \eqref{New_tau}:
\begin{equation}
\tilde{\tau} = -\frac{1}{2} \frac{\tilde{\chi}}{L(\tilde{t})} \tilde{w} \frac{\partial \tilde{p}}{\partial \tilde{r}} ,
 \label{Nobel_taudefNorm}
\end{equation}
where $\tilde{\chi}$ \eqref{rstar_defn} is given by
\begin{equation} \label{final_ss_rstar}
\tilde{\chi} (\tilde{r}) = 1 - \left( 1 - \tilde{r}\right)^\beta , \quad \beta\geq 1.
\end{equation}

As such the fluid mass balance equation \eqref{Nobelfluidmass}, alongside the global balance equation \eqref{Nobel_fluidbalance1}, become:
\begin{equation}
\frac{\partial \tilde{w}}{\partial \tilde{t}} - \frac{L^\prime (\tilde{t})}{L(\tilde{t})} \tilde{r} \frac{\partial \tilde{w}}{\partial \tilde{r}} + \frac{1}{\tilde{r} L(\tilde{t})} \frac{\partial}{\partial \tilde{r}} \left(\tilde{r} \tilde{w} \tilde{v}\right) + \tilde{q}_l = 0 ,
 \label{Nobel_fluidmassNorm}
\end{equation}
\begin{equation}
\int_0^1 \tilde{r} \left[ L^2 (\tilde{t}) \tilde{w}(\tilde{r},\tilde{t}) - L^2 (0) \tilde{w}_* (\tilde{r})\right] \, d\tilde{r} + \int_0^{\tilde{t}} \int_0^1 \tilde{r} L^2 ( s ) \tilde{q}_l (\tilde{r} , s ) \, d\tilde{r} \, ds = \int_0^{\tilde{t}} L( s) \tilde{Q}_0 (s) \, ds .
 \label{Nobel_globalbalanceNorm}
\end{equation}

The elasticity equation \eqref{newPre2} takes the form:
\begin{equation}
\tilde{p}(\tilde{r},\tilde{t}) = -\frac{1}{L(\tilde{t})} \int_0^1 \left[ \frac{\partial \tilde{w}}{\partial \eta} - k_1 L(\tilde{t}) \tilde{\tau}(\eta,\tilde{t}) \right] {\cal M}\left[ \tilde{r} , \eta \right] \, d\eta ,
 \label{Nobel_ElastNorm1}
\end{equation}
alongside associated inverse \eqref{Nobel_InvElast}:
\begin{equation}
\tilde{w}(\tilde{r},\tilde{t}) + k_1 L(\tilde{t}) \int_{\tilde{r}}^1 \tilde{\tau}(s,\tilde{t}) \, ds = \frac{4}{\pi^2 }L(\tilde{t}) \left[ \int_0^1 \frac{\partial \tilde{p}(y,\tilde{t})}{\partial y} {\cal K}(y,\tilde{r}) \, dy + \sqrt{1-\tilde{r}^2} \int_0^1 \frac{\eta \tilde{p}(\eta,\tilde{t})}{\sqrt{1-\eta^2}} \, d\eta \right] ,
 \label{NormElast2}
\end{equation}
where the kernel ${\cal K}$ is given by \eqref{Almighty_Kernel_K}.
By evaluating the asymptotic limit of \eqref{NormElast2} at the crack tip, it can be shown that:
\begin{equation}
\tilde{w}_0 + k_1 \tilde{w}_0 \tilde{p}_0 = \frac{4\sqrt{2}}{\pi^2} L(\tilde{t}) \int_0^1 \frac{\eta \tilde{p}(\eta ,\tilde{t})}{\sqrt{1-\eta^2}} \, d\eta ,
 \label{NewElast}
\end{equation} 
which replaces the standard integral definition of the stress intensity factor.

The boundary conditions for the problem \eqref{Nobel_SourceIntense}-\eqref{Nobel_TipBC} are now given by:
\begin{equation}
\lim_{\tilde{r}\to 0} \tilde{r} \tilde{w} \tilde{v} = \tilde{Q}_0 ,
\label{NQ0}
\end{equation}
\begin{equation}
\tilde{w} (1 , \tilde{t}) = 0 , \quad \tilde{q}(1,\tilde{t})=0 ,
\end{equation}
where the system has initial conditions \eqref{Nobel_InitCond}:
\begin{equation}
\tilde{w}(\tilde{r},0) = \tilde{w}_*(r) , \quad L(0) = L_0 ,
\end{equation} 

The crack tip asymptotics \eqref{wasym1}, \eqref{pasym1}, \eqref{vasym1} now take the form:
\begin{equation}
\tilde{w}(\tilde{r},\tilde{t}) = \tilde{w}_0 (\tilde{t}) \sqrt{1-\tilde{r}} + \tilde{w}_1 (\tilde{t}) \left(1-\tilde{r}\right) + \tilde{w}_2 (\tilde{t}) \left(1-\tilde{r}\right)^{\frac{3}{2}}\log\left(1-\tilde{r}\right)+\hdots , \quad \tilde{r}\to 1,
 \label{wasym2}
\end{equation}
\begin{equation}
\tilde{p}(\tilde{r},\tilde{t}) = \tilde{p}_0 (\tilde{t}) \log\left(1-\tilde{r}\right) + \tilde{p}_1 (\tilde{t}) + \tilde{p}_2 (\tilde{t}) \sqrt{1-\tilde{r}} + \tilde{p}_3 \left(1-\tilde{r}\right)\log\left(1-\tilde{r}\right) + \hdots , \quad \tilde{r}\to 1,
 \label{pasym2}
\end{equation}
\begin{equation}
\tilde{v}(\tilde{r},\tilde{t}) = \tilde{v}_0 (\tilde{t}) + \tilde{v}_1 (\tilde{t}) \sqrt{1-\tilde{r}} + \hdots , \tilde{r}\to 1 ,
 \label{vasym2}
\end{equation}

We note that we can rewrite the parameter $\bar{\omega}$ \eqref{Nobel_StessIntense}$_3$ as:
\begin{equation}
\tilde{\omega} = \frac{\tilde{p}_0}{\pi (1-\nu ) - \tilde{p}_0} , \quad 0 < \tilde{p}_0 < \pi (1-\nu ),
 \label{Nobel_OmegaNorm}
\end{equation}
as such, the first term of the aperture asymptotics at the fracture tip are given by \eqref{w02}, \eqref{Nobel_w03}:
\begin{equation}
\tilde{w}_0 (\tilde{t}) = \sqrt{L(\tilde{t})} \frac{1+\tilde{\omega}}{\sqrt{1+4(1-\nu)\tilde{\omega}}} \tilde{K}_{Ic}  = \sqrt{L(\tilde{t})} \left[ \tilde{K}_I + \tilde{K}_f \right] ,
\label{w024}
\end{equation}
while the stress intensity factors \eqref{Nobel_StessIntense} are described by:
\begin{equation}
\tilde{K}_I = \frac{\tilde{K}_{Ic}}{\sqrt{1+4(1-\nu)\tilde{\omega}}} , \quad \tilde{K}_f = \frac{\tilde{K}_{Ic} \tilde{\omega}}{\sqrt{1+4(1-\nu) \tilde{\omega}}} ,
 \label{Nobel_StressIntenseNorm}
\end{equation}

The Steffan condition \eqref{Nobel_SpeedEq}, utilizing the Poiseuille equation \eqref{Npv} and terms from the asymptotic representation \eqref{wasym2}-\eqref{pasym2}, can be expressed as:
\begin{equation}
\frac{dL}{d\tilde{t}} = \tilde{v}_0 (\tilde{t}) = -\frac{1}{L(\tilde{t})} \lim_{\tilde{r}\to 1} \tilde{w}^2 \frac{\partial \tilde{p}}{\partial \tilde{r}} = \frac{\tilde{w}_0^2 \tilde{p}_0}{L(\tilde{t})} ,
\label{Steffan1}
\end{equation}
Utilizing \eqref{w024}, we can rewrite this condition as follows:
\begin{equation}
\frac{1}{\tilde{K}_{Ic}^2} \tilde{v}_0 = \tilde{p}_0 F \left(\tilde{p}_0 \right) ,
 \label{NewRel00k1}
\end{equation}
where:
\begin{equation}
F\left( \tilde{p}_0 \right) = \frac{\pi^2 \left(1-\nu\right)^2}{\left[ \pi\left(1-\nu\right)+\left(3-4\nu\right)\tilde{p}_0\right]\left[\pi\left(1-\nu\right)-\tilde{p}_0\right]} . 
\label{Fdef2}
\end{equation}
Noting the above definition, we can rewrite \eqref{w024} in the form:
\begin{equation}
\tilde{w}_0 (\tilde{t}) = \tilde{K}_{Ic} \sqrt{L(\tilde{t}) F(\tilde{p}_0)} ,
 \label{NewRel00k2}
\end{equation}
Further, by integrating \eqref{Steffan1}, we can obtain a formula for the crack length:
\begin{equation}
L(\tilde{t}) = \sqrt{L^2 (0) + 2\int_0^{\tilde{t}} \tilde{w}_0^2 (s) \tilde{p}_0 (s) \, ds } .
\label{Norm_CrackLength1}
\end{equation}

\subsection{The self-similar formulation}

As we are incorporating the effect of tangential traction, it is not possible to obtain a self-similar solution of power-law type. Instead, an exponential variant must be obtained, similar to that utilized in \cite{Spence1985}. We formulate utilize the following separation of variables:
\begin{equation}
\tilde{Q}_0 (\tilde{t}) = \hat{Q}_0 e^{2\alpha \tilde{t}} .
\label{SSdef1Nobel}
\end{equation}
This can be obtained by assuming the parameters take the form:
$$
\tilde{w}(\tilde{r},\tilde{t}) = \sqrt{L_0 \hat{Q}_0} e^{\alpha \tilde{t}} \hat{w}(\tilde{r}) , \quad L(t) = L_0^{\frac{3}{2}} \sqrt{\hat{Q}_0} e^{\alpha \tilde{t}} , \quad \tilde{p}(\tilde{r},\tilde{t}) = \hat{p}(\tilde{r}) ,
$$
\begin{equation}
\tilde{q}_l (\tilde{r},\tilde{t}) = \alpha \sqrt{L_0 \hat{Q}_0} e^{\alpha \tilde{t}} \hat{q}_l (\tilde{r}), \quad \tilde{v}(\tilde{r},\tilde{t}) = \sqrt{\frac{\hat{Q}_0}{L_0}} e^{\alpha \tilde{t}} \hat{v}(\tilde{r}) , \quad \tilde{q}(\tilde{r},\tilde{t}) = \hat{Q}_0 e^{2\alpha \tilde{t}} \hat{q}(\tilde{r}) ,
\end{equation}
$$
\tilde{K}_{\{Ic,I,f\}} = \left(L_0 \hat{Q}_0\right)^{\frac{1}{4}} e^{\frac{\alpha \tilde{t}}{2}} \hat{K}_{\{Ic,I,f\}} , \quad \tilde{\tau}(\tilde{r},\tilde{t})=\hat{\tau}(\tilde{r}),
$$
where:
\begin{equation}
\hat{v}_0 = \hat{v} (1) , \quad L_0 = \sqrt{\frac{\hat{v}_0}{\alpha}} .
\label{L0def}
\end{equation}

Under this scheme, the Poiseuille equation provides the following relation for the fluid velocity \eqref{NobelPVInit}:
\begin{equation}
\hat{v}(\tilde{r}) = -\hat{w}^2 (\tilde{r}) \frac{d \hat{p}(\tilde{r})}{d \tilde{r}},
\label{PvSS5}
\end{equation}
The tangential traction \eqref{New_tau} is now given as
\begin{equation}
\hat{\tau}(\tilde{r}) = -\frac{\tilde{\chi}(\tilde{r})}{2L_0} \hat{w}(\tilde{r}) \frac{d\hat{p}}{d\tilde{r}} ,
\end{equation}
where the term $\tilde{\chi}$  is given by \eqref{final_ss_rstar}.

As such the fluid mass balance equation \eqref{Nobelfluidmass}, alongside the global balance equation \eqref{Nobel_fluidbalance1}, become:
\begin{equation}
\hat{w} - \tilde{r} \frac{d\hat{w}}{d\tilde{r}} + \frac{1}{\hat{v}_0 \tilde{r}}\frac{d}{d\tilde{r}} \left(\tilde{r}\hat{w}\hat{v}\right) + \hat{q}_l = 0 ,
 \label{Nobel_SSFluidMass5}
\end{equation}
\begin{equation}
3\int_0^1 \tilde{r} \hat{w}(\tilde{r}) \, d\tilde{r} + \int_0^1 \tilde{r} \hat{q}_l (\tilde{r}) \, d\tilde{r}  = \frac{1}{\hat{v}_0} ,
 \label{SSbalance5}
\end{equation}
The elasticity equation \eqref{newPre2} takes the form:
\begin{equation}
\hat{p}(\tilde{r}) = - \frac{1}{L_0}\int_0^1 \left[ \frac{d\hat{w}}{d\eta} + \frac{k_1}{2}\tilde{r}(\eta)\hat{w}(\eta)\frac{d\hat{p}}{d\eta} \right] {\cal M}\left[\tilde{r},\eta\right] \, d\eta ,
\end{equation}
the associated inverse \eqref{Nobel_InvElast} is simplified following the approach from \cite{Peck2018a}, to become:
\begin{equation}
\hat{w}(\tilde{r}) = \frac{4}{\pi^2} L_0 \left[ \int_0^1 \frac{d\hat{p}}{dy} {\cal K}(y,\tilde{r}) \, dy + \sqrt{1-\tilde{r}^2}\int_0^1 \frac{\eta \hat{p}(\eta)}{\sqrt{1-\eta^2}} \, d\eta  \right] + \frac{k_1}{2} \int_{\tilde{r}}^1 \tilde{\chi}(s) \hat{w}(s) \frac{d\hat{p}}{ds} \, ds ,
 \label{Nobel_SSElasticity8}
\end{equation}
By evaluating the asymptotic limit of \eqref{Nobel_SSElasticity8} at the crack tip, it can be shown that:
\begin{equation}
\hat{w}_0 + k_1 \hat{w}_0 \hat{p}_0 = \frac{4\sqrt{2}}{\pi^2} L_0 \int_0^1 \frac{\eta \hat{p}(\eta)}{\sqrt{1-\eta^2}} \, d\eta ,
 \label{Nobel_SSElasticity9}
\end{equation}
which replaces the standard integral definition of the stress intensity factor.

Meanwhile, the source intensity and boundary conditions are given by:
\begin{equation}
\lim_{\tilde{r}\to 0} \tilde{r} \hat{w} \hat{v} = 1,
\end{equation}
\begin{equation}
\hat{w}(1) = 0 , \quad \hat{q}(1) = 0 ,
\end{equation}
The crack tip asymptotics \eqref{wasym1}, \eqref{pasym1}, \eqref{vasym1} now take the form:
\begin{equation}
\hat{w}(\tilde{r}) = \hat{w}_0 \sqrt{1-\tilde{r}} + \hat{w}_1 \left(1-\tilde{r}\right)+\hat{w}_2\left(1-\tilde{r}\right)^{\frac{3}{2}}\log\left(1-\tilde{r}\right) + \hdots , \quad \tilde{r}\to 1,
 \label{wasym8}
\end{equation}
\begin{equation}
\hat{p}(\tilde{r}) = \hat{p}_0 \log\left(1-\tilde{r}\right) + \hat{p}_1 + \hat{p}_2 \sqrt{1-\tilde{r}} + \hat{p}_3 \left(1-\tilde{r}\right)\log\left(1-\tilde{r}\right) + \hdots , \quad \tilde{r}\to 1,
 \label{pasym8}
\end{equation}
\begin{equation}
\hat{v}(\tilde{r}) = \hat{v}_0 + \hat{v}_1 \sqrt{1-\tilde{r}} + \hdots , \quad \tilde{r}\to 1,
 \label{vasym8}
\end{equation}
where:
\begin{equation}
\hat{v}_0 = \hat{w}_0^2 \hat{p}_0 , \quad \hat{v}_1 = \frac{\hat{w}_0^2 \hat{p}_2 + 4\hat{w}_0 \hat{w}_1 \hat{p}_0}{2} , \quad L_0 = \hat{w}_0 \sqrt{\frac{\hat{p}_0}{\alpha}}. 
 \label{L0rel}
\end{equation}

We note that we can rewrite the parameter $\tilde{\omega}$ \eqref{Nobel_StessIntense}$_3$ as:
\begin{equation}
\hat{\omega} = \frac{\hat{p}_0}{\pi (1-\nu ) - \hat{p}_0} .
 \label{SSomega5}
\end{equation}
As such, the first term of the aperture asymptotics at the fracture tip \eqref{Nobel_w03}-\eqref{Nobel_w0344} is given by:
\begin{equation}
\hat{w}_0 = \sqrt{L_0} \frac{1+\hat{\omega}}{\sqrt{1+4(1-\nu)\hat{\omega}}} \hat{K}_{Ic}  =  \sqrt{L_0}\left[ \hat{K}_I + \hat{K}_f \right] = \hat{K}_{Ic} \sqrt{L_0 \hat{F}(\hat{p}_0 )},
\label{w0ss5}
\end{equation}
with $\hat{F}$ being simplified from \eqref{FdefOld}, to give:
\begin{equation}
\hat{F}\left( \hat{p}_0 \right) = \frac{\pi^2 \left(1-\nu\right)^2}{\left[ \pi\left(1-\nu\right)+\left(3-4\nu\right)\hat{p}_0\right]\left[\pi\left(1-\nu\right)-\hat{p}_0\right]} . 
\label{Fdef}
\end{equation}
while the stress intensity factors \eqref{Nobel_StessIntense} are described by:
\begin{equation}
\hat{K}_I = \frac{\hat{K}_{Ic}}{\sqrt{1+4(1-\nu)\hat{\omega}}} , \quad \hat{K}_f = \frac{\hat{K}_{Ic} \hat{\omega}}{\sqrt{1+4(1-\nu) \hat{\omega}}} ,
 \label{SSKI5}
\end{equation}
Noting the definition of $\hat{F}(\hat{p}_0)$ from \eqref{Fdef}, relationship \eqref{w0ss5} immediately yields:
\begin{equation}
(4\nu-3)\hat{p}_0^2 + 2\pi (1-\nu)(1-2\nu) \hat{p}_0 + \pi^2 (1-\nu)^2 \left[1-\frac{L_0 \hat{K}_{Ic}^2}{\hat{w}_0^2}\right] = 0.
 \label{SSp05}
\end{equation} 
Finally, asymptotic analysis of \eqref{Nobel_SSFluidMass5} and \eqref{Nobel_SSElasticity8} reveals that, provided the fluid leak-off at the crack tip is finite $\hat{q}_l(1)<\infty$, the second asymptotic coefficients can be obtained using the relations:
\begin{equation}
\hat{w}_1 = \frac{2\hat{p}_0}{2 - k_1 \hat{p}_0} \left[ L_0 - \frac{k_1}{4} \hat{q}_l (1) \right], \quad \hat{p}_2 = \frac{\hat{p}_0}{\hat{w}_0} \left[ \hat{q}_l (1) - 4\hat{w}_1 \right].
\end{equation}

\end{appendix}
\end{document}